%% file: main.tex
\documentclass[letterpaper,twocolumn,10pt]{article}
\usepackage{usenix-2020-09}
\usepackage[available,functional,reproduced]{usenixbadges}

\usepackage{tikz}
\usepackage{amsmath}

\input{macro}

\begin{document}

\date{}

\title{\Large \bf Automated Side Channel Analysis of Media Software with Manifold Learning
\thanks{The extended version of the USENIX Security 2022 paper~\cite{yuan2022automated}.}}

\author{
  {\rm Yuanyuan Yuan, Qi Pang, Shuai Wang\thanks{Corresponding author.}}\\
  The Hong Kong University of Science and Technology\\
  \textit{\{yyuanaq,  qpangaa, shuaiw\}@cse.ust.hk}
}

\twocolumn
\maketitle

\begin{abstract}
The prosperous development of cloud computing and machine learning as a service
has led to the widespread use of \ms\ to process confidential media data. This
paper explores an adversary's ability to launch side channel analyses (SCA)
against \ms\ to reconstruct confidential media inputs. Recent advances in
representation learning and perceptual learning inspired us to consider the
reconstruction of media inputs from side channel traces as a
\textit{cross-modality manifold learning} task that can be addressed in a
unified manner with an autoencoder framework trained to learn the mapping
between media inputs and side channel observations. We further enhance the
autoencoder with \textit{attention} to localize the program points that make the
primary contribution to SCA, thus automatically pinpointing information-leakage
points in \ms. We also propose a novel and highly effective defensive technique
called \textit{perception blinding} that can perturb media inputs with
perception masks and mitigate manifold learning-based SCA.

Our evaluation exploits three popular \ms\ to reconstruct inputs in image,
audio, and text formats. We analyze three common side channels --- cache
bank, cache line, and page tables --- and userspace-only cache set accesses
logged by standard \pp. Our framework successfully reconstructs high-quality
confidential inputs from the assessed \ms\ and automatically pinpoint their
vulnerable program points, many of which are unknown to the public. We further
show that perception blinding can mitigate manifold learning-based SCA with
negligible extra cost.
\end{abstract}

\input{introduction}
\input{background}

\input{overview}
\input{design}

\input{attack-setup}

\input{evaluation}

\input{real-attack}

\input{discussion}

\input{related}
\input{conclusion}


\bibliographystyle{plain}
\bibliography{bib/main}

\begin{appendix}
\input{hypothesis}
\input{blinding-appendix}
\input{attack-setup-appendix}
\input{real-attack-appendix}
\input{representation}

\input{case-study}
\input{mitigation}

\input{noise}

\end{appendix}

\end{document}

%% file: macro.tex
\usepackage{graphicx}
\usepackage[many]{tcolorbox}

\usepackage{mathtools,amssymb,latexsym,amsfonts,stmaryrd}
\usepackage{pbox}
\usepackage{xfrac}
\usepackage{booktabs}
\usepackage{xcolor}
\usepackage{threeparttable}
\usepackage{amsthm}
\usepackage{hyperref}
\usepackage{multirow}
\usepackage{tikz}
\usepackage{mathrsfs}
\usepackage{mathpartir}
\usepackage{algorithm}
\usepackage{url}
\usepackage{syntax}
\usepackage{framed}
\usepackage[noend]{algpseudocode}
\usepackage{flushend}
\usepackage{listings}
\usepackage{moresize}
\usepackage{wrapfig}
 
\usepackage{seqsplit}
\usepackage{alltt}
\usepackage{xspace}
\usepackage{enumitem}
\usepackage{pifont}

\DeclareMathAlphabet{\mathcal}{OMS}{cmsy}{m}{n}

\input{defs.tex}

\usepackage{thmtools}

\declaretheoremstyle[spaceabove=\topsep,notefont=\normalfont\itshape]{mystyle}

\newcommand{\revise}[2]{{\color{red}{\ifx&#1&\else- #1\fi}} {\color{ForestGreen}{\ifx&#2&\else+ #2\fi}}}%
\renewcommand{\revise}[2]{#2}%

\usetikzlibrary{arrows,patterns, decorations.pathreplacing}

\newcommand{\F}{Fig.}
\newcommand{\E}{Eq.}
\newcommand{\T}{Table}
\newcommand{\appx}{Appx.}
\renewcommand{\S}{Sec.}

\newcommand{\ms}{media software}

\newcommand{\MS}{Media Software}
\newcommand{\ignore}[1]{}

\newcommand{\ffmpeg}{\texttt{FFmpeg}}
\newcommand{\libjpeg}{\texttt{libjpeg}}
\newcommand{\victima}{\textbf{Victim$_{1}$}}
\newcommand{\victimb}{\textbf{Victim$_{2}$}}
\newcommand{\hunspell}{\texttt{Hunspell}}
\newcommand{\pp}{\texttt{Prime+Probe}}
\newcommand{\spy}{\texttt{spy}}
\newcommand{\victim}{\texttt{victim}}
\newcommand{\pin}{\texttt{Pin}}

\lstdefinestyle{base}{
  moredelim=**[is][\color{red}]{@}{@},
  escapeinside={<@}{@>}
}

\usepackage{amssymb}
\usepackage{pifont}

\newcommand\DejaVuttfamily{%
  \fontfamily{DejaVuSansMono-TLF}\selectfont }

\lstdefinestyle{base}{
  moredelim=**[is][\color{red}]{@}{@},
  escapeinside={<@}{@>}
}

\lstdefinelanguage
   [x64]{Assembler}     
   [x86masm]{Assembler} 
   {morekeywords={CDQE,CQO,CMPSQ,CMPXCHG16B,JRCXZ,LODSQ,MOVSXD, %
                  POPFQ,PUSHFQ,SCASQ,STOSQ,IRETQ,RDTSCP,SWAPGS, %
                  rax,rdx,rcx,rbx,rsi,rdi,rsp,rbp, %
                  r8,r8d,r8w,r8b,r9,r9d,r9w,r9b}} 

\usepackage{listings}
\usepackage{fancyvrb}
\usepackage{color}
\definecolor{lightgray}{rgb}{.9,.9,.9}
\definecolor{darkgray}{rgb}{.4,.4,.4}
\definecolor{purple}{rgb}{0.65, 0.12, 0.82}
\definecolor{commentgreen}{RGB}{63,127,95}

\colorlet{myPurple}{blue!40!red}
\definecolor{myOrange}{RGB}{255,192,0}

\lstdefinelanguage{Solidity}{
  keywords={len,delete,int,void,payable, public, event, contract, typeof, new, true, false, catch, function, return, null, catch, switch, var, if, in, while, do, else, case, break,struct,const,socklen_t,sa_familty_t,char,sockaddr},
  keywordstyle=\color{violet}\bfseries,
  ndkeywords={class, export, boolean, throw, implements, import, this},
  ndkeywordstyle=\color{darkgray}\bfseries,
  identifierstyle=\color{black},
  sensitive=false,
  comment=[l]{//},
  escapeinside={(*@}{@*)},          
  morecomment=[s]{/*}{*/},
  commentstyle=\color{commentgreen}\ttfamily,
  stringstyle=\color{red}\ttfamily,
  morestring=[b]',
  morestring=[b]"
}

\newcommand{\rnum}[1]{\uppercase\expandafter{\romannumeral #1\relax}}

\usepackage{xcolor}

\algnewcommand{\LeftComment}[1]{\Statex \(\triangleright\) #1}

\definecolor{pptbrown}{RGB}{132,60,12}
\definecolor{pptgreen}{RGB}{56,87,35}
\definecolor{pptblue}{RGB}{0,176,240}
\definecolor{pptred}{RGB}{192,0,0}

\let\OLDthebibliography\thebibliography
\renewcommand\thebibliography[1]{
  \OLDthebibliography{#1}
  \setlength{\parskip}{0pt}
  \setlength{\itemsep}{0pt plus 0.1ex}
}

\definecolor{pptgreen}{RGB}{84,130,53}
\definecolor{pptred}{RGB}{176,35,24}

\definecolor{pptgreen1}{RGB}{78,173,91}
\definecolor{pptred1}{RGB}{192,0,0}

\definecolor{pptyellow1}{RGB}{255,242,204}
\definecolor{pptgreen2}{RGB}{226,240,217}

\newcommand\sbullet[1][.5]{\mathbin{\vcenter{\hbox{\scalebox{#1}{$\bullet$}}}}}

%% file: defs.tex
\usepackage{amsmath, listings, amsthm, amssymb, proof, xspace}



%% file: introduction.tex
\section{Introduction}
\label{sec:introduciton}

Side channel analysis (SCA) infers program secrets by analyzing the target
software's influence on physical computational characteristics, such as the
execution time, accessed cache units, and power consumption. Practical SCA
attacks have been launched on real-world crypto
systems~\cite{yarom2017cachebleed,Liu15,Wu12} to recover crypto keys. With the
adoption of cloud computing and machine learning as a service (MLaaS), \ms, a
type of application software used for processing media files like images and
text, is commonly involved in processing private data uploaded to cloud (e.g.,
for medical diagnosis).
Existing works have exploited \ms\ with extensive manual efforts or reconstruct
only certain media data~\cite{xu2015controlled,hahnel2017high,yuan2021private}.
However, the community lacks a systematic and thorough understanding of SCA
attack vectors for \ms\ and of the ways that private user inputs of various
types (e.g., images or text) can be reconstructed in a unified and automated
manner. Hence, this is the first study toward \ms\ of various input formats to
assess how their inputs, which represent private user data, can be leaked via
SCA in a fully automatic way.

Recent advances in representation learning and perceptual
learning~\cite{bengio2013representation,zhu2018image} inspired us to recast SCA
of \ms\ as a cross-modality manifold learning task in which an
autoencoder~\cite{hinton1994autoencoders} is used to learn the mapping between
confidential media inputs and the derived side channel traces in an end-to-end
manner. The autoencoder framework can learn a low-dimensional joint manifold of
media data and side channel observations to capture a highly expressive
representation that is generally immune to noise.

Our proposed autoencoder framework is highly flexible. It converts side channel
traces into latent representations with an encoder module $\phi_{\theta}$, and
the media data in image, audio and text formats can be reconstructed by
assembling decoders $\psi_{\theta}$ that correspond to various media data
formats to $\phi_{\theta}$. Furthermore, by enhancing encoder $\phi_{\theta}$
with attention~\cite{woo2018cbam}, the autoencoder framework can automatically
localize program points that make primary contributions to the reconstruction of
media inputs. That is, the attention mechanism delivers a ``bug detector'' to
locate program points at which information can leak.

Further, the observation that manifold learning captures key perceptions of
high-dimensional data in a low-dimensional space~\cite{zhu2018image} inspired us
to propose the use of \textit{perception blinding} to mitigate manifold
learning–based SCA. Well-designed perception blinding ``dominates'' the
projected low-dimensional perceptions and thus confines adversaries to only
generate media data perceptually bounded to the mask. In contrast, \ms\ that is
typically used to process data bytes of media data experiences no extra
difficulty in processing the blinded data and recovering the original outputs.

Our evaluation exploits \ms, including \libjpeg~\cite{libjpeg},
\ffmpeg~\cite{ffmpeg}, and \hunspell~\cite{hunspell}, widely used to process
media data in image, audio, and text formats. We assess these \ms\ with regard
to a common threat model in which \textit{trace-based}
attackers~\cite{wang2017cached,kim2019make,cacheaudit,brotzman2019casym} can log
a trace of CPU cache banks, cache lines, or OS page-table entries accessed
during the execution of \ms. Moreover, we also launch standard
\pp\ attack~\cite{Tromer10} in userspace-only scenarios and use the logged cache
side channels to reconstruct media data. We conduct qualitative and quantitative
evaluations of six datasets that represent daily media data whose user privacy
can be violated if leaked to adversaries. Our findings show that user inputs can
be reconstructed automatically and that the recovered media content, such as
images or text, shows considerable (visual) similarity to user inputs. The
attention modules facilitate localizing program points that incur input leakage;
some have been disclosed
before~\cite{xu2015controlled,hahnel2017high,yuan2021private}, but many, to the
best of our knowledge, were previously unknown. Further, we find that perception
blinding is highly effective in mitigation of manifold learning–based SCA. We
also demonstrate the noise resiliency of our attack, and how oblivious
RAM~\cite{goldreich1987towards,shi2020path} can mitigate our attack, though it
incurs high cost and becomes impractical in real-life usage. In summary, this
thorough study makes the following contributions:

\begin{itemize}
\item Advances in cross-modality manifold learning inspired us to advocate SCA
  of media software as a supervised task that learns a joint
  manifold of media data and side channel traces. High-quality media data can be
  reconstructed from side channel traces in a noise-resilient manner without
  knowledge of the underlying \ms\ implementation or media data formats.

\item We enhance autoencoder with attention to localize program points that make
  notable contributions to information leakage. Furthermore, we design a
  low-cost perception-blinding technique that effectively mitigates the proposed
  SCA exploitation.

\item Our evaluation subsumes widely used \ms\ used to process images, audio,
  and text. We demonstrate that high-quality user inputs in various formats can
  be reconstructed and that perception blinding predominantly impedes our SCA.
  Our attention-based error-localization technique confirms some program points
  that have been reported as vulnerable and flags many previously unknown
  problems in \ms.
%
\end{itemize}

To facilitate result verification and future research, we released all code and
data generated in this research at~\cite{snapshot}.

%% file: background.tex
\section{Background}
\label{sec:background}

We introduce the high-level procedure of launching SCA in which program inputs
are assumed confidential. Let a deterministic and terminating program be $P$.
Executing an input $i \in I$ can be modeled as $P: I \rightarrow R$, where $R$
denotes the program behavior during the runtime. Although modern computer
architectures prohibit attackers from directly recording $R$ and inferring input
$i \in I$, attackers can leverage various \textit{side channels}, which map the
runtime behavior of $R$ into an adversarial observation $O$ of certain
properties (e.g., cache status) in the execution context of $P$. The attacker's
view can be represented as $view: R \rightarrow O$, where given side channel
observation $O$, the attackers leverage composite inverse function $(view \circ
P)^{-1}: O \rightarrow I$ to map $O$ back to input $i \in I$.

Promising progress has been made by logging (high-resolution) side channels such
as accessed cache line, cache bank, or page table entries in an automated
manner~\cite{chiang2015virtual,Wu12,Liu15,xu2015controlled,hahnel2017high}.
Nevertheless, reconstruction of $i$ from logged side channels requires attackers
to infer the composite inverse function $(view \circ P)^{-1}: O \rightarrow I$.
Recovery of such mappings requires an in-depth understanding of how program
secrets are propagated (i.e., secret information flow), which could require
considerable manual efforts~\cite{xu2015controlled,hahnel2017high} or conducting
formal analysis~\cite{wang2017cached,cacheaudit,bortzman2018casym}. Note that
high-resolution side channels (e.g., cache line access) usually contain millions
of records, but only a tiny portion $o^{*}$ is indeed
\textit{input-dependent}~\cite{wang2019identifying}.

\smallskip
\noindent \textbf{SCA on \MS.}~Despite the widespread adoption of MLaaS to
process users' private data, the SCA of media software has not been thoroughly
examined. For instance, \ms\ is commonly used to process X-ray images because it
allows cost-efficient disease diagnosis with cloud resources. However, the
leakage of such images on the cloud (e.g., via cache-based side
channels~\cite{liu2016cache}) involves a high risk of violating patient privacy.
An immense demand exists to gain insights into the extent of privacy problems in
media software, given its pervasive use in processing private data. Therefore,
we examine real-world \ms\ used to process media data such as photos and daily
conversations.

\smallskip
\noindent \textbf{Threat Model and Attack Scenarios.}~This study reconstructs
confidential inputs of \ms\ from side channels. We thus reasonably assume that
different inputs of targeted media software can induce distinguishable memory
access traces. Otherwise, no information regarding inputs would be leaked.

Profiled
SCA~\cite{heuser2012intelligent,maghrebi2016breaking,cagli2017convolutional,hettwer2018profiled,kim2019make,hettwer2020encoding}
commonly assumes that side channel logs have been prepared for training and data
reconstruction. For our scenario, we generally assume a standard
\textit{trace-based} attacker. We assume that a trace of system side channel
accesses made by the victim software has been prepared for use. Our evaluated
system side channels include cache line, cache bank, and page table entries. The
feasibility of logging such fine-grained information has been demonstrated in
real-world
scenarios~\cite{xu2015controlled,yarom2017cachebleed,hahnel2017high,disselkoen2017prime},
and this assumption has been consistently made by many previous
works~\cite{wang2017cached,Doychev16,gorka2017side,jan2018microwalk,brotzman2019casym,wang2019identifying}.
In this study, we use Intel Pin~\cite{pin} to log memory access traces and
convert them into corresponding side channel traces (see \S~\ref{sec:attack}).

We also benchmark userspace-only scenarios where attackers can launch
\pp\ attack~\cite{Tromer10} to log cache activities when \ms\ is processing a
secret input. We use Mastik~\cite{yarom2016mastik}, a micro-architectural
side channel toolkit, to launch ``out-of-the-box'' \pp\ and log victim's L1I and
L1D cache activities. We pin victim process and \spy\ process on the same CPU
core; see attack details in \S~\ref{subsec:real-world}.

Exploiting new side channels is \textit{not} our focus. We demonstrate our
attack over commonly-used side channels. This way, our attack is shown as
practically approachable, indicating its high impact and threats under
real-world scenarios.
Unlike previous SCA on \ms~\cite{xu2015controlled,hahnel2017high} or on crypto
libraries~\cite{Zhang12}, we do \textit{not} require a ``white-box'' view (i.e.,
source code) of victim software. We automatically analyze
\ms\ \textit{executables} with different input types. As will be discussed in
\S~\ref{sec:evaluation}, we launch manifold learning to reconstruct media data
with excellent (visual) similarity to user inputs. Many studies have only
flagged program points of information leakage with (unscalable) abstract
interpretation or symbolic
execution~\cite{wang2017cached,brotzman2019casym,cacheaudit}. Direct
reconstruction of media data is beyond the scope of such formal method-based
techniques, and these studies did not propose SCA mitigation.

%% file: overview.tex
\section{A Manifold View on SCA of \MS}
\label{sec:manifold}

This study recasts the SCA of \ms\ as a cross-modality manifold learning task
that can be well addressed with supervised learning. We train an
autoencoder~\cite{hinton1994autoencoders} that maps side channel observations
$O$ to the media inputs $I$ of \ms. Our threat model (\S~\ref{sec:background})
assumes that attackers can profile the target \ms\ and collect side channel
traces derived from many inputs. Therefore, our autoencoder framework is trained
to learn from historical data and implicitly forms a low-dimensional joint
manifold between the side channel logs and media inputs. We first introduce the
concept of manifold, which will help to clarify critical design decisions of our
framework (see \S~\ref{sec:design}).

\smallskip
\noindent \textbf{Manifold Learning.}~~The use of manifold underlies the
feasibility of dimensionality reduction~\cite{lee2007nonlinear}. The key premise
of manifold is the \textit{manifold hypothesis}, which states that real-world
data in high-dimensional space are concentrated near a low-dimensional
manifold~\cite{fan2019spherereid}. That is, real-world data often lie in a
manifold $\mathcal{M}$ of much lower dimensionality $d$, which is embedded in
its high-dimensional space $\mathcal{R}$ of dimensionality $D$ ($d \ll D$).
\textit{Manifold learning} aims to find a projection $f: \mathcal{R} \rightarrow
\mathcal{M}$ that converts data $x \in \mathcal{R}$ into $y$ in an intrinsic
coordinate system of $\mathcal{M}$.\footnote{``Intrinsic coordinate'' denotes
  the coordinate system of the low-dimensional manifold space for each
  high-dimensional data sample~\cite{lin2008riemannian,costa2004geodesic}.}
$f^{-1}$ projects $f(x) \in \mathcal{M}$ back onto representation $x$ in the
high-dimensional space $\mathcal{R}$.

PCA~\cite{abdi2010principal} is a linear manifold learning algorithm that aims
to find $\mathcal{M}$ by extracting ``principal components'' of data
points~\cite{bengio2013representation}. However, most real-world manifolds are
nonlinear, and manifold learning algorithms (e.g., ISOMAP) are proposed to
project data $x$ onto nonlinear $\mathcal{M}$~\cite{balasubramanian2002isomap}.

Manifold learning views high-dimensional media data $x \in \mathcal{R}$ as a
composite of perceptually meaningful contents that are shown as robust to noise
or other input
perturbations~\cite{fefferman2016testing,zhu2016generative,zhu2018image}.
Manifold learning algorithms extract expressive representations of
high-dimensional data such as images, audio, and
text~\cite{chang2003manifold,he2005face}, which explains why AI models can make
accurate predictions pertaining to high-dimensional
data~\cite{bengio2013representation}. It is shown that data of the same class
(e.g., face photos) generally lie in the same manifold, whereas data of
different classes (face vs. vehicle photos) are concentrated on separate
manifolds in low-dimensional space~\cite{thorstensen2009manifold}. manifold
learning clarifies the inherent difficulty of designing \textit{universal}
encoding and generative models applicable to high-dimensional data from
different manifolds. The manifold hypothesis has been verified theoretically and
empirically in a comprehensive
manner~\cite{fefferman2016testing,zhu2016generative,zhu2018image}.
\appx~\ref{sec:misc} presents our exploration on the validity of manifold
hypothesis.

\smallskip
\noindent \textbf{Parametric Manifold.}~Most manifold learning schemes adopt
non-parametric approaches. Despite the simplicity, non-parametric approaches
cannot be used to project \textit{new} data points in $\mathcal{R}$ onto
$\mathcal{M}$. Recent advances in deep neural networks, particularly
autoencoders, have enabled a parametric nonlinear manifold projection
$f_{\theta}: \mathcal{R} \rightarrow \mathcal{M}$~\cite{zhu2018image}. Manifold
learning can thus process unknown data points of high-dimensional media
data~\cite{zhu2018image,zhu2016generative,fu2015zero,holden2015learning,li2017unsupervised,bengio2013representation}
and facilitate downstream tasks like face recognition~\cite{fan2019spherereid}.

\begin{figure}[!ht]
    \centering
    \includegraphics[width=0.7\linewidth]{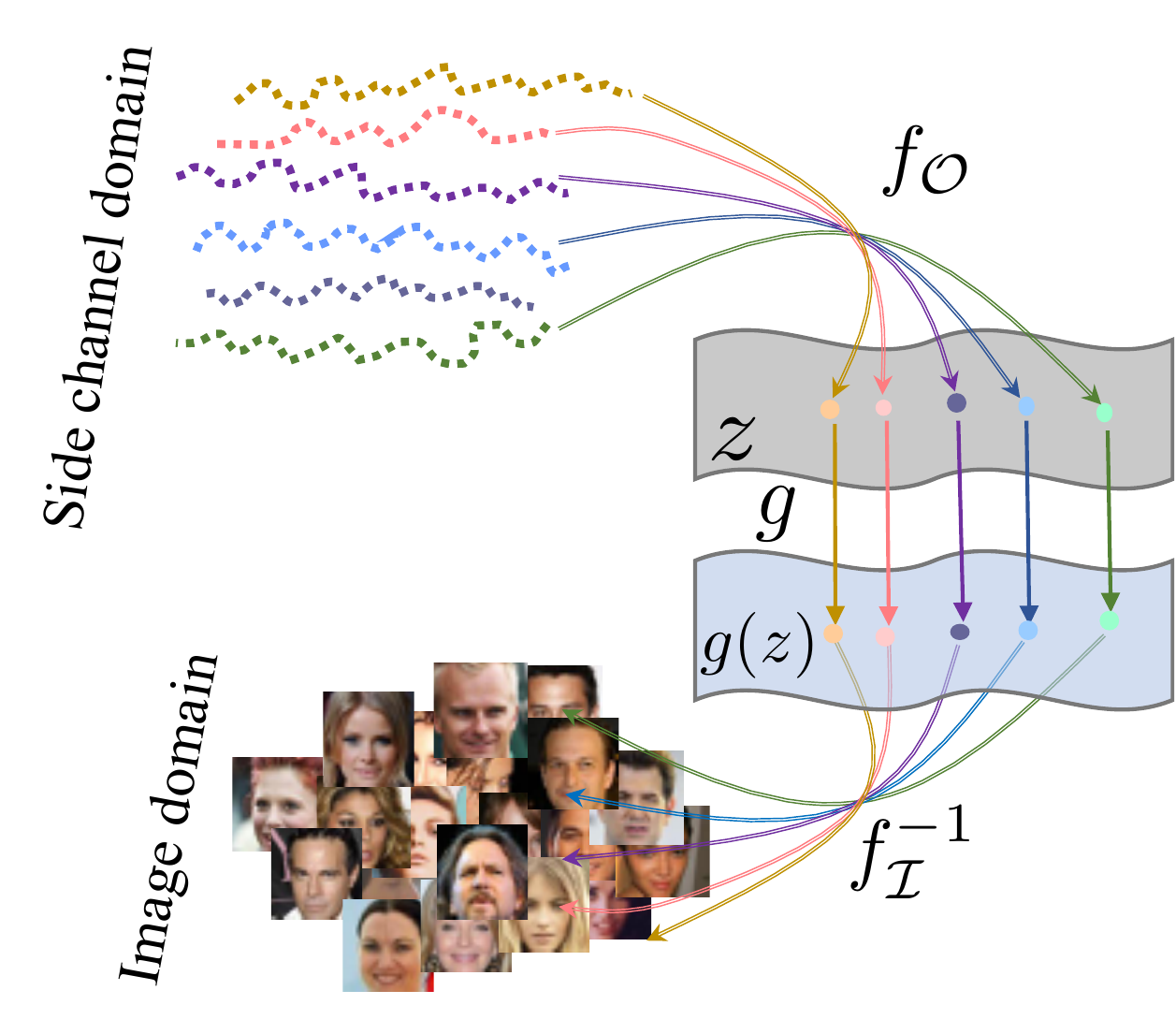}
    \caption{Mapping between side channels and images via a low-dimensional
      joint manifold $\mathcal{M}_{\mathcal{I},\mathcal{O}} = \mathcal{I} \times
      \mathcal{O}$.}
    \label{fig:cross-modality}
\end{figure}

\begin{figure*}[!ht]
    \centering
    \includegraphics[width=0.95\linewidth]{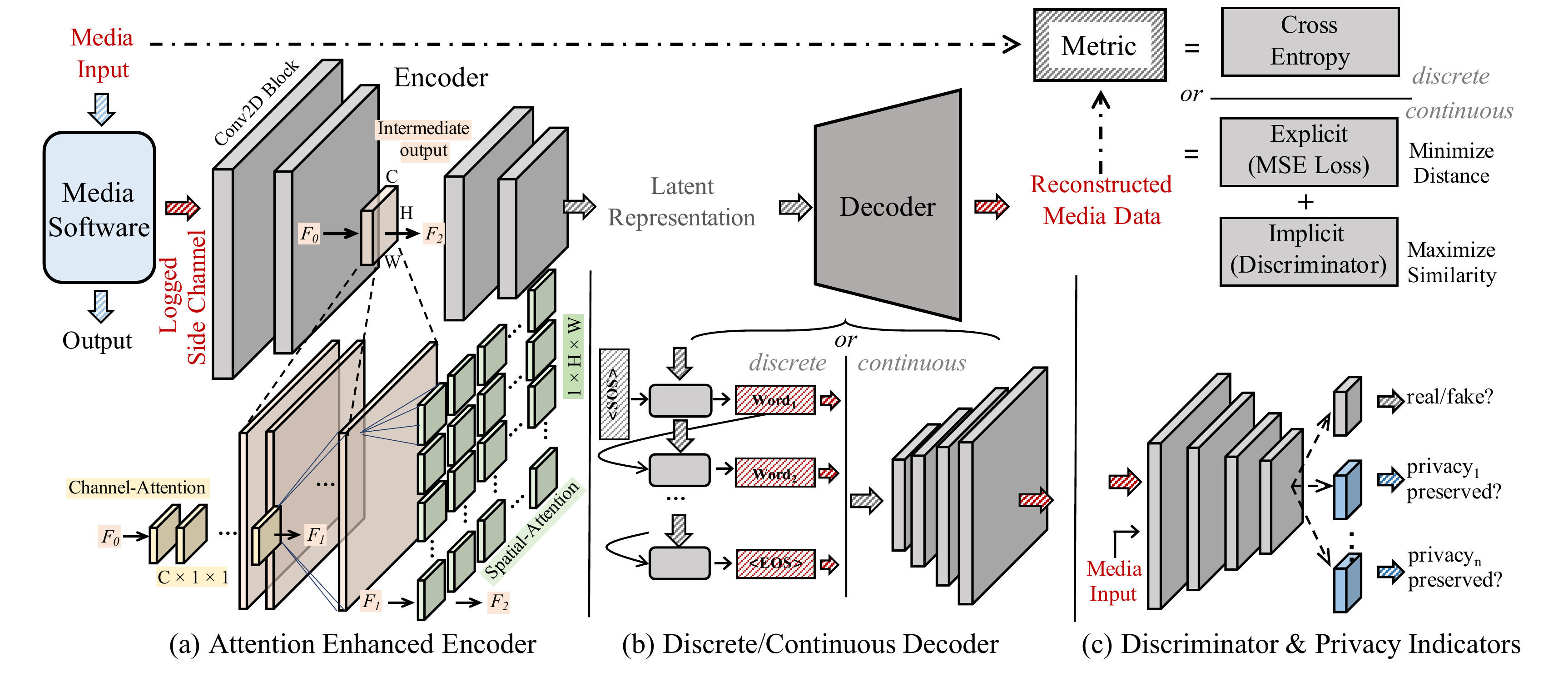}
    \caption{Reconstructing media data of different types with a unified
      autoencoder framework.}
    \label{fig:workflow}
\end{figure*}

\subsubsection*{High-Level Research Overview}

Processing media data has an observable influence on the underlying computing
environment; it thus induces side channel traces that can be logged by attackers
to infer private inputs. Previous SCA studies, from a holistic view, attempted
to (manually) map side channel logs to \textit{data bytes} in media data
(similar to how \ms\ treats media data)~\cite{xu2015controlled,hahnel2017high};
reconstruction of media data in a per-pixel manner is thus error-prone and
likely requires expertise and manual efforts.

The success of manifold learning in tasks like image editing and cross-modality
reconstruction~\cite{zhu2016generative,zhu2018image} led us to construct a
joint, \textit{perception-level} connection between side channel logs and
high-dimensional media inputs.\footnote{Perception-level connection means
  constraints on data bytes formed by perceptual contents (e.g., gender, hair
  style) in media data are extracted from side channels.} Therefore, instead of
deciding value of each byte, reconstructing media data is recast into exploring
the manifold of media data that satisfies the perception-level combinatory
constraints.

Overall, we view SCA as a cross-modality high-dimensional data reconstruction
task that is addressed with joint manifold learning in this
work~\cite{zhu2018image}. Aligned with the notations in \S~\ref{sec:background},
let the \ms\ inputs be $I$; the attacker's observation on executing each input
$i \in I$ can be represented as $(view \circ P): I \rightarrow O$, where $o \in
O$ denotes the observation of side channel traces. Let $F$ be the composite
function $view \circ P$. According to the manifold hypothesis, we assume that
$I$ and $O$ also lie in the unknown manifolds $\mathcal{I}$ and $\mathcal{O}$,
respectively. As mentioned in our threat model (\S~\ref{sec:background}), we
assume that side channel observations depend on the inputs of \ms; therefore,
the entire joint dataset $\{i_{i}, o_{i}\}$ formed by the $i$th media input
$i_{i} \in I$ and the corresponding $i$th observation $o_{i} \in O$ lies in a
joint manifold

$$\mathcal{M}_{\mathcal{I},\mathcal{O}} = \{(i, F(i)) | i \in \mathcal{I}, F(i)
\in \mathcal{O} \}$$

\noindent where $(i, F(i))$ is described with the regular high-dimensional
coordinate system. Since $I$ and $O$ also lie in the corresponding manifolds
$\mathcal{I}$ and $\mathcal{O}$, the data points in
$\mathcal{M}_{\mathcal{I},\mathcal{O}}$ should be equivalently described using
an intrinsic coordinate system $(z, g(z))$. Hence, we assume the existence of a
homomorphic mapping $(f_{\mathcal{I}}, f_{\mathcal{O}})$ over $(z, g(z))$ such
that $z = f_{\mathcal{O}}(o)$ and $g(z) = f_{\mathcal{I}} \circ F^{-1}(o)$.
$f_{\mathcal{O}}$ maps side channel observation $\mathcal{O}$ onto the intrinsic
coordinate $z$, whereas $f_{\mathcal{I}}$ maps high-dimensional media data $I$
onto $g(z)$. Note that $g$ denotes the diffeomorphism (i.e., an isomorphism of
two manifolds) between the $\mathcal{I}$ and $\mathcal{O}$
manifolds~\cite{zhu2018image}. Hence, instead of computing $F^{-1} = (view \circ
P)^{-1}: O \rightarrow I$ to map the side channel observation back onto the
media inputs, we leverage the joint manifold to constitute the following
composite function:

\begin{equation}
\begin{aligned}
F^{-1}(o) = f^{-1}_{\mathcal{I}} \circ g \circ f_{\mathcal{O}}(o)
\label{formula:composition}
\end{aligned}
\end{equation}

$i \in I$ can thus be reconstructed using the inverse composite function
$f^{-1}_{\mathcal{I}} \circ g \circ f_{\mathcal{O}}$ over the joint manifold
$\mathcal{M}_{\mathcal{I} , \mathcal{O}}$. \F~\ref{fig:cross-modality} provides
a summary and presents a schematic view of how $I$ and $O$ of high-dimensional
data are mapped via $\mathcal{M}_{\mathcal{I},\mathcal{O}}$.

The feasibility of using neural networks, especially autoencoders, to facilitate
parametric manifold learning has been discussed~\cite{zhu2018image,
  zhu2016generative, holden2015learning, martinez2019studying}. Accordingly, we
train an autoencoder by encoding side channel traces $O$ onto the latent space
with encoder $\phi_{\theta}$ and by generating media data $I$ with decoder
$\psi_{\theta}$ from the latent space. Therefore, \E~\ref{formula:composition}
can be learned in an end-to-end
manner~\cite{zhu2018image,bengio2013representation}. Holistically,
$\phi_{\theta}$ and $\psi_{\theta}$ correspond to $f_{\mathcal{O}}$ and
$f^{-1}_{\mathcal{I}}$, respectively, whereas $g$ is implicitly constructed in
the encoded latent space.




%% file: design.tex
\section{Framework Design}
\label{sec:design}

We describe the design of our autoencoder in \S~\ref{subsec:design-attack}.
\S~\ref{subsec:design-attention} clarifies the usage of attention to localize
code fragments inducing information leakage. \S~\ref{subsec:design-defense}
introduces perception blinding to mitigate our SCA.

\input{design-sca}
\input{design-attention}
\input{design-mitigation}

%% file: design-sca.tex
\subsection{SCA with Autoencoder}
\label{subsec:design-attack}

We propose a general and highly-flexible design in which an autoencoder is used
to facilitate SCA of various media data, including images, audio and text. The
autoencoder framework~\cite{hinton1994autoencoders} defines a parametric
feature-extracting function $f_{\theta}$, named \textit{encoder}, that enables
the projection of the input $x$ onto a latent vector $h = \phi_{\theta}(x)$.
Similarly, autoencoder frameworks also use $\psi_{\theta}$ as a \textit{decoder}
that reconstructs input $\hat{x}$ from a latent vector $\hat{x} =
\psi_{\theta}(h)$. A well-trained autoencoder framework gradually identifies a
parameter vector $\theta$ to minimize the reconstruction error as follows:

$$L(\theta) = \sum_{t} L(x_{t},\psi_{\theta} \circ \phi_{\theta}(x_{t}))$$

\noindent where $x_{t}$ is a training sample. Minimal errors can be found with
statistical methods like stochastic gradient descent.

The first row of \F~\ref{fig:workflow} depicts the workflow. We clarify that our
focus is \textit{not} to propose novel model architectures; rather, we show that
high-quality inputs can be synthesized by assembling standard models, which
indicates severity and effectiveness of our attack. We now discuss the
high-level workflow and present the model structures and training details in
\S~\ref{subsec:implementation}.
Given a logged side channel trace $o \in O$, encoder $\phi_{\theta}(o)$ converts
$o$ into the corresponding latent representation. We prepare three decoders
$\psi^{i}_{\theta}$, $\psi^{a}_{\theta}$, $\psi^{t}_{\theta}$ to reconstruct
these types of media data (i.e., image, audio, and text) from the encoded latent
representation. We pair encoder $\phi_{\theta}(o)$ with each $\psi^{*}_{\theta}$
and train the assembled pipeline for our customized objective functions
$L(\theta)$. Our proposed framework is task-agnostic. Generating media data of
various types requires only assembling corresponding decoders to the unified
encoder $\phi_{\theta}$.

\smallskip
\noindent \textbf{Encoder $\phi_{\theta}$.}~A logged side channel trace will
first be folded into a $K \times N \times N$ matrix (see
\T~\ref{tab:data-statistics} for the detailed configuration of each trace). We
then feed this matrix as the input of encoder $\phi_{\theta}$. The encoder
$\phi_{\theta}$ comprises several stacked 2D convolutional neural networks
(CNNs). For the current implementation, $\phi_{\theta}$ converts the
high-dimensional inputs into latent vectors of 128 dimensions, given that the
dimensions of our media inputs are all over 10K. See
\appx~\ref{sec:input} for clarification on how side channels, including both
Intel Pin- and \pp-logged records, are represented and processed by
$\phi_{\theta}$.
Moreover, we find that increasing the dimension of latent vectors (i.e., from
128 to 256) does not make an observable improvement. This observation is
consistent with the manifold hypothesis~\cite{lee2007nonlinear}, such that only
\textit{limited} ``perceptions'' exist in normal media data. In contrast,
reducing the number of dimensions (e.g., 32) makes the outputs (visually) much
worse. However, users who strive to recover media data of lower-dimensions can
configure our framework with smaller latent vectors (e.g., 32 dimensions).

\F~\hyperref[fig:workflow]{2(a)} shows that we enhance encoder $\phi_{\theta}$
with attention. Indeed, we insert one attention module between every two stacked
CNN layers in the encoder. Attention generally improves output quality of
autoencoder~\cite{vaswani2017attention}. More importantly, attention facilitates
localizing program points of information leakage. We elaborate on
\F~\hyperref[fig:workflow]{2(a)} in \S~\ref{subsec:design-attention}.

\smallskip
\noindent \textbf{Decoder $\psi^{*}_{\theta}$.}~~We categorize the media data
exploited by this study into two types: continuous and discrete. Image and audio
data are represented as a continuous floating-point matrix and reconstructed by
$\psi^{i}_{\theta}$ and $\psi^{a}_{\theta}$ in a continuous manner. In contrast,
textual data comprise word sequences, and because there is no ``intermediate
word,'' textual data are regarded as sequences of discrete values and handled by
$\psi^{t}_{\theta}$.

As shown in \F~\hyperref[fig:workflow]{2(b)}, we use a common approach to
stacking 2D CNNs to design $\psi^{i}_{\theta}$. A 2D CNN has several
convolutional kernels; each kernel focuses on one feature dimension of its input
and captures the spatial information of this feature dimension. Images can thus
be reconstructed from vectors in the low-dimensional latent space with stacked
2D CNNs, as each 2D CNN upsamples from the output of the previous layer. For
audio data, we first convert raw audio into the log-amplitude of Mel spectrum
(LMS), a common 2D representation of audio data. As will be shown in
\F~\ref{fig:audio}, audio data are represented as 2D images, in which the x-axis
denotes time and the y-axis denotes the log scale of amplitudes at different
frequencies. Herein, like $\psi^{i}_{\theta}$, $\psi^{a}_{\theta}$ uses stacked
2D CNNs to process each converted 2D image, gradually upsamples from the latent
representation, and reconstruct the LMSs of the audio data. Because the LMSs
usually are not in square-shape, we append a fully connected layer to transform
the shape of the reconstructed LMSs. These LMSs are then converted to raw audio
losslessly.

Textual data, however, are reconstructed sequentially ``word by word'' due to
their discrete nature. As shown in \F~\hyperref[fig:workflow]{2(b)}, to
reconstruct a sentence from the latent space of a side channel trace $o$, a
single word is gradually inferred based on words already inferred from sentence
$i$. Following a common practice of training sequence-to-sequence autoencoders,
we add a start-of-sequence (SOS) token before each sentence $i$ and an
end-of-sequence (EOS) token after $i$. Then, given a side channel trace $o$ that
corresponds to unknown text $i$, the trained decoder $\psi^{t}_{\theta}$ starts
from the SOS token and predicts a word $w \in i$ sequentially until it yields
the EOS token. From a holistic perspective, the trained model projects a
sentence $i$ into a low-dimensional manifold space of \textit{word dependency},
which facilitates the gradual inference of each word $w$ on $i$.

\smallskip
\noindent \textbf{Designing Objective Functions.}~As depicted in the first row
of \F~\ref{fig:workflow}, for discrete data (i.e., text), each decode step is a
multi-class classification task where the output is classified as one element in
a pre-defined dictionary. Thus, we use \textit{cross entropy} as the training
objective. For continuous data, we design the training objective $L_{\theta}$
\textit{composing both explicit and implicit metrics}. We now introduce each
component in detail.

\smallskip
\noindent \underline{Explicit Metrics}~A common practice in training an
autoencoder is to explicitly assess the point-wise distance between the
reconstructed media input $i'$ and reference input $i$ with metrics such as MSE
loss, L1 loss, and KL
divergence~\cite{kullback1951information,chernov2000study}. The autoencoder will
be guided to gradually minimize the point-wise distance $L_{\theta}(i,
\psi_{\theta} \circ \phi_{\theta}(o))$ during training. Nevertheless, a major
drawback of such explicit metrics is that the loss of each data point is
calculated \textit{independently} and contributes \textit{equally} to update
$\theta$ and minimize $L_{\theta}$. Our preliminary study (see
\F~\ref{fig:oversmooth} in \appx~\ref{sec:case}) shows that such explicit metrics suffer
from ``over-smoothing''~\cite{saito2017statistical}, a well-known problem that leads to quality degradation
of the reconstructed data.

\smallskip
\noindent \underline{Implicit Metrics}~Another popular approach is to assess the
``distributed similarity''~\cite{saito2017statistical} of reconstructed $i'$ and
reference input $i$. Viewing the general difficulty of extracting the distribution
of arbitrary media data, a common practice is to leverage a neural discriminator
$D$. Discriminator $D$ and decoder $\psi_{\theta}$ play a zero-sum game, in
which $D$ aims to distinguish the reconstructed input $i'$ from normal media
data $i$. In contrast, decoder $\psi_{\theta}$ tries to make its output $i'$
indistinguishable with $i$ to fool $D$. Although this paradigm generally
alleviates the obstacle of ``over-smoothing''~\cite{saito2017statistical}, it
creates the new challenge of mode collapse; that is, $\psi_{\theta}$ generates
realistic (albeit very limited) $i'$ from any inference inputs. From a holistic
perspective, the use of a discriminator mainly ensures that the reconstructed
$i'$ is near $i$ from a \textit{distribution} perspective; no guarantee is
provided from the view of a single data point.

\begin{table}[t]
  \caption{Privacy-aware indicators. \T~\ref{tab:data-statistics} introduces each dataset.}
  \label{tab:indicator}
  \centering
  \resizebox{0.85\linewidth}{!}{
  \begin{tabular}{l|c}
    \hline
     \textbf{Dataset} & \textbf{Indicator} \\
    \hline
     CelebA & Is the celebrity's identity preserved? \\
    \hline
     ChestX-ray & Is the disease information preserved? \\
    \hline
     SC09 & Is the speaker's identity preserved? \\
    \hline
     Sub-URMP & Is the musical instrument's type preserved? \\
    \hline
  \end{tabular}
  }
\end{table}

\smallskip
\noindent \underline{Privacy-Aware Indicators}~In addition to the two standard
objective functions mentioned above, we further take into account a set of
\textit{privacy-aware indicators}. As shown in \F~\hyperref[fig:workflow]{2(c)},
we extend discriminator $D$ such that it checks whether the reconstructed
outputs preserve the ``privacy'' in an explicit manner. \T~\ref{tab:indicator}
lists the privacy indicators used in our framework, which correspond to
exploited media data of different types. For instance, for face photos (CelebA),
we specify checking the identity. Hence, the enhanced discriminator $D$ serves
as a classifier to check whether the identity of the person is preserved, which
thus forces decoder $\psi_{\theta}$ to decode the identity information in the
zero-sum game. A specific fully connected layer is appended to the discriminator
$D$ in accordance with each privacy indicator.

\smallskip
\noindent \textbf{Comparison with Generative Model-Based
  SCA~\cite{yuan2021private}.}~~One contemporary study~\cite{yuan2021private}
uses generative models (e.g., GANs~\cite{goodfellow2014generative}) to
conduct SCA towards image libraries by capturing image distribution from
side channels. Nevertheless, their work is particularly designed to recover images
instead of proposing a general and flexible framework to exploit media software of
various input types.
In addition, we explicitly use privacy indicators when designing objective
functions, while~\cite{yuan2021private} focuses on polishing the \textit{visual
  appearance} of the reconstructed images.

\begin{figure}[!ht]
    \centering
    \includegraphics[width=0.55\linewidth]{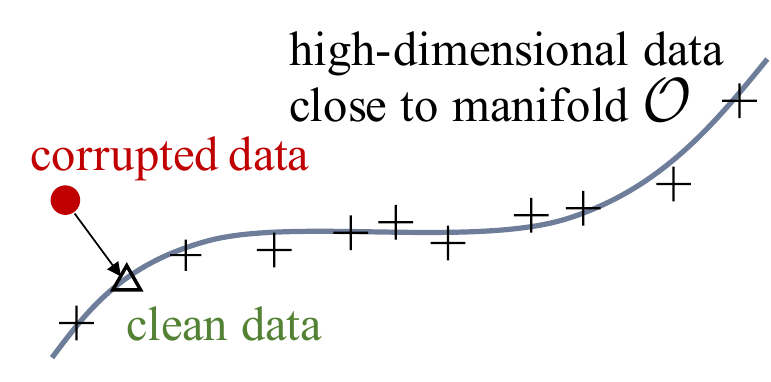}
    \caption{Denoising corrupted data during manifold learning.}
    \label{fig:denoising}
\end{figure}

\smallskip
\noindent \textbf{Noisy Side Channel.}~~Reconstructing media data from noisy
side channels is of particular importance, because adversaries often face
considerable noise in real-world attack scenarios. Manifold learning features
denoising by design, the schematic view of which is presented in
\F~\ref{fig:denoising}.
Overall, manifold learning forces side channel traces $O$ to concentrate near
the learned low-dimensional manifold $\mathcal{O}$, where a corrupted
high-dimensional data point $\tilde{o}$ (\textcolor{pptred}{$\sbullet[1.25]$} in
\F~\ref{fig:denoising}) should typically remain \textit{orthogonal} to the
manifold $\mathcal{O}$~\cite{bengio2013representation}. Thus, when the decoder
$\psi_{\theta}$ learns to reconstruct media data $i \in I$ from the
representations lying on the joint manifold, corrupted $\tilde{o}$ can be fixed
by first being projected onto the $\Delta$ in the manifold for denoising; and
$i$ can then be reconstructed from the $\Delta$~\cite{holden2015learning}.

%% file: design-attention.tex
\subsection{Fault Localization with Neural Attention}
\label{subsec:design-attention}

Some studies have detected software vulnerabilities that lead to side channel
attacks~\cite{cacheaudit,wang2017cached,bortzman2018casym,wang2019identifying}.
However, we note that such studies typically use heavyweight program-analysis
techniques, such as abstract interpretation, symbolic execution, and constraint
solving. Thus, performing scalable program analysis of real-world media software
could prove a great challenge, given that such media software usually contains
complex program structures (e.g., nested loops) and a large code base.
Furthermore, the primary focus of previous studies has been crypto libraries
(e.g., OpenSSL~\cite{openssl}), whose ``sensitive data'' are private key bytes
or random numbers. In contrast, modeling potentially lengthy media data with
various strictly defined formats could impose a further challenge (e.g.,
symbolizing such complex input formats) that may require the incorporation of
domain-specific knowledge.

Inspired by advances in program neural
smoothing~\cite{she2019neuzz,she2020neutaint} and SCA based on neural
networks~\cite{yuan2021private,picek2018performance,kim2019make,yang2018convolutional},
we seek to overcome question ``which program point leaks side channel
information'' by answering the following question:

\begin{quote}
``Which records on a logged side channel trace contribute most to the
  reconstruction of media data?''\footnote{See \appx~\ref{sec:input} for trace representation.}
\end{quote}

Although answering the former question often requires rigorous and unscalable
static analysis, the second question can be addressed smoothly by extending the
encoder $\phi_{\theta}$ with \textit{attention}~\cite{woo2018cbam}, a
well-established mechanism that improves the representation of interest by
telling the neural network where and upon what to focus. In particular, by
enhancing the autoencoder with attention, our framework \textit{automatically}
flags side channel logs that make a primary contribution to input
reconstruction. These logs are \textit{automatically} mapped to the
corresponding memory access instructions. We can then \textit{manually} identify
the corresponding ``buggy'' source code. For the last step, our current
experiments rely on symbol information in the assembly programs to first
identify corresponding functions in source code, and then narrow down to code
fragments inducing input leakage.


Despite attention being a standard mechanism to boost deep learning
models~\cite{vaswani2017attention,woo2018cbam}, attention in our new scenario
acts like a ``bug detector'' to principally ease localizing vulnerable program
points. In contrast to program analysis-based
approaches~\cite{cacheaudit,wang2019identifying,wang2017cached,bortzman2018casym},
our solution is highly scalable and incurs no extra cost during exploitation.
Moreover, it analyzes software in a black-box setting that is agnostic to
\ms\ implementation details or input formats.

\F~\hyperref[fig:workflow]{2(a)} depicts the enhanced trace encoder with
attention. An attention module (we follow the design in~\cite{woo2018cbam} given
its simplicity and efficiency) is inserted within every two stacked CNN layers.
Let the intermediate input of a CNN layer as $C \times H \times W$, the
\colorbox{pptyellow1}{``Channel-Attention''} module $A_{channel}$ processes each
segment of $1 \times H \times W$ data points from $C$ channels and tells the
encoder ``where'' to focus on by assigning different weights to each segment.
The \colorbox{pptgreen2}{``Spatial-Attention''} module $A_{spatial}$ processes
each segment of $C \times 1 \times 1$ records and advises the encoder ``what to
locate'' by assigning different weight on each record. From a holistic
perspective, attention module $A_{channel}$ projects a coarse-grained focus on
potentially interesting segments, while $A_{spatial}$ further identifies
interesting side channel records in a segment.\footnote{It is well accepted that
  a CNN is organized in the form of $\texttt{num_channels} \times \texttt{width}
  \times \texttt{height}$. Therefore, we name two attention components as
  $A_{channel}$ and $A_{spatial}$, which are aligned with the convention.}



%% file: design-mitigation.tex
\subsection{Mitigation with Perception Blinding}
\label{subsec:design-defense}

This section presents mitigation against manifold learning–enabled SCA
(\S~\ref{subsec:design-attack}). Consistent with our attack and
fault-localization, mitigation is also agnostic to particular \ms\ and input
types.
We only need perturb the media input $I$ with pre-defined perception blinding
masks.

We first introduce blinding images of the human face, and then explain how to
extend perception blinding toward other input types. \appx~\ref{sec:blind-appx}
presents the workflow of perception blinding in real systems and discusses
application scope.

\smallskip
\noindent \textbf{A Working Example.}~As introduced in \S~\ref{sec:background},
manifold learning casts images of the human face into a set of perceptually
meaningful representations; typical representations include hair style, age, and
skin color. Hence, we define a \textit{universal} mask $i_{mask}$ of human face,
such that by perturbing arbitrary images $i$ of human face with $i_{mask}$, the
produced output $i_{blinded}$ will be primarily projected to the same intrinsic
coordinates $z_{mask}$ in the manifold space $\mathcal{M}$. To use perception
blinding, users only need to pick \textit{one} mask $i_{mask}$ to blind all
input images $i$.
Consequently, adversaries are restricted to the generation of media data
perceptually correlated to $z_{mask}$. Particularly, to perturb $i$, we add
$i_{mask}$ as follows:

$$i_{blinded} = \alpha \times i \oplus \beta \times i_{mask}$$

\noindent where we require $\beta \gg \alpha$ and $\alpha + \beta = 1$. 
Perceptual contents of $i_{mask}$ thus ``dominates'' the projected low-dimensional
perceptions in $\mathcal{M}$. Let $P(i_{blinded})$ be the output of \ms\ after
processing $i_{blinded}$, and the user can recover the desired output by
subtracting $P(i_{mask})$ from the output as follows:

$$P(i_{private}) = \frac{1}{\alpha} \times (P(i_{blinded}) \ominus \beta \times
P(i_{mask}))$$

\noindent where $P(i_{private})$ is the desired output, and $P(i_{mask})$ can be
pre-computed. $\oplus$ and $\ominus$ directly operate $i \in I$ of various formats, as
will be defined later in this section.
Because typical operations of \ms\ (e.g., compression) are \textit{independent}
of the perceptual meaning of media inputs, the proposed blinding scheme
introduces no extra hurdle for \ms. In contrast, as shown in
\S~\ref{subsec:eval-mitigation}, SCA based on manifold learning can be mitigated
in a highly effective manner.

\begin{table*}[t]
\caption{Side channels derived from a memory access made by victim \ms\ using
  address $addr$.}
\label{tab:attack}
\centering
\resizebox{0.90\linewidth}{!}{
\begin{tabular}{l|l}
\hline
{\bf Side Channel Name}  & {\bf Side Channel Record Calculation} \\
\hline
CPU Cache Bank Index~\cite{yarom2017cachebleed}    & $addr \gg L$ where $L$, denoting cache bank size, is usually 2 on modern computer architectures. \\
CPU Cache Line Index                               & $addr \gg L$ where $L$, denoting cache line size, is usually 6 on modern computer architectures. \\
OS Page Table Index~\cite{hahnel2017high,Yarom14}  & $addr \mathbin{\&} (\sim M)$ where $M$, denoting \texttt{PAGE\_MASK}, is usually 4095 on modern computer architectures. \\
\hline
\end{tabular}
}
\end{table*}

\begin{table*}[t]
  \centering
  \caption{Statistics of side channel traces and media software. There is \textit{no} overlapping between training and testing data.}
	\label{tab:data-statistics}
	\resizebox{1.01\linewidth}{!}{
  \begin{tabular}{l|c|c|c|c|c|c}
    \hline
   \textbf{Dataset} & \textbf{Information} & \textbf{Training Split} & \textbf{Testing Split}   & \textbf{Trace Length} & \textbf{Matrix Encoding}  & \textbf{Media Software}             \\ 
    \hline
     CelebA~\cite{liu2015faceattributes} & Large-scale celebrity face photos & 162,770 & 19,962 & $338,123 \pm 14,264$  & $6 \times 256 \times 256$                      & \texttt{libjpeg} (ver. 2.0.6)   \\ 
     ChestX-ray~\cite{wang2017chestx} & Hospital-scale chest X-ray images & 86,524 & 25,596     & $329,155 \pm 10,186$ & $6 \times 256 \times 256$                      &        LOC: 103,273                \\
    \hline                                                                                     
     SC09~\cite{warden2018speech} & Human voice of saying number 0--9 & 18,620 & 2,552          & $1,835,067 \pm 103,328$ & $8 \times 512 \times 512$                    & \texttt{ffmpeg} (ver. 4.3)     \\
     Sub-URMP~\cite{li2018creating} & Sound clips of 13 instruments & 71,230 & 9,575            & $1,678,485 \pm 36,122$ & $8 \times 512 \times 512$                    &        LOC: 1,236,079             \\
    \hline
     COCO~\cite{lin2014microsoft} & Image captions & 414,113 & 202,654                          &  $77,796 \pm 14$ & $6 \times 128 \times 128$                       & \texttt{hunspell} (vers. 1.7.0)   \\ 
     DailyDialog~\cite{li2017dailydialog} & Sentences of daily chats & 11,118 & 1,000           & $77,799 \pm 102$ & $6 \times 128 \times 128$                       &           LOC: 39,096               \\
    \hline
  \end{tabular}
  }
\end{table*}

\smallskip
\noindent \textbf{Requirement of $i_{mask}$.}~Comparable to how RSA blinding is
used to mitigate timing channels~\cite{brumley2005remote}, perception blinding
is specifically designed to mitigate manifold learning-based SCA. We require
that $i_{mask}$ must lie in the same low-dimensional manifold with the private
data. Thus, $i_{mask}$ must manifest high \textbf{perception correlation} with
media software inputs $i_{private} \in I$. This shall generally ensure two
properties: 1) the privacy (in terms of certain perceptions, such as gender and
skin color) in $i_{private}$ can be successfully ``covered'' by $i_{mask}$, and
2) $i_{mask}$ imposes nearly no information loss on recovering $P(i_{private})$
from $P(i_{blinded})$ except a mild computational cost due to mask
operations.
%
Considering \F~\ref{fig:denoising}, when violating this requirement of
\textit{perception correlation}, for instance, such as by using random noise to
craft $i_{mask}$, the intrinsic coordinate of the original input ($\Delta$) can
likely drift to a ``corrupted input'' (\textcolor{pptred}{$\sbullet[1.25]$})
that is mostly orthogonal to the manifold of $I$. As explained in
\S~\ref{subsec:design-attack}, due to the inherent noise resilience of manifold
learning, crafting such a corrupted input can cause less challenge to attackers
when recovering $i$ from the low-level manifold space. Although $i_{private}$ is
of low weight in $i_{blinded}$, it can still be reconstructed to some extent, as
will be shown in \F~\ref{fig:blinding} of \S~\ref{subsec:eval-mitigation}.

\smallskip
\noindent \textbf{Extension to other data types.}~For image and audio data, we
recommend using a normal image $i \in I$ as the mask $i_{mask}$. Intuitively, by
amplifying $i_{mask}$ with a large coefficient $\beta$ in generating
$i_{blinded}$, $i_{mask}$ is presumed to dominate the perceptual features in
$i_{private}$. Hence, we stealthily hide the private perceptual features of
$i_{private}$ in $i_{blinded}$, whose contents are difficult for adversaries to
disentangle without knowing $i_{mask}$. For textual data, we recommend inserting
notional words of high frequency to blind $i_{private}$. We present empirical
results on how various choices of $i_{mask}$ can influence the mitigation
effectiveness in \S~\ref{subsec:eval-mitigation}.

\smallskip
\noindent \textbf{Implementation of Operators $\oplus$ And $\ominus$.}~For image
and audio data, we use floating-point number addition and subtraction to
implement $\oplus$ and $\ominus$. Textual data are discrete: considering that
\ms\ often manipulates textual data at the word level, simply ``adding'' or
altering words in the input text will likely trigger some error handling
routines of the corresponding \ms, which is not desirable.
\S~\ref{subsec:design-attack} clarifies that our autoencoder framework
essentially captures the ``word dependency'' between words in a sentence;
accordingly, we define the $\oplus$ operation as inserting words in a sentence,
whereas the $\ominus$ operation is implemented to remove previously inserted
words. As shown in \S~\ref{subsec:eval-mitigation}, this strategy effectively
breaks the word dependency in the original text.

%% file: attack-setup.tex
\section{Attack Setup}
\label{sec:attack}

We leverage three high-resolution side channels, as shown in
\T~\ref{tab:attack}. As clarified in our threat model (\S~\ref{sec:background}),
these side channel are commonly adopted in previous works.
See \appx~\ref{sec:attack-appendix} for detailed setup of these side channels.
We clarify that exploiting new side channels is \textit{not} our focus. We use
common side channels in the era of cloud computing, implying the severity and
effectiveness of our proposed attack. The resolution when performing attacks on
those side channels are 4B, 64B, and 4096B, respectively. Higher-resolution side
channels should enable recovering media data with more vivid details. Media data
of better quality, however, does not necessarily enhance privacy stealing (e.g.,
determining whether chest X-Ray images indicate pneumonia). See quantitative
evaluation of privacy inference in \S~\ref{subsubsec:eval-quantitative}.

For evaluation in \S~\ref{sec:evaluation}, we use \pin~\cite{pin} to collect
memory access traces and map each trace into three side channel traces following
mapping rules in \T~\ref{tab:attack}. \S~\ref{subsec:real-world} further
demonstrates attack in an userspace-only scenario, i.e., we collect cache side
channels via \pp~\cite{Tromer10}.

\smallskip
\noindent \textbf{Media Software and Media
  Dataset.}~\T~\ref{tab:data-statistics} reports evaluated media software and
statistics of side channel traces. We pick \ms\ consistent with previous
works~\cite{xu2015controlled,hahnel2017high,yuan2021private}. All \ms\ are
complex real-world software, e.g., \ffmpeg\ contains 1M LOC. In contrast, crypto
libraries are usually much succinct, e.g., x86 core implementation of AES in
recent \texttt{OpenSSL} has about 3K LOC.
We prepare two common datasets for each \ms\ to comprehensively evaluate our
attack. All datasets contain daily media data that, once exposed to adversaries,
would result in privacy leakage. We compile all three media software into 64-bit
binary code using \texttt{gcc} on a 64-bit Ubuntu 18.04 machine. See
\appx~\ref{sec:attack-appendix} for details of these software and datasets.

\subsection{Implementation}
\label{subsec:implementation}

We implement our framework in Pytorch (ver. 1.4.0). We use the Adam optimizer
with learning rate as $0.0002$ for all models. Batch size is 64. For continuous
decoders, we set the loss function as $\lambda L_{\textit{exlicit}} +
L_{\textit{implicit}} + \sum_{i=1}^{n}L_{\textit{privacy}}$, where $\lambda =
50$ and $n$ is the number of privacy-aware indicators.
We ran experiments on Intel Xeon CPU E5-2683 with 256 GB RAM and one Nvidia
GeForce RTX 2080 GPU. For experiments based on \pp-logged traces
(\T~\ref{tab:data-statistics}), the training is completed at 100 epochs and
takes less than 24 hours. For experiments using \pp-logged traces, training and
takes shorter time (see \textbf{Running Time} in
\appx~\ref{subsec:real-attack-details}). \T~\ref{tab:data-statistics} reports
the dataset size and training/testing splits. See our released
codebase~\cite{snapshot} for result verification.

%% file: evaluation.tex
\begin{figure*}[!ht]
    \centering
    \includegraphics[width=1.0\linewidth]{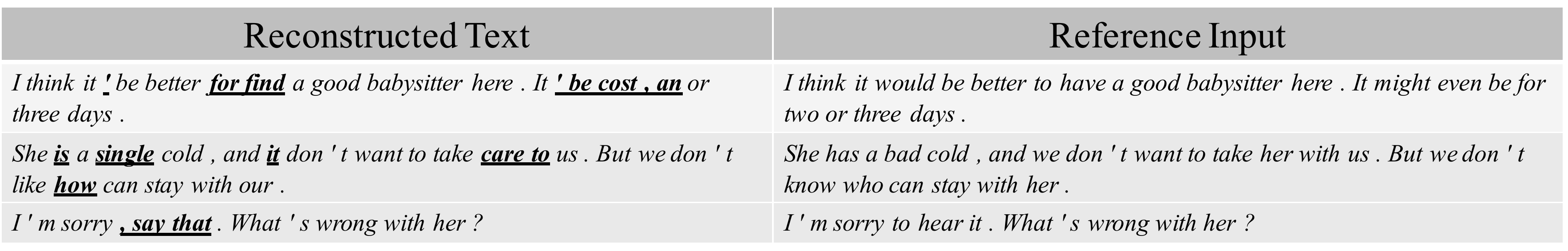}
    \caption{Qualitative evaluation of DailyDialog. We mark
      \underline{\textit{\textbf{inconsistent reconstructions}}}.}
    \label{fig:text}
\end{figure*}

\begin{figure}[!ht]
    \centering
    \includegraphics[width=1.0\linewidth]{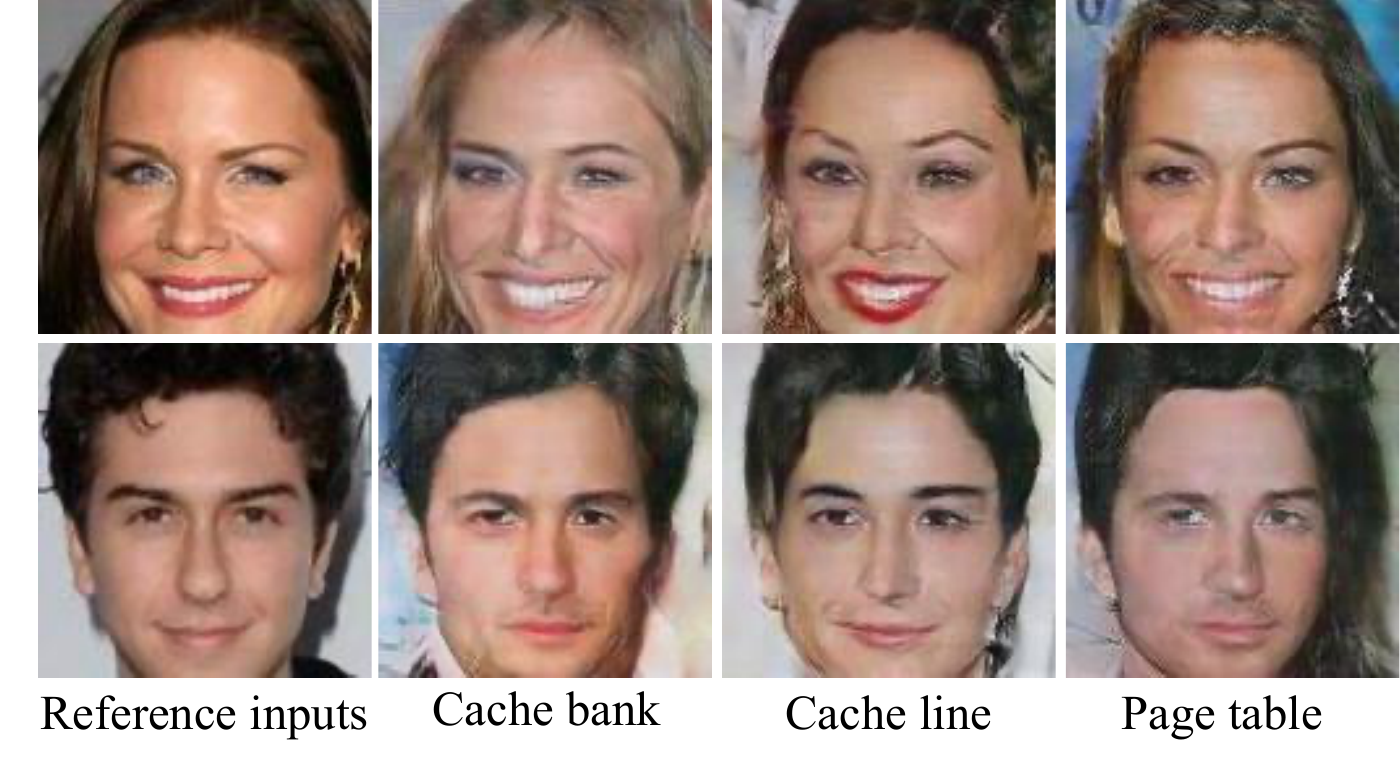}
    \caption{Qualitative evaluation of CelebA.}
    \label{fig:face}
\end{figure}


\section{Evaluation}
\label{sec:evaluation}

We present the SCA exploitation toward media software in
\S~\ref{subsec:eval-attack}. We discuss program points that induce information
leakage in \S~\ref{subsec:eval-localization}, and demonstrate the effectiveness
of perception blinding in \S~\ref{subsec:eval-mitigation}.

\subsection{Side Channel Attack}
\label{subsec:eval-attack}

This section reports evaluation results of our attack. Due to the limited space,
some setups are reported in \appx~\ref{sec:attack-appendix}. We present more
evaluation results in \appx~\ref{sec:case}.

\subsubsection{Qualitative Evaluation}
\label{subsubsec:eval-qualitative}

This section presents and compares the reconstructed media data with the
reference inputs in terms of various settings.
\F~\ref{fig:face} demonstrates that the reconstructed images and the references
show highly aligned visual appearances, including gender, eyebrow shapes, skin
color, and hair styles. Images constructed from different side channels manifest
comparable visual quality.
\F~\ref{fig:text} further reports the text reconstruction results of daily
dialogs by comparison with the reference inputs. The reconstructed sentences,
although are not fully aligned with the reference, still retain considerable
correct contents and the original intents.

We interpret the overall qualitative evaluation results, in terms of images and
text, as highly encouraging. We present reconstructed chest X-ray images,
sub-URMP/SC09 audio data, and COCO text in \appx~\ref{sec:case}. Promising
results can be consistently observed.

\subsubsection{Quantitative Evaluation}
\label{subsubsec:eval-quantitative}

\input{table}

\smallskip
\noindent \textbf{Image Data.}~For CelebA, we leverage commercial face
recognition APIs, Face++~\cite{facepp}, to decide whether a reconstructed face
and its reference input can be considered as from the \textit{same person} with
over 99.9\% confidence scores. We thus launch a de-anonymization attack of
user identity with reconstructed images. \T~\ref{tab:face} reports the
evaluation results; for all three exploited side channels, more than 43\% of the
reconstructed faces can be correctly matched to their reference inputs, showing
a high success rate of face matching. Only 2\% of the reconstructed images are
deemed as ``non-face,'' which indicates the negligible chance of generating
corrupted faces. Due to the limited space, we report the quantitative evaluation
of chest X-ray in \appx~\ref{sec:case}.

Our attack achieves plausible accuracy. The quantitative results are \textit{not}
noticeably affected by differences in side channels, which indicates that face
matching evaluation extracts representative attributes from images for matching.
%
As mentioned in \S~\ref{sec:attack}, the three side channels manifest different
resolutions: although higher-resolution side channels enable reconstruction of
more vivid images, this does not necessarily promote privacy stealing. However,
enabled by manifold learning-based autoencoder and our objective functions which
explicitly account for privacy indicators (\T~\ref{tab:indicator}),
privacy-related factors are extracted in the reconstructed images across side
channels of various resolutions. Similar observations are made for media data of
other formats; our discussion follows.

\smallskip
\noindent \textbf{Audio Data.}~~\T~\ref{tab:voice} reports the voice matching
results for SC09. Using the reconstructed voice commands (number 0--9), we train
two classifiers for speaker identity and command 0--9
classification.\footnote{Please refer to \appx~\ref{sec:attack-appendix} for
  details of these classifiers.} The evaluation results largely outperform the
baseline (i.e., random guessing). With a total of 184 speakers, we achieve
greater than 20\% accuracy in matching correct speaker identities across all
settings. We also exceed 20\% accuracy in content matching (0--9). We observed
decreasing accuracy in speaker identity matching, which is reasonable given that
the cache bank side channel only ``kicks off'' two least significant bits, while
cache line and page table side channels retain less amount of information.
\appx~\ref{sec:case} reports the matching rate of musical instruments in
\T~\ref{tab:audio1}, which yields mostly consistent and promising findings.

\smallskip
\noindent \textbf{Text Data.}~To reconstruct text data, we gradually predict
each word based on previously-predicted words in the sentence. Hence, for the
quantitative evaluation, we adopt an attack strategy mostly aligned
with~\cite{carlini2019secret} to measure the average accuracy of word-level
prediction accuracy. \T~\ref{tab:text} reports the evaluation results for the
COCO and Dailydialog datasets. To prepare a baseline for comparison, we feed a
random input to the decoder $\psi_{\theta}$ instead of using the latent vector
of an input side channel trace. As expected, our exploitation of both datasets
achieves much greater accuracy than the baseline regarding all side channels.

\input{eval-mitigation}

%% file: table.tex
\begin{table}[t]
  \caption{CelebA face image matching evaluation.}
  \label{tab:face}
  \centering
\resizebox{0.8\linewidth}{!}{
  \begin{tabular}{l|c|c|c}
    \hline
            & \textbf{Cache bank} & \textbf{Cache line} & \textbf{Page table} \\
    \hline
               same face & 45.4\% & 43.5\% & 44.5\% \\ 
    \hline
               non-face  & 2.0\% & 2.0\% & 2.1\% \\ 
    \hline
  \end{tabular}
  }
\end{table}

\begin{table}[t]
  \caption{SC09 human voice matching evaluation.}
  \label{tab:voice}
  \centering
\resizebox{0.9\linewidth}{!}{
  \begin{tabular}{l|c|c|c}
    \hline
            & \textbf{Cache bank} & \textbf{Cache line} & \textbf{Page table} \\
    \hline
     ID accuracy & 29.1\% & 28.8\% & 23.2\% \\ 
    \hline
    Content accuracy & 21.6\% & 24.2\% & 22.6\% \\ 
    \hline
  \end{tabular}
  }
\end{table}

\begin{table}[t]
  \caption{Text data inference evaluation.}
  \label{tab:text}
  \centering
\resizebox{1.0\linewidth}{!}{
  \begin{tabular}{l|c|c|c|c}
    \hline
     Dataset & \textbf{Cache bank} & \textbf{Cache line} & \textbf{Page table} & \textbf{Baseline} \\
    \hline
     COCO Caption & 43.4\% & 42.6\% & 42.1\%   & 0.0000\% \\ 
    \hline
     Daily Diolgue & 38.1\% & 37.4\% & 37.6\%  & 0.0183\% \\ 
    \hline
  \end{tabular}
  }
\end{table}

%% file: eval-mitigation.tex
\begin{figure}[t]
\centering
\begin{lstlisting}
static void idct32(int *coeffs, int @col_limit@) {
  <@int \textcolor{red}{limit} = min(H, \textcolor{red}{col_limit} + 4);@>
  for (int i = 0; i < H; i++) 
    TR_32(src, src, H, H, @limit@);
}
static void TR_32(int *dst, int *src, int dstep,
                  int sstep, int @end@) {
  int o_32[16] = { 0 };
  for (int i = 0; i < 16; i++)
    // loading pre-calculated matrix ``transform''
    <@for (int j = 1; \textcolor{red}{j} < \textcolor{red}{end}; j += 2)@>
      <@\textbf{o_32[i] += transform[\textcolor{red}{j}][i] * src[\textcolor{red}{j} * sstep];}@>
  // TR_16 calls TR_8, and TR_8 calls TR_4.
  <@TR_16(e_32, src, 1, 2 * sstep, SET, \textcolor{red}{end}/2);@>
}
\end{lstlisting}
\caption{Vulnerable code components in \ffmpeg. We mark variables depending on \ffmpeg's input in \textcolor{red}{red}, and \textbf{bold} input-dependent memory accesses (line 12).}
\label{fig:ffmpeg-vul}
\end{figure}

\begin{table}[t]
  \caption{Localized program points in \libjpeg.}
  \label{tab:libjpeg}
  \centering
\resizebox{1.0\linewidth}{!}{
  \begin{tabular}{c|c|c|c}
    \hline
     \textbf{Module}            & \textbf{\#Functions} & \textbf{Frequency}    & \textbf{Sample Func. Names} \\
     \hline
     \multirow{2}{*}{MCU}       &\multirow{2}{*}{2}    & \multirow{2}{*}{7,060} & \texttt{encode_mcu_huff}\\
                                &                      &                       & \texttt{decode_mcu}  \\
    \hline
     Transform                  &1                     & 5866                  & \texttt{jtransform_execute_transform}\\
     \hline
     \multirow{2}{*}{IDCT}      &\multirow{2}{*}{13}   & \multirow{2}{*}{4,027} & \texttt{jpeg_idct_15x15}\\
                                &                      &                       & \texttt{jsimd_idct_ifas}  \\
     \hline
     \multirow{2}{*}{Upsample}  &\multirow{2}{*}{15}   & \multirow{2}{*}{2,033} & \texttt{h2v1_merged_upsample}\\
                                &                      &                       & \texttt{h2v1_fancy_upsample}  \\
     \hline
     \multirow{2}{*}{Decompress}&\multirow{2}{*}{6}   & \multirow{2}{*}{1,352}  & \texttt{tjDecompress2}\\
                                &                      &                       & \texttt{tjDecompressHeader3}  \\
     \hline
     \multirow{2}{*}{Dump}      &\multirow{2}{*}{4}   & \multirow{2}{*}{8,23}   & \texttt{write_bmp_header}\\
                                &                     &                        & \texttt{start_input_bmp} \\
   \hline
  \end{tabular}
  }
\end{table}

\begin{table}[t]
  \caption{Localized program points in \ffmpeg.}
  \label{tab:ffmpeg}
  \centering
\resizebox{1.0\linewidth}{!}{
  \begin{tabular}{c|c|c|c}
    \hline
    \textbf{Module}         &  \textbf{\#Functions}    &\textbf{Frequency}            & \textbf{Sample Func. Names} \\
    \hline
    Encode                  & 50+   & 10K+         & \texttt{encode_frame}\\
    \hline
    Decode                  & 50+   & 10K+         & \texttt{decode_frame}\\
    \hline
    Filter                  & 50+   & 10K+         & \texttt{filter_frame}\\
    \hline
    \multirow{2}{*}{IDCT}   & \multirow{2}{*}{10+}   & \multirow{2}{*}{5K+}          & \texttt{idct_idct_32x32_add_ce}\\
                            &                       &                    &              \texttt{iadst_idct_16x16_add_c}  \\
    \hline
    \multirow{2}{*}{Dump}   & \multirow{2}{*}{10+}   & \multirow{2}{*}{5K+}          & \texttt{wav_write_trailer}\\
                            &                       &                    &              \texttt{wav_write_header}  \\
    \hline
  \end{tabular}
  }
\end{table}

\begin{table}[t]
  \caption{Localized program points in \hunspell.}
  \label{tab:hunspell}
  \centering
\resizebox{1.0\linewidth}{!}{
  \begin{tabular}{c|c|c|c}
    \hline
    \textbf{Module}            &  \textbf{\#Functions} &\textbf{Frequency}    & \textbf{Sample Func. Names} \\
    \hline
    Interface                  & 1    & 1,333 & \texttt{pipe_interface} \\
    \hline
    \multirow{2}{*}{Parser}    & \multirow{2}{*}{4}   & \multirow{2}{*}{1,230} & \texttt{next_token}, \texttt{alloc_token}\\
                               &                       &                      & \texttt{get_parser}, \texttt{TextParser::init} \\
    \hline
    Look up                    & 2                    & 213  &  \texttt{check}, \texttt{insertion_sort} \\
    \hline
    \multirow{2}{*}{Insert}    & \multirow{2}{*}{5}   & \multirow{2}{*}{1,076} & \texttt{putdic}, \texttt{allocate_string} \\
                               &                       &                      & \texttt{chenc}, \texttt{allocate_char_vector}  \\
    \hline
  \end{tabular}
  }
\end{table}

\begin{figure*}[!ht]
    \centering
    \includegraphics[width=0.95\linewidth]{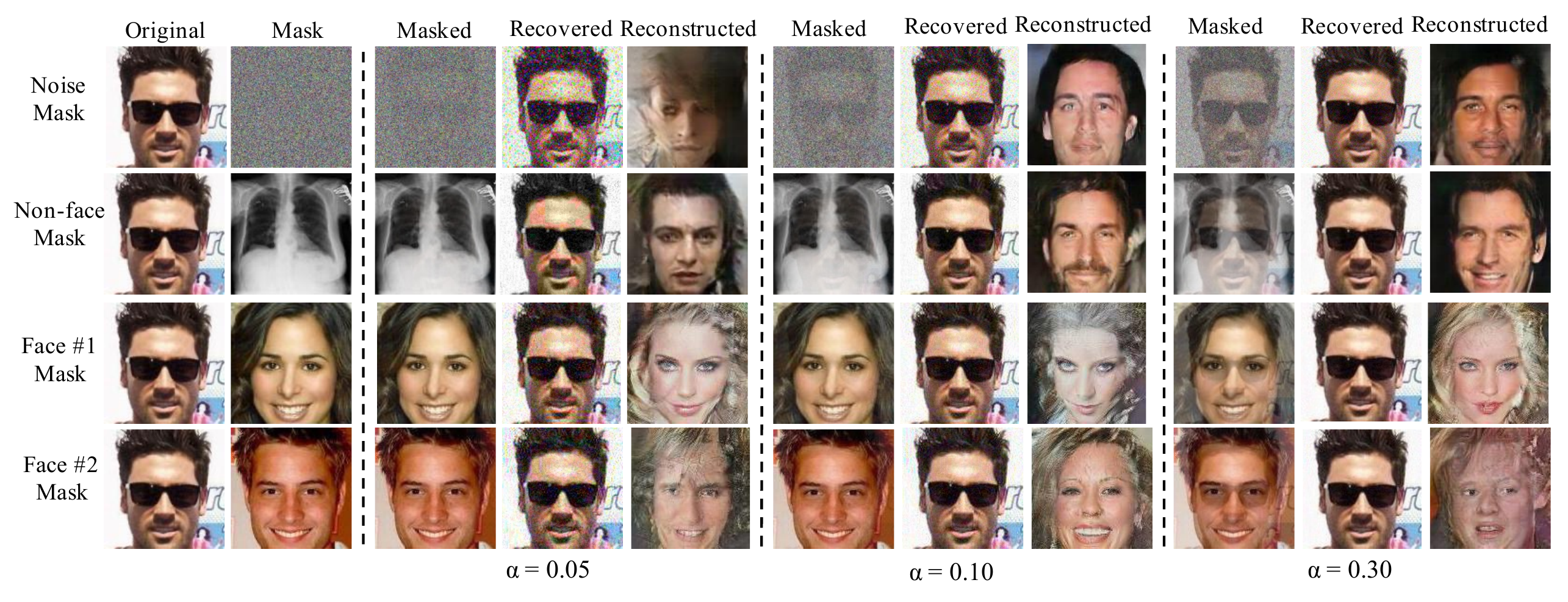}
    \caption{Qualitative evaluation results of perception blinding.}
    \label{fig:blinding}
\end{figure*}

\subsection{Program Point Localization}
\label{subsec:eval-localization}

We now discuss the localized buggy code of each \ms\ using attention. We present
a representative buggy code fragment of \ffmpeg\ in \F~\ref{fig:ffmpeg-vul}. We further
present representative buggy code fragments of \libjpeg\ and \hunspell\ in
\appx~\ref{sec:case}. We also list \textbf{all} localized program points in term
of assembly code in~\cite{snapshot}.

\smallskip
\noindent \textbf{\texttt{libjpeg}.}~We analyze 2,000 media inputs from the
CelebA and Chest X-ray datasets. \T~\ref{tab:libjpeg} reports the localization
results of \libjpeg. For instance, we identify 7,060 side channel points from
4,000 traces, which can be mapped back to two functions
(\texttt{encode_mcu_huff} and \texttt{decode_mcu}) performing minimum coded unit
(MCU)-related operations. Similarly, we find information leakage points in
modules related to decompression, IDCT, and also output dumping.

\T~\ref{tab:data-statistics} shows that \textit{one} trace has approximately
400K data points. In other words, \T~\ref{tab:libjpeg} reveals that a tiny
portion of ``informative'' points on a side channel trace make a primary
contribution to information reconstruction. Given an image compressed in JPEG,
\texttt{libjpeg} decompresses the image into a bitmap. It is pointed out that
the decompression process introduces side channels. IDCT-related functions that
were noted by~\cite{xu2015controlled} are automatically re-discovered by us. In
addition, we identify functions in other image transformation routines (e.g.,
MCU, unsampling) and output dumping routines that leak inputs. We manually
inspected the corresponding implementation of \libjpeg\ and confirmed our
findings. Note that our approach is automated and treats the entire
\libjpeg\ software as a blackbox, whereas previous
studies~\cite{xu2015controlled,hahnel2017high,lee2020off} could rely on expert
knowledge to first localize the vulnerable program points before launching SCA.

\smallskip
\noindent \textbf{\ffmpeg.}~We use 2,000 inputs from SC09 and Sub-URMP as the
inputs of \ffmpeg. Our findings, as reported in \T~\ref{tab:ffmpeg}, can be
mapped to five modules of \ffmpeg, each of which contains many functions.
\ffmpeg\ processes audio inputs with audio sampling (the sampling frequency is
set as 4,000 in our experiments). \F~\ref{fig:ffmpeg-vul} presents a vulnerable
program component in the IDCT module of \ffmpeg\ where the input ``taints''
certain variables (marked in \textcolor{red}{red}) and eventually influences the
memory accesses at line 12. Given different input values, different memory cells
are visited at line 12, resulting in access to different cache units or page
table entries. In addition, \texttt{TR_32}, \texttt{TR_8} and \texttt{TR_4} also
suffer from similar patterns (line 14).
To the best of our knowledge, side channel issues on \ffmpeg\ are rarely
studied, but the findings in \ffmpeg\ are conceptually similar to other \ms; for
example, IDCT algorithms and output dumping in both \ffmpeg\ and \libjpeg\ are
flagged as vulnerable by us.

\smallskip
\noindent \textbf{\hunspell.}~We use 2,000 inputs from each dataset (i.e.,
DailyDialog and COCO) to run \hunspell\ and analyze the logged side channels.
\T~\ref{tab:hunspell} reports the results, where information leakage points are
found from the interface, parser, and also spell checking . In fact, previous
works~\cite{xu2015controlled,lee2020off} have pointed out side channel issues of
\hunspell. \hunspell\ performs spell check, where a dictionary of words is
maintained as a hash table. \hunspell\ iterates each word $w$ in the input
sentence to check if $w$ is in the hash table, thus deciding the correctness of
its spelling. When checking each $w$, \hunspell\ computes the hash value of $w$
and looks up the corresponding hash bucket of words. This would lead to a
sequence of memory accesses, which can be potentially used to map back to word
$w$. Note that while previous works attacking \hunspell\ assumes the knowledge
of the dictionary~\cite{xu2015controlled,lee2020off} before attack, such
pre-knowledge is \textit{not} needed for our attack. Instead, we use side
channel traces and their corresponding sentences fed to \hunspell\ as the
training data to implicitly learn a mapping in the low-dimensional joint
manifold space.
\cite{xu2015controlled,lee2020off} reports that functions \texttt{lookup} and
\texttt{add_word} primarily leak inputs. Our manual confirmation shows that our
findings (e.g., \texttt{putdic}, \texttt{chenc}, \texttt{check},
\texttt{insertion_sort}) indeed invoke \texttt{lookup} and \texttt{add_word}
functions. We also find that the parser and interface (we use Linux utility
\texttt{echo} to feed \hunspell) of \hunspell\ also influence side channels,
both of which are not disclosed by previous works.

\smallskip
\noindent \textbf{Confirmation with the Developers.}~We have reported our
localized program points to the developers. By the time of writing, the FFmpeg
developers confirmed our findings. Nevertheless, they mentioned that
software-level fixing is undesirable, given the difficulty of writing side
channel-free code and the incurred extra performance penalty. From his
perspective, OS-level or hardware-level fixing seems more practical.

\subsection{Mitigation with Perception Blinding}
\label{subsec:eval-mitigation}

We benchmark the mitigation effectiveness in terms of quantitative and
qualitative analysis. We also discuss how different masks can influence the
mitigation.

\subsubsection{Qualitative Evaluation}
\label{subsubsec:eval-qualitative}

We report qualitative evaluation by comparing the reference inputs with the
reconstructed inputs after applying blinding. Due to the limited space,
\F~\ref{fig:blinding} only reports the perception blinding over a private face
image $i_{private}$ in terms of different settings. The original image
$i_{private}$ is presented in the ``Original'' column, and applied perception
masks are presented in the ``Mask'' column. For each masked image $i_{blinded}$,
the adversarial recovered images are presented in the ``Reconstructed'' columns,
and the final media software outputs after unblinding are given in the
``Recovered'' columns.

``Noise mask'' (the first row) and ``non-face mask'' (the second row) do not
seem helpful in blinding $i_{private}$ because features such as face orientation
are still preserved in the reconstructed images. However, the use of real face
images as the mask, as shown in the third and fourth columns, gives promising
results to blind key perceptual-level contents like hair color and skin color.
Overall, after blinding, the adversary-reconstructed images seem to show a
correlation with $i_{mask}$ instead of $i_{private}$. This is intuitive; as
clarified in \S~\ref{subsec:design-defense}, a large coefficient $\beta$ is
assigned to $i_{mask}$ such that the perception contents of $i_{mask}$ largely
determine the projected intrinsic coordinate in the manifold. This way, the
reconstructed images incline to manifest the perception of $i_{mask}$.

Additionally, although a small $\alpha$ value (e.g., 0.05) introduces a
non-trivial amount of noise in the final output, outputs of much better quality
can be recovered when $\alpha$ is set to 0.10 or even higher. We thus recommend
that users adopt a reasonably high $\alpha$ when constituting $i_{blinded}$.
More reconstructed cases are given in \appx~\ref{sec:mitigation}.

\begin{table}[t]
  \caption{Face matching results after blinding in terms of (cache bank/cache line/page table).}
  \label{tab:mitigation-face}
  \centering \resizebox{1.01\linewidth}{!}{
  \begin{tabular}{l|c|c|c}
    \hline
      Mask      & $\alpha = 0.05$ & $\alpha = 0.1$ & $\alpha = 0.3$ \\
    \hline
      Noise     & 27.5/28.6/27.8\% & 25.2/26.9/28.2\% & 26.6/27.5/29.0\% \\
    \hline
      Non-face  & 28.8/28.8/26.5\% & 26.2/27.6/27.4\% & 28.7/31.4/26.2\% \\
    \hline
      Face\#1   & 1.4/1.2/2.4\% & 1.8/1.4/2.7\% & 2.0/1.5/3.1\% \\
    \hline
      Face\#2   & 0.6/1.3/1.6\% & 0.7/1.7/1.9\% & 1.2/1.6/2.2\% \\
    \hline
  \end{tabular}
  }
\end{table}

\begin{table}[t]
  \caption{Mitigating COCO text inference attack in terms of (cache bank/cache
    line/page table). $\alpha = 0.05$, $\alpha = 0.1$, $\alpha = 0.3$ denote
    each word are appended with 19, 9, and 2 masks, respectively.}
  \label{tab:mitigation-text1}
  \centering
\resizebox{1.0\linewidth}{!}{
  \begin{tabular}{l|c|c|c}
    \hline
     \textbf{Mask} & \textbf{$\alpha = 0.05$} & \textbf{$\alpha = 0.1$} & \textbf{$\alpha = 0.3$} \\
    \hline
     ``man''     & 0.39/0.40/0.36\% & 0.68/0.69/0.67\% & 2.46/2.45/2.13\% \\ 
    \hline
     ``sitting'' & 0.16/0.16/0.22\% & 0.30/0.30/0.39\% & 1.36/1.40/1.39\%  \\ 
    \hline
  \end{tabular}
  }
\end{table}

\begin{table*}[t]
  \caption{Quantitative evaluation results using cache side channels logged by \pp. We also provide the processing time (\textit{ms}) when launching \pp\ (normal $\rightarrow$ with \pp).}
  \label{tab:pp}
  \centering
\resizebox{0.95\linewidth}{!}{
  \begin{tabular}{l|c|c|c|c|c|c}
    \hline
     & \multicolumn{2}{c|}{\ffmpeg\ \& SC09 voice matching} & \multicolumn{2}{c|}{\libjpeg\ \& CelebA face matching/non-face} & \multicolumn{2}{c}{\hunspell\ \& DDialog text matching} \\
    \hline
                & \textbf{Intel} & \textbf{AMD}         & \textbf{Intel} & \textbf{AMD} & \textbf{Intel} & \textbf{AMD} \\
    \hline                                                                              
     L1I Cache & 12.6\% (36 $\rightarrow$ 580)  & 11.6\% (10 $\rightarrow$ 420) & 38.0/0.85\% (5 $\rightarrow$  18)& 35.9/1.2\% (2 $\rightarrow$  22) & 33.9\% (60 $\rightarrow$ 130) & 33.2\% (23 $\rightarrow$ 60)  \\
    \hline                                                                              
     L1D Cache & 81.8\% (36 $\rightarrow$ 590)  & 15.7\% (10 $\rightarrow$ 420) & 36.9/0.80\% (5 $\rightarrow$ 20) & 33.9/0.90\% (2 $\rightarrow$ 24) & 32.2\% (60 $\rightarrow$ 130) & 31.8\% (23 $\rightarrow$ 60) \\
    \hline
  \end{tabular}
  }
\end{table*}

\subsubsection{Quantitative Evaluation}
\label{subsubsec:mitigation-qualitative}

We launch quantitative evaluation following the procedures in
\S~\ref{subsec:eval-attack}. The perception blinding of ``Noise'' and
``Non-face'' masks, as shown in \T~\ref{tab:mitigation-face}, reduces the
average success rates of face matching from approximately 44\%
(\T~\ref{tab:face}) to 27.4\%, but still has non-negligible privacy leakage.
Compared with ``Face\#1'' and ``Face\#2'', \F~\ref{fig:blinding} shows that
images reconstructed from ``Noise''- and ``Non-face''-blinded images manifest
better visual similarity with the reference inputs. In addition,
\T~\ref{tab:mitigation-face} reports that ``Face\#1'' and ``Face\#2'' exhibit
much better mitigation (less than 3.1\% matching rates) in terms of quantitative
metrics. We find that the value of $\alpha$ does \textit{not} notably influence
the results but is still positively correlated with privacy leakage. Overall,
for images, we mask the perception contents using blinding. However, the privacy
indicator for this scenario, i.e., celebrity's identity, is \textit{not} simply
a linear sum of all perception contents. Overall, identity recognition depends
on subtle features of a human face: changing $\alpha$ not necessarily impedes
capturing informative features.

\T~\ref{tab:mitigation-text1} reports the mitigation results of \hunspell\ using
COCO. We use two notional words of high frequency, ``man'' and ``sitting'', for
blinding. We insert $N$ notional words after each word in an input sentence,
where $N$ ($\frac{1}{\alpha}-1$) is 19, 9, and 2 given different $\alpha$. When
more notional words are used, we observe a higher decrease in inference
accuracy. However, with blinding, the inference accuracy decreases from more
than 40.0\% (see \T~\ref{tab:text}) to less than 2.5\% (close to baseline; see
\T~\ref{tab:text}) even two notional words are inserted after each normal word.
Different from masking images, the privacy of text is assessed by word
dependency (introduced in \S~\ref{subsubsec:eval-quantitative}). $\alpha$
decides \#notional words inserted to break word dependency. Therefore, the
results changes notably w.r.t.~values of $\alpha$.

In sum, our quantitative evaluation demonstrates the effectiveness of our
proposed mitigation despite differences in the media data formats or exploited
side channels. See Appendix~\ref{sec:mitigation} for more results; for instance,
blinding chest X-ray images can drastically reduce the disease diagnosis F1
score from an average of 0.73 (see \T~\ref{tab:xray}) to less than 0.1.

%% file: real-attack.tex
\subsection{Real-World Attack with \pp}
\label{subsec:real-world}
This section explores collecting cache access traces via a practical cache
attack, \pp~\cite{Tromer10,Zhang12}, in \textit{userspace-only scenarios}. To do
so, we conduct an end-to-end experiment, by leveraging
Mastik~\cite{yarom2016mastik}, a micro-architectural side channel toolkit, to
perform \pp\ and log victim's access toward L1D and L1I cache. We use Linux
\texttt{taskset} to pin the victim software and the \spy\ process on the same
CPU core. We launch experiments on both Intel Xeon CPU and AMD Ryzen CPU. See
\appx~\ref{subsec:real-attack-details} for setup details of this end-to-end
attack. We also clarify how cache side channels are represented and processed by
our encoder in \appx~\ref{sec:input}.

The quantitative evaluation results, as reported in \T~\ref{tab:pp}, are
generally encouraging. Attacks toward \libjpeg\ and \hunspell\ manifest high
accuracy comparable with attacks over \pin-logged traces
(\S~\ref{subsubsec:eval-quantitative}). While \pp\ logs relatively noisier cache
side channels, our attack shows promising noise resilience, as trace encoder
(and manifold learning by design) is noise resilient. Further, the logged side
channels are \textit{sparse}; given only a few records are secret-dependent,
noise introduced by \pp\ and other workloads do not primarily impede our attack.
See further discussion on noise resilience in \appx~\ref{sec:input}.

\ffmpeg\ reports high attack accuracy (over 80\%) on Intel L1D cache but lower
accuracy for other settings. As will be reported in
\T~\ref{tab:pp-trace-statistics} (\appx~\ref{subsec:real-attack-details}), the
logged side channel traces are unstable (and challenging to comprehend), where
stddev is about half of the average trace length. To verify the high attack
accuracy on Intel L1D cache, we manually checked all the reconstructed 2,552
audio clips (also uploaded at~\cite{snapshot} for the reference). The
reconstructed audio clips on Intel L1D cache manifest high quality. However,
while the original audio clips of the same class are produced by different
persons (and sound very different), all reconstructed audio clips of the same
class sound \textit{indistinguishable}. This indicates that the trained model
stealthily simplifies the task of reconstruction into a task of ten
class-conditional generation (recall this dataset has ``0--9'' labels). We then
manually checked the collected traces: we find that for this particular case,
Intel L1D cache ``amplifies'' the distance of inter-class traces while reduces
the distance of intra-class traces. As a result, intra-class differences are not
well learned using training data. However, since our quantitative metrics only
check if the reconstructed audio can be classified correctly, the attack accuracy
is high, indicating privacy leakage. We use attention
(\S~\ref{subsec:design-attention}) and compared all the localized code
components contributing side channels: we report that localized functions are
the \textit{same} on Intel/AMD CPUs. However, they manifest different
frequencies, which result in this subtle model decaying. We confirm that this
stealthy issue \textit{only} occurs for this case. To solve this issue (and
therefore reconstruct diverse outputs within the same class), users can opt for
more complex models or larger training data, if needed.

%
%
Overall, inspired by recent work~\cite{Weber2021} exploring timing-based
microarchitectural side channels, we deem it an interesting future work to
benchmark the microarchitectural side channel differences. Our tool can
automatically check whether side channels are informative enough to reconstruct
secrets, when it largely outperforms the baseline.


We report the slowdown incurred by \pp\ attack in \T~\ref{tab:pp}. Overall,
\ms\ are highly complex, and processing media data can usually induce a large
volume of cache accesses. This way, frequent cache misses due to \pp\ can cause
a reasonably high slowdown. \appx~\ref{subsec:real-attack-details} gives further
discussion regarding this point. We also present qualitative evaluation results
in \appx~\ref{subsec:real-attack-details}. In short, the reconstructed media
data (e.g., images) manifest fairly high visual quality. For instance, the
reconstructed CelebA face photos retain many correct perception features, such
as face orientation, skin color, hair color and hair styles.

\begin{table}[t]
  \caption{Attack PathOHeap.}
  \label{tab:oram}
  \centering
\resizebox{0.5\linewidth}{!}{
  \begin{tabular}{l|c|c}
    \hline
     Function & \texttt{IDCT} & \texttt{MCU}\\
    \hline
     w/o ORAM & 40.3\% & 38.0\% \\ 
    \hline
     with ORAM & 0.2\% & 0.2\% \\ 
    \hline
  \end{tabular}
  }
\end{table}

\subsection{Mitigation Using ORAM}
\label{subsec:oram}

Besides perception blinding, this section assesses other mitigations. Existing
mitigations aim at adding randomness, making it constant, or directly masking
inputs. Nearly all of them are particularly designed to protect crypto
software~\cite{tiri2003securing,coron2009efficient,brumley2005remote}.
Raccoon~\cite{rane2015raccoon} proposes general mitigation using software
obfuscation; however, its implementation is not available. That said, oblivious
RAM (ORAM)~\cite{goldreich1987towards} conceal memory access sequences of a
program by continuously shuffling data as they are accessed. We study whether a
representative ORAM, PathOHeap~\cite{shi2020path}, can mitigate our attack. Due
to the limited space, we report the key evaluation results in \T~\ref{tab:oram}.
PathOHeap takes several hours to process one memory access made by \libjpeg. We
thus focus on two critical functions localized by our framework (see
\S~\ref{subsec:eval-localization}), \texttt{IDCT} and \texttt{MCU},
separately.\footnote{Focusing on functions with known information leakage (i.e.,
  \texttt{IDCT})~\cite{xu2015controlled,hahnel2017high} demonstrates a
  ``white-box'' attacker using our technique.} We measure attack success rates
with and w/o first converting memory traces using PathOHeap.

Cache line side channels derived from either \texttt{IDCT} or \texttt{MCU} are
sufficient for attack. Nevertheless, ORAM eliminates information leak: memory
access traces, after processed by PathOHeap, do \textit{not} depend on input
images. We report that our autoencoder does not even reach convergence during
training, and the reconstructed images (using poorly trained autoencoder) show
indistinguishable and meaningless visual appearances. The non-zero result (i.e.,
$0.2\%$) is because that many face photos look like an ``average'' face. In
other words, $0.2\%$ implies the \textit{baseline} of face matching.

\smallskip
\noindent \textbf{Comparison.}~PathOHeap is very costly: while \libjpeg\ can
process an image within 100ms, PathOHeap takes several hours to convert the
corresponding memory trace. The obfuscator, Racoon~\cite{rane2015raccoon}, has
an average overhead of $16.1\times$. In contrast, perception-blinding delivers
negligible extra cost (i.e., processing masked data using \ms\ once) albeit its
mitigation is specific for manifold learning-based SCA. This underlines the key
novelty of our technique.

%% file: discussion.tex
\begin{table}[t]
  \caption{Quantitative evaluation results (same face/non-face) of face images
    reconstructed from noisy side channels.}
  \label{tab:face-noise}
  \centering
\resizebox{1.0\linewidth}{!}{
  \begin{tabular}{c|c|c|c|c}
    \hline
    \multirow{2}{*}{Setting} & Noise  & \multirow{2}{*}{\textbf{NA}}      & \multirow{2}{*}{\textbf{Low}} & \multirow{2}{*}{\textbf{High}} \\
    & Insertion Scheme &  &  &  \\
    \hline
    \multirow{3}{*}{\shortstack{\pin\\logged\\trace}}  & \textbf{Gaussian} & 43.5/2.0\% & 33.8/2.1\%  & 28.0/1.5\% \\ 
      & \textbf{Shifting} & 43.5/2.0\% & 42.9/1.7\%  & 39.1/1.8\% \\
      & \textbf{Removal} & 43.5/2.0\% & 30.0/1.9\%  & 29.3/4.3\% \\ 
    \hline
    \multirow{3}{*}{\shortstack{\pp\\logged trace}}  & \textbf{Leave out} & 36.9/0.8\% & 36.8/1.1\%  & 36.8/1.1\% \\ 
      & \textbf{False hit/miss} & 36.9/0.8\% & 36.4/1.2\%  & 36.1/1.2\% \\ 
      & \textbf{Wrong order} & 36.9/0.8\% & 36.8/1.0\%  & 36.7/1.0\% \\ 
    \hline
    \multirow{3}{*}{\shortstack{Workload under\\\pp}}  & \textbf{Bzip2} & 36.9/0.8\% & \multicolumn{2}{c}{27.6/1.0\%} \\ 
      & \victima\ & 36.9/0.8\% & \multicolumn{2}{c}{30.7/1.1\%} \\ 
      & \victimb\ & 36.9/0.8\% & \multicolumn{2}{c}{29.0/1.0\%} \\ 
    \hline
  \end{tabular}
  }
\end{table}

\subsection{Noise Resilience}
\label{subsec:noise}

We have discussed the general immunity to noise of manifold learning in
\F~\ref{fig:denoising}. This section empirically assesses our attacks under
various scenarios where noise is introduced in side channels. We summarize our
noise insertion schemes in \T~\ref{tab:face-noise}. The first three schemes are
launched to mutate cache line access traces logged by \pin, whereas the latter
three mutate the cache set hit/miss records logged via \pp. \textbf{NA} means no
noise is inserted, whereas \textbf{Low/High} denote to what extent side channel
logs are perturbed (see \appx~\ref{sec:noise} for details).
We also benchmark how real-world workload, i.e., by launching \texttt{bzip2} or
another victim software (e.g., \libjpeg) on the same CPU core, can undermine our
attack. \victima\ and \victimb\ represent launching another victim software on
the same core and processing the same or different inputs.

Due to the limited space, we only report the quantitative evaluation results of
\libjpeg\ on CelebA in \T~\ref{tab:face-noise}. See \appx~\ref{sec:noise} for
other quantitative and qualitative results: despite the applied noise, many
perceptual features are still retained in the reconstructed data (e.g., face
photos), illustrating capability of privacy stealing under noisy scenarios.

The reconstructed images are more resilient toward \textbf{Shifting}.
\textbf{Removal} and \textbf{Gaussian} noise, by extensively leaving out or
perturbing data points on the logged trace (e.g., \textbf{Removal/High} removes
\textit{half} records on a trace), show greater influence on data
reconstruction. As for noise inserted in \pp\ logged side channel records, none
of them primarily affect the attack accuracy. We note that \pp\ logged side
channel traces, even without applying these noise insertion schemes, are of high
stddev. That is, our autoencoder will be trained with more ``diverse'' side
channel logs, which enhance the robustness but undermines accuracy. Similar
findings are obtained in launching extra real-world workloads. Please refer to
\appx~\ref{sec:noise} for further evaluation on noise resilience, and our
analysis on noise resilience from the trace encoder structure perspective and
training data perspective.


%% file: related.tex
\section{Related Work}
\label{sec:related}

\smallskip
\noindent \textbf{Side Channel Analysis.}~Kocher proposes to use timing side
channel to exploit crypto systems~\cite{kocher1996timing}. To date, side
channels have been used to exploit crypto systems under different
scenarios~\cite{al2013lucky,bang2016string,dong2018shielding,Wu12},
including trusted computing environments like Intel
SGX~\cite{lee2020off,brasser2017software,moghimi2017cachezoom,schwarz2017malware}.
\cite{Brumley:Boneh:2005} demonstrates that timing side channel can be launched
remotely through network. The CPU cache are particularly exploited given its
indispensable role in boosting modern computing
platforms~\cite{hahnel2017high,oren2015spy,yarom2017cachebleed,Liu15}.
Controlled side channel assumes an adversarial-controlled OS to log page table
access of victim software~\cite{xu2015controlled}. 
DNNs have been used to infer secret keys from crypto
libraries~\cite{heuser2012intelligent,maghrebi2016breaking,hettwer2020encoding}.
These works, usually referred to as ``profiled SCA'', share the same assumption
with our research that models are trained using historical data. Most existing
DNN-based SCA focuses on attacking crypto systems; they typically perform
low-level \textit{bit-wise classification} to gradually infer key bits. In
contrast, we show that attackers in black-box scenarios can use manifold
learning to reconstruct media data of various types in an end-to-end
manner.~\cite{kwon2020improving,wu2020remove} also use autoencoders in the
context of SCA. However, they use autoencoder to denoise side channel traces as
a \textit{preprocessing} step for SCA of crypto software.

\smallskip
\noindent \textbf{Countermeasures.}~Software-based techniques include
constant-time techniques which ensure that software behavior is independent with
its confidential
data~\cite{coppens2009practical,raj2009resource,schwarz2018keydrown,kopf2007transformational,molnar2005program}.
Techniques have also been proposed to blind secrets or randomize side channel
access
patterns~\cite{askarov2010predictive,braun2015robust,crane2015thwarting,hu1992reducing,kopf2010vulnerability,rane2015raccoon}.
ORAM~\cite{goldreich1996software,liu2015ghostrider,stefanov2013path,shi2020path}
translates memory access into identical or indistinguishable traces, which can
provably eliminate many side channels but incur high performance penalty.
Program analysis methods such as information flow
tracking~\cite{ctgrind,myers1999jflow}, model
checking~\cite{almeida2016verifying}, type
system~\cite{agat2000transforming,sabelfeld2003language}, abstract
interpretation~\cite{kopf2012automatic,cacheaudit,wang2019identifying}, and
constraint solving~\cite{wang2017cached,bortzman2018casym} are used to check
crypto software and detect side channels. In contrast, our study delivers a
\textit{neural attention}-based approach to detecting code fragments inducing
information leakage. Hardware-based countermeasures include randomizing side
channel access or enforcing fine-grained resource
isolation~\cite{wang2006covert,wang2008novel,wang2007new,liu2016cache,page2005partitioned}.
Compared with system- and hardware-based countermeasures, software-based
approaches usually do not require to modify the underlying hardware design.
Nevertheless, software-based countermeasures are generally high cost and low
scalable in analyzing real-world software.

%% file: conclusion.tex
\section{Conclusion}
\label{sec:conclusion}

This research proposes SCA for \ms. We perform cross-modality manifold learning
to reconstruct media data from side channel traces. We also use attention to
localize program points leading information leakage. We design perception
blinding to mitigate the proposed SCA. Our evaluation on real-world media
software reports promising results.

%% file: hypothesis.tex
\begin{figure}[!ht]
    \centering
    \includegraphics[width=1.01\linewidth]{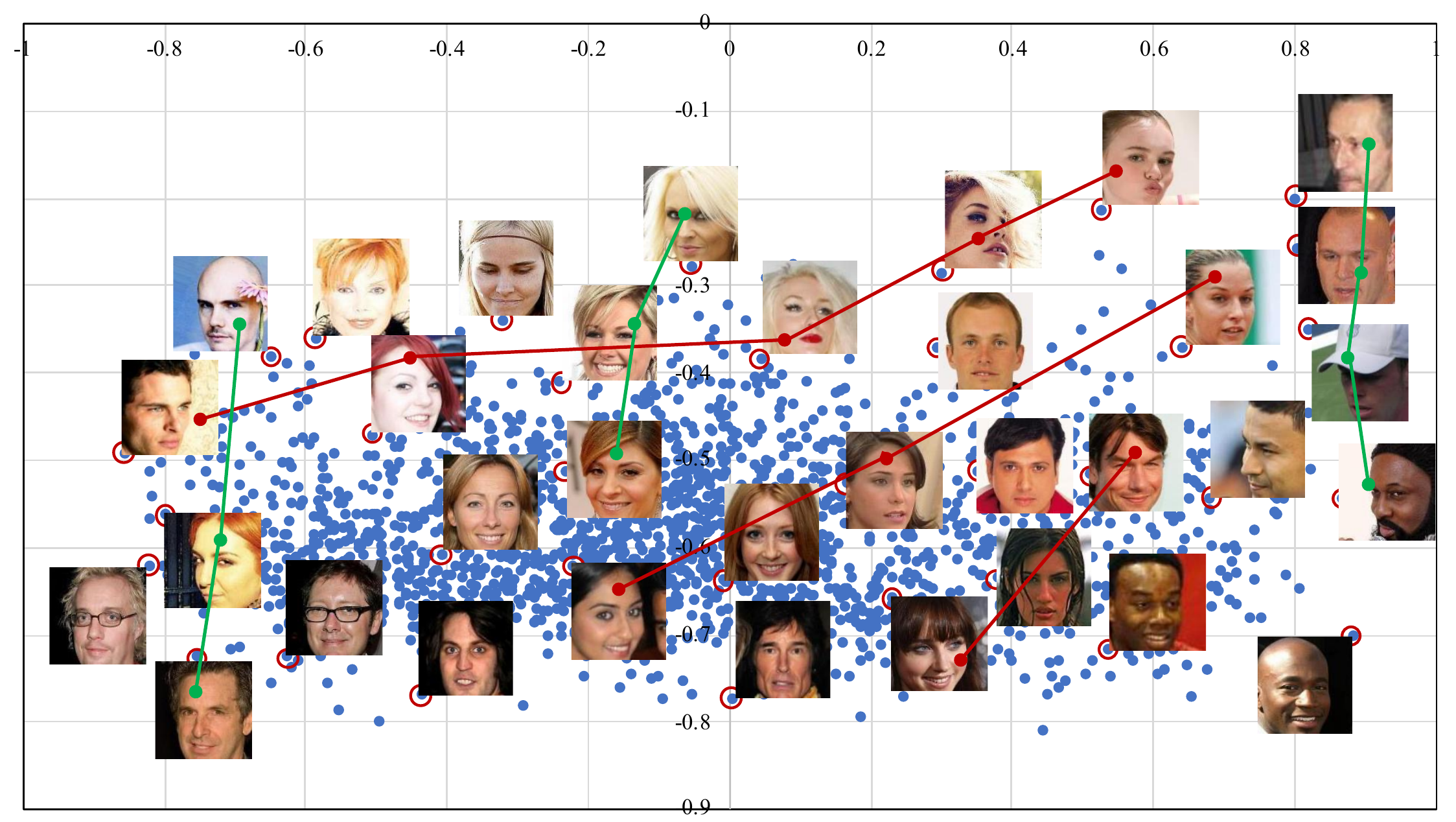}
    \caption{Project face photos to a two-dimensional manifold.}
    \label{fig:manifold-face}
\end{figure}

\begin{figure}[!ht]
  \centering
  \includegraphics[width=0.8\linewidth]{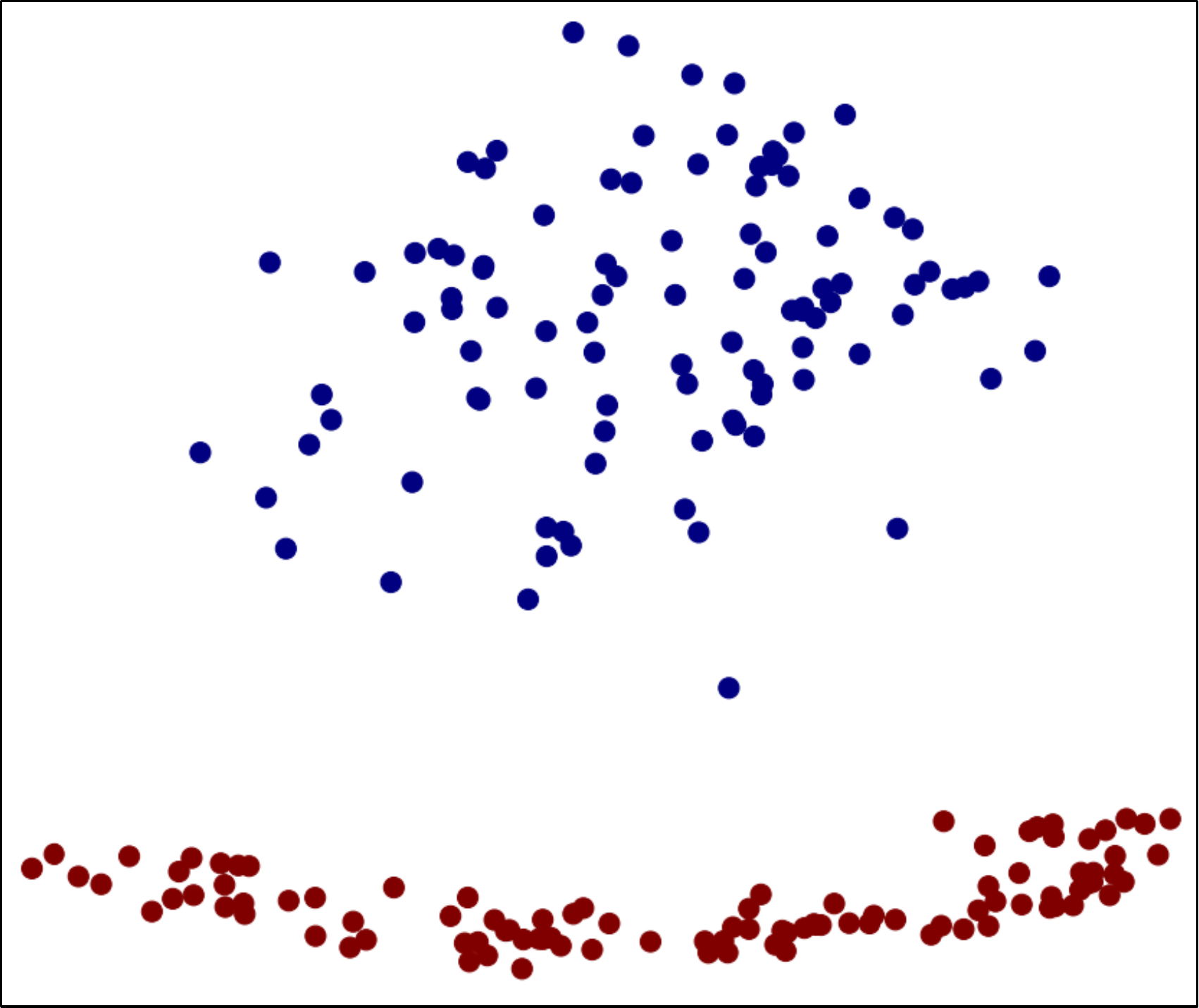}
  \caption{Applying PCA on two classes of images from MNIST dataset.}
  \label{fig:pca}
\end{figure}

\section{Empirical Exploration of Manifold Hypothesis}
\label{sec:misc}

This section empirically explores the validity of the manifold hypothesis. In
\F~\ref{fig:manifold-face}, a set of real-world face images is projected onto
manifold $\mathcal{M}_{img}$ of two dimensions. To draw this projection, we
adjust our autoencoder framework (see \S~\ref{sec:design}) to convert each face
image into a latent representation of two dimensions. We observe that the images
are generally distinguished by skin and hair colors, and face orientations are
roughly decomposed into two orthogonal directions (\textcolor{pptgreen1}{green}
and \textcolor{pptred1}{red} lines). For instance, the photos at the bottom
right and bottom left are grouped together with similar hair colors but are
further differentiated due to differences in skin colors and face directions.
Overall, we interpret that \F~\ref{fig:manifold-face} provides clear evidence
that our autoencoder frameworks can capture key perception features of complex
high-dimensional media data. In addition, because 2D projections are highly
condensed for real-world large datasets, we expect to find greater
discrimination for higher dimensions (e.g., 64).

In fact, recent studies have shown the effectiveness of conducting image editing
by first projecting image samples into the manifold
space~\cite{zhu2016generative,shen2020interpreting}. Editing images in the
high-dimensional space involves a large search space $[0,255]$ for each pixel;
the random selection of pixel values in the range $[0,255]$ struggles to create
realistic images because arbitrary editing could ``fall off'' the manifold of
natural images. In contrast, manifold learning facilitates sampling within
$\mathcal{M}$, and the perceptually meaningful contents in $\mathcal{M}$ confine
the manipulations to generate mostly realistic
images~\cite{zhu2016generative,shen2020interpreting}.

Furthermore, as introduced in \S~\ref{sec:manifold}, the manifold learning
hypothesis also assumes that after casting high-dimensional data into their
low-dimensional space (i.e., manifolds), data of different classes can be well
separated whereas data of the same class generally lies in the same
manifold~\cite{thorstensen2009manifold}. Accordingly, our preliminary study
confirmed this hypothesis: as shown in \F~\ref{fig:pca}, after launching
dimensionality reduction with manifold learning, two classes of MNIST images are
well separated into distinct groups.

%% file: blinding-appendix.tex
\begin{figure*}[t]
  \centering
  \includegraphics[width=0.8\linewidth]{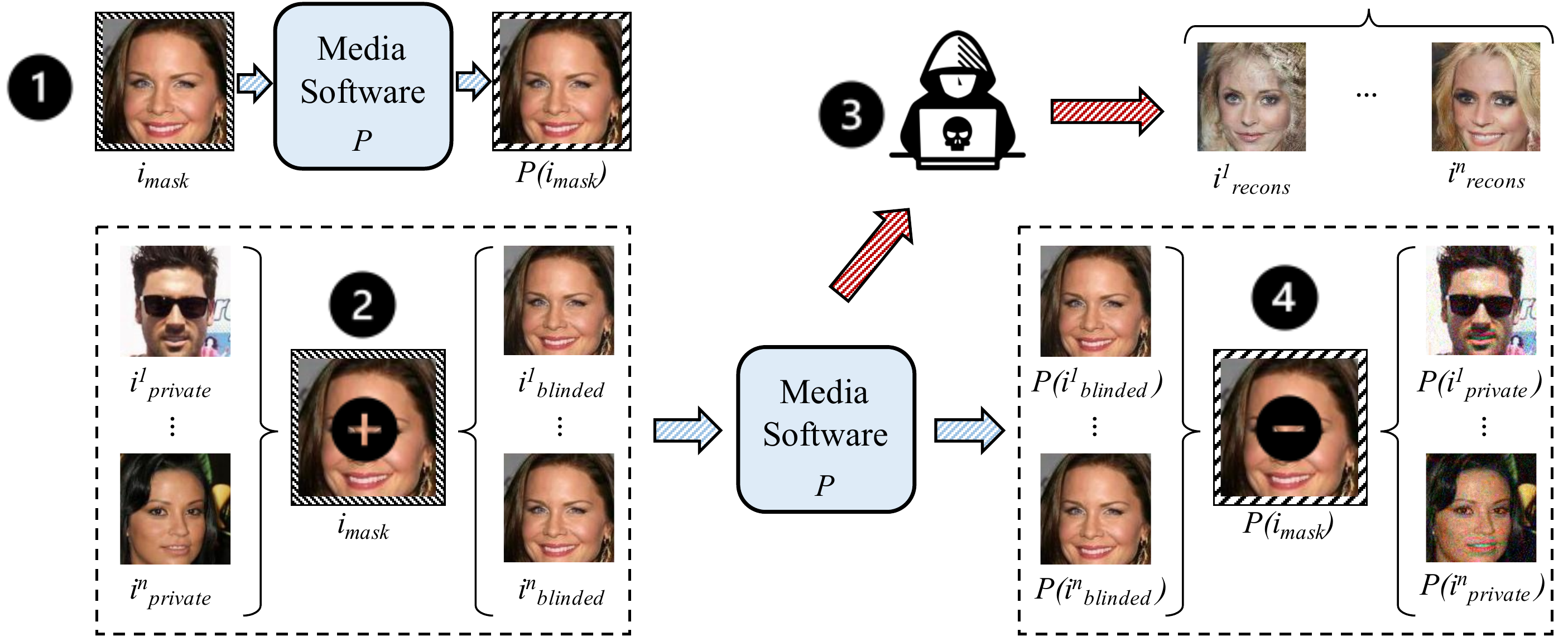}
  \caption{Workflow of perception blinding.}
  \label{fig:blind-pipe}
\end{figure*}

\section{Discussion about Perception Blinding}
\label{sec:blind-appx}

\S~\ref{subsec:design-defense} introduces perception blinding, as an effective
mitigation to defeat our proposed SCA. In this appendix section, we further
clarify the usage of perception blinding by discussing the workflow of applying
perception blinding in real systems. We then give discussions about its usage
and other considerations.

\smallskip
\noindent \textbf{Workflow.}~\F~\ref{fig:blind-pipe} depicts the detailed
workflow. As already clarified in \S~\ref{subsec:design-defense}, to use
perception blinding, users only need to pick \textit{one} mask $i_{mask}$ to
blind all input data.
\ding{182} A user first picks this \textit{universal} mask $i_{mask}$ and
processes it with the to-be-protected \ms\ $P$ to pre-compute $P(i_{mask})$.
\ding{183} Then, before processing any private input $i_{private}$ with $P$, a
user should ``$\oplus$'' it with $i_{mask}$ locally to get $i_{blinded}$, and
then feed $i_{blinded}$ to $P$. \ding{184} As introduced in
\S~\ref{subsec:design-defense}, $i_{recons}$ reconstructed by the attacker will
mostly retain perception contents of $i_{mask}$. \ding{185} In contrast, to get
the desired output $P(i_{private})$, a user only needs to ``$\ominus$'' the
pre-computed $P(i_{mask})$ from $P(i_{blinded})$.

\smallskip
\noindent \textbf{Application Scope.}~By masking media data, perception blinding
is designed to effectively mitigate manifold learning-based SCA (see evaluation
in \S~\ref{subsec:eval-mitigation}). We notice some existing side channel
attacks toward \ms~\cite{hahnel2017high,xu2015controlled} which extensively
relies on manual efforts. Perception blinding may \textit{not} be effective to
mitigate those works, given those attacks generally conduct data byte-level
inference. If data bytes can be \textit{losslessly} recovered, privacy leakage
could still occur, though such data byte-level reconstruction generally involves
considerable \textit{manual efforts}~\cite{hahnel2017high,xu2015controlled}. In
contrast, as will be reported in \appx~\ref{subsec:real-attack-details}, our
well-trained framework takes less than 10 seconds to reconstruct 2,000 media
inputs. Also, our proposed perception blinding is not applicable to protect
crypto libraries. Unlike media data, the feasibility of defining and extracting
``perceptions'' over private keys is unclear. Standard blinding techniques
(e.g., RSA blinding~\cite{brumley2005remote}) should be used to prevent crypto
systems from SCA.

\smallskip
\noindent \textbf{``White-Box'' Attackers.}~Readers may wonder what if a
``white-box'' attacker trains autoencoder with the dataset that includes
$i_{mask}$. We clarify that the mitigation should still be effective, whose
reasons are threefold. First, since users \textit{randomly} choose masks, it's
impossible for attackers to guess what the mask is. Second, if attackers use a
training dataset containing $i_{mask}$ but don't know which one it is, our
perception blinding still works. Third, suppose this particular $i_{mask}$ is
known, attackers will need to mask all training data and re-train another model,
leading to a high cost. Note that ``$\ominus$'' $i_{mask}$ from $i_{recons}$
will \textit{not} get $i_{private}$, because ``$\ominus$'' operates on
\textit{byte-level} rather than \textit{perception-level}. To understand this,
suppose $i_{private}$ is a human face photo toward left, whereas the human face
in $i_{mask}$ is toward right. As a result, face recovered in $i_{recons}$
should be toward right as well. It's easy to see that ``$\ominus$'' $i_{mask}$
from $i_{recons}$ will not flip the face direction.

%% file: attack-setup-appendix.tex
\section{Attack Setup}
\label{sec:attack-appendix}
We give a general introduction of each side channel adopted in our attack as
follows:

\smallskip
\noindent \textbf{Cache Bank.}~Cache bank generally denotes the minimal storage
unit of modern CPU caches and has been exploited by real-world
attacks~\cite{yarom2017cachebleed}. Assume a program memory address has $N$
bits, then typically the upper $N - L$ bits map a memory access to its
corresponding cache bank access. In this research, we adopt a common setting for
most CPUs on the market where $L$ is 2. That is, given a memory address $addr$,
its cache bank index can be computed as $addr \gg 2$.

\smallskip
\noindent \textbf{Cache Line.}~Similar to cache bank, attackers can also log all
the accessed cache line indexes for exploitation~\cite{hahnel2017high,Yarom14}.
For modern CPUs whose cache-line size is usually 64, the cache line index of a
memory address $addr$ can be computed as $addr \gg 6$. 

\smallskip
\noindent \textbf{Page Table.}~Program virtual memory accesses are converted
into corresponding physical memory accesses by querying the OS page table. Given
an adversarial-controlled OS, a practical high-resolution side channel is to
keep track of all accessed page table entries made by the media
software~\cite{xu2015controlled}. Given a memory access with address $addr$, the
induced page table index can be calculated by masking $addr$ with
\texttt{PAGE\_MASK} $m$: $addr \mathbin{\&} (\sim m)$, where $m$ set as
4095~\cite{pagemask}.

\smallskip
\noindent \textbf{Media Software and Inputs.}~We use \libjpeg, a widely-used
image processing library, to process two image datasets CelebA and Chest X-ray.
CelebA comprises 10,000 different celebrities, with 20 images of each. Chest
X-ray comprises real-life X-ray images that are used for screening and diagnosis
of many lung diseases. In addition, we use a popular media software, \ffmpeg, to
process two audio datasets, SC09 and Sub-URMP. Both datasets are commonly-used
in benchmarking audio synthesis and analysis
research~\cite{chen2017deep,duan2019cascade,chang2019deep}. SC09 denotes
real-world ``speech commands.'' It contains single spoken words from 0–-9 by
various speakers in real-life scenarios. Sub-URMP consists of audio recordings
of 13 musical instruments, such as double bass, viola, and violin. \hunspell, a
popular spell checker used by commercial software, such as Google Chrome,
OpenOffice, and LibreOffice, is also exploited in this research. We use two
datasets, COCO and DailyDialog, as the inputs of \hunspell. COCO is a popular
large-scale dataset that contains images and text descriptions. Each image is
associated with three to eight manually-annotated sentences. Each sentence
contains a number of object and attribute names that denote how humans would
describe a photograph of a real-world scenario. The DailyDialog dataset contains
real-life multi-turn dialogues, which reflect daily communication and cover
various topics.

\smallskip
\noindent \textbf{Tools Used in Evaluations.}~We leverage Face++~\cite{facepp},
a commercial face recognition API, to calculate the similarity score between
reconstructed face photos and reference inputs. We clarify that the
implementation details of Face++ is \textit{not} disclosed. To decide the
diseases of reconstructed X-ray images and reference images, we reuse the tool
provided by the champion of CheXpert competition~\cite{CheXpert}. This tool can
categorize the disease of X-ray images and localize corresponding lesions.

When launching quantitative evaluation to measure the privacy leakage of
reconstructed audio data, we launch an experiment to train two classifiers for
speaker identity and command 0--9 matching. This denotes a typical setup of
identity de-anonymization attack~\cite{yuan2021private}. We now report the
details of two classifiers. To classify the content of each human voice, we
reuse the architecture of our ``Privacy-Aware Indicator'' in
\S~\ref{subsec:design-attack}. The classifier is trained on reference audio
recordings to achieve over $98\%$ testing accuracy. We then use the trained
classifier to classify the reconstructed audio. Note that the identities of
training and testing splits are \textit{not} overlapped, and we modify
the training objective based on a face recognition model~\cite{schroff2015facenet}.
When evaluating, the accuracy is calculated as how many reconstructed audio
recordings yield a similarity score that is higher than $50\%$ with its
reference audio (i.e., the correct match).

In addition, when quantitatively evaluating the musical instrument types of
reconstructed audio using the Sub-URMP dataset, we also reuse the architecture
of ``Privacy-Aware Indicator'' and train the classifier on reference audio to
achieve a testing accuracy higher than $98\%$. Then, the trained classifier is
used to classify the reconstructed audio clips.

%% file: real-attack-appendix.tex
\section{\pp\ Attack Details}
\label{subsec:real-attack-details}

In this section, we clarify the detailed setup of our \pp\ attack. We
consistently assume that attackers can first train the autoencoder framework
with cache side channel locally collected via \pp\ toward victim software,
but do \textit{not} need the source code. Then, the attacker launches \pp\
toward the target \ms\ to log cache side channels when it is processing an
unknown input. The unknown input will be reconstructed from the logged
cache side channels.

\smallskip
\noindent \textbf{Probing the L1I and L1D Caches.}~We use
Mastik~\cite{yarom2016mastik} to launch \pp\ attack towards L1D cache and L1I
cache on Intel Xeon and AMD Ryzen CPU. We use Linux \texttt{taskset} to pin the
\victim\ software and a \spy\ process to the same CPU core. We take a common
assumption~\cite{Tromer10} that attackers know when the victim \ms\ begins and
ends to process an unknown input. The \spy\ process primes and probes the cache.
Technically, there is another ``coordinator'' process on the same core which
computes victim process's cache activities and logs the cache side channels to
disk. Nevertheless, according to our observation, this coordinator process has
generally consistent cache access patterns, and therefore, its mostly fixed
cache access does not interfere with our well-trained autoencoder.

The thresholds of deciding cache hit and cache miss are 120 CPU cycles on Intel
Xeon CPU and 100 CPU cycles on AMD Ryzen CPU. \pp\ is performed in the following
manner:

\textsc{Prime:}~The \spy\ process fills all cache sets.

\textsc{Idle:}~The attacker logs the access time of all cache sets for the
previous \pp\ iteration. As a result, the idle phase \textit{interval} equals
the duration of performing one file I/O operation. Meanwhile, the cache is
utilized by the \victim.

\textsc{Probe:}~The \spy\ process refills all cache sets and times the duration
to refill the same cache sets to learn how \victim\ accesses cache sets.

For a cache set, the logged cache status flip, from hit to miss, indicates at
least one cache access of \victim. We are thus particularly interested in
logging such status flip. We name such cache status flips as ``cache activity''
in the rest of this paper. Whenever a cache activity is observed, we record the
cache activities of the all cache sets into a vector $V$, whose length equals to
the number of cache sets. $V[i] = 1$ indicates there is a cache activity in
$i$-th cache set. In other words, the $i$-th cache set is accessed at least one
time by the \victim. If no cache activity is observed from any cache set, we
omit to generate a new $V$.


\begin{figure*}[t]
  \centering 
   \includegraphics[width=1.0\linewidth]{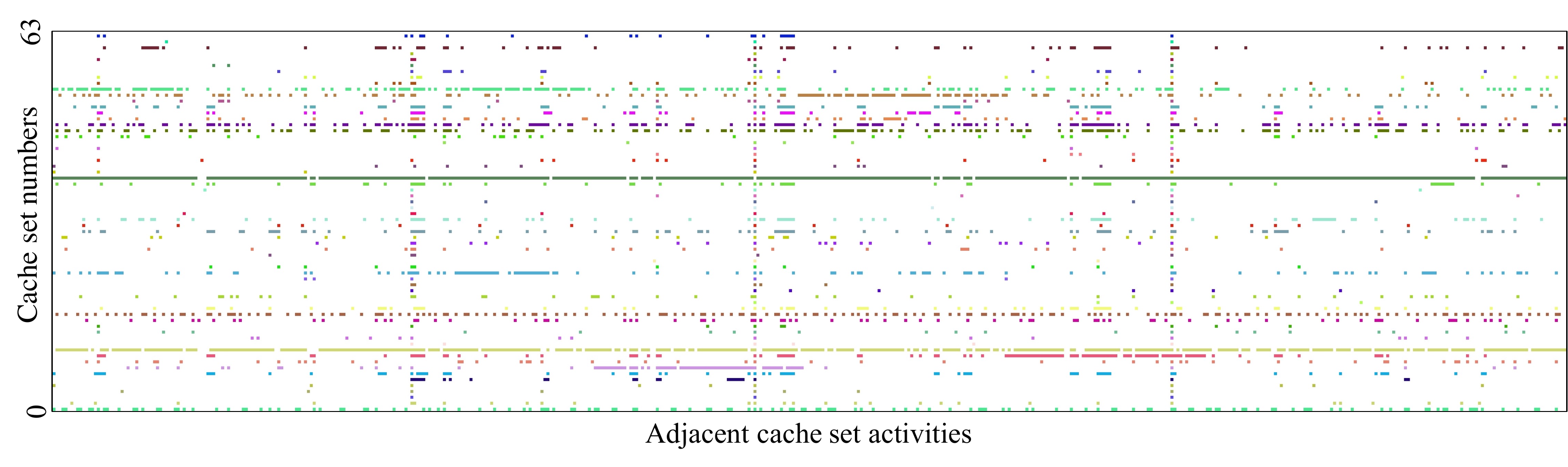}
   \caption{Cache set activities of \ffmpeg\ logged by \pp\ on Intel CPU L1D
     cache. Colored blocks denote cache access activities, i.e., ``hit'' flips
     to ``miss''. We use different colors to differentiate different cache sets.}
  \label{fig:ffmpeg-dcache-intel}
\end{figure*}

\begin{table*}[t]
  \caption{Statistics (\textit{mean}$\pm$\textit{stddev}) and matrix encoding of
    cache side channel traces logged by \pp. Each trace is a sequence of vector
    with length 64. Note that the \#training and \#test splits in each of the
    evaluation setting are the same as statistics reported in
    \T~\ref{tab:data-statistics}.}
  \label{tab:pp-trace-statistics}
  \centering
\resizebox{0.85\linewidth}{!}{
  \begin{tabular}{l|c|c|c|c|c|c}
    \hline
     & \multicolumn{2}{c|}{\libjpeg ($\times 8$)} & \multicolumn{2}{c|}{\ffmpeg ($\times 4$)} & \multicolumn{2}{c}{\hunspell ($\times 8$)} \\
    \hline
                & \textbf{Intel} & \textbf{AMD} & \textbf{Intel} & \textbf{AMD} & \textbf{Intel} & \textbf{AMD} \\
    \hline                                                                              
    \multirow{2}{*}{L1I Cache} &  $2719 \pm 745$  & $ 1114 \pm 290 $ & $44315 \pm 23374$ & $ 7013 \pm 2540 $ & $4834 \pm 1788$ & $ 1865 \pm 432 $ \\
                                & $1 \times 256 \times 256$ & $1 \times 256 \times 256$ & $8 \times 512 \times 512$ & $2 \times 512 \times 512$ & $4 \times 256 \times 256$ & $2 \times 256 \times 256$ \\
    \hline                                                                              
    \multirow{2}{*} {L1D Cache} &  $780 \pm 137$  & $ 5750 \pm 931 $ & $8837 \pm 3050$ & $ 46698 \pm 35163 $ & $3739 \pm 164$ & $ 9732 \pm 1498 $ \\
                                 & $1 \times 256 \times 256$ & $4 \times 256 \times 256$ & $2 \times 512 \times 512$ & $8 \times 512 \times 512$ & $4 \times 256 \times 256$ & $8 \times 256 \times 256$ \\
    \hline
  \end{tabular}
  }
\end{table*}

\smallskip
\noindent \textbf{Cache Set Activities Logged by
  \pp.}~\F~\ref{fig:ffmpeg-dcache-intel} presents a recorded sequence of vectors
$V$ on L1D cache of Intel Xeon CPU when \ffmpeg\ is processing an audio input.
Here, each row represents a cache set and each column is a recorded $V$. The
color block indicates there is a cache activity.
Consequently, the logged side channel trace of L1 cache is in the format of a
binary matrix, where bit $1$ represents a cache activity. The binary matrix is
taken as the input of our framework in an end-to-end manner. See
\appx~\ref{sec:input} for clarification on representation of side channel traces
and how they are fed into the encoding step.
Furthermore, the code of launching \pp\ is released at~\cite{snapshot} for
result verification and benefiting future research.

\smallskip
\noindent \textbf{Statistics of Logged Cache Side
  Channels.}~\T~\ref{tab:pp-trace-statistics} presents the mean, the standard
deviation and the matrix encoding of logged side channel traces under different
settings. Overall, the standard deviation of trace length is seen as high. This
indicates the difficulty of exploiting real-world \ms. In general, different
inputs can lead to the execution of different code paths, which further induce
unaligned cache accesses. To ease the deviation of \pp, we take a simple but
effective trick to repeatedly execute \ms\ with the same input for $t$ times and
concatenate the logged side channels as one input of our framework. The value of
$t$ is set to 8 for \libjpeg\ and \hunspell, and 4 for \ffmpeg.
In particular, when benchmarking \ffmpeg, the stddev can grow to half of the
average trace length. We view this explains the relatively low attack accuracy
of attacking \ffmpeg\ using cache side channels recovered by \pp: as shown in
\T~\ref{tab:pp}, attack accuracies toward \ffmpeg\ are generally below 15\%
except the Intel CPU \& L1D Cache setting.

\smallskip
\noindent \textbf{Workload.}~We have reported the workload in \T~\ref{tab:pp},
where the incurred execution slowdown is generally within 2 to 10 times
compared with the normal execution. Overall, all three \ms\ benchmarked in this
study is much complex than crypto libraries like AES. With more cache accesses
occurring during execution, cache misses incurred by \pp\ attack can presumably
slowdown the execution of victim software in an effective way. In particular,
our attack on \ffmpeg\ incurs relatively larger slowdown. This is reasonable:
\ffmpeg, as the most complex software among these three, produce a large volume
of cache accesses, which, accordingly increase the slowdown due to numerous
cache misses. Nevertheless, we still point out that even for the \ffmpeg\ case,
the average slowdown incurred by \pp\ is about 500 milliseconds. We have
demonstrated the highly effective \pp\ attack results toward these \ms\ in
\S~\ref{subsec:real-world}. To reduce the slowdown and deliver more stealthy
attacks, attackers can prolong the idle phase of \pp.

\smallskip
\noindent \textbf{Running Time.}~In general, running time primarily includes two
aspects: 1) online side channel logging and 2) offline model training. To
collect side channels for training, we spend 54 hours for \libjpeg, 36 hours for
\ffmpeg, and 100 hours for \hunspell\ on one CPU core. We note that processing
and converting the logged lengthy traces into ``Tensor'', the legitimate input
format of PyTorch, takes several extra hours. After excluding those file
processing cost, the logging duration is 9 hours, 12 hours, 27 hours for
\libjpeg, \ffmpeg\ and \hunspell, respectively. As a research prototype, we use
Python for those tedious file processing task. To speed up, users can re-write
those codes in C, if needed. The offline model training takes 10 hours.

The ``training phases'' using cache side channels collected by \pp\ can likely
take \textit{several hours to several days}: for instance, Zhang et
al.~\cite{Zhang12} spend six to 46 hours for the training phase when launching
\pp\ to extract crypto keys. Media software is generally more complex than crypto
libraries. We attribute our promising training cost (comparable
to~\cite{Zhang12}) to: 1) recent advances in neural networks and the underlying
high-quality deep learning framework PyTorch, and 2) better hardware
acceleration since we use one Nvidia GeForce RTX 2080 GPU for training.
The ``testing phase'' of \pp, i.e., reconstructing the unknown input by using a
cache side channel trace logged by \pp, takes less than 10 seconds to
reconstruct in total 2,000 media inputs.

\begin{figure*}[t]
\centering 
    \includegraphics[width=1.0\linewidth]{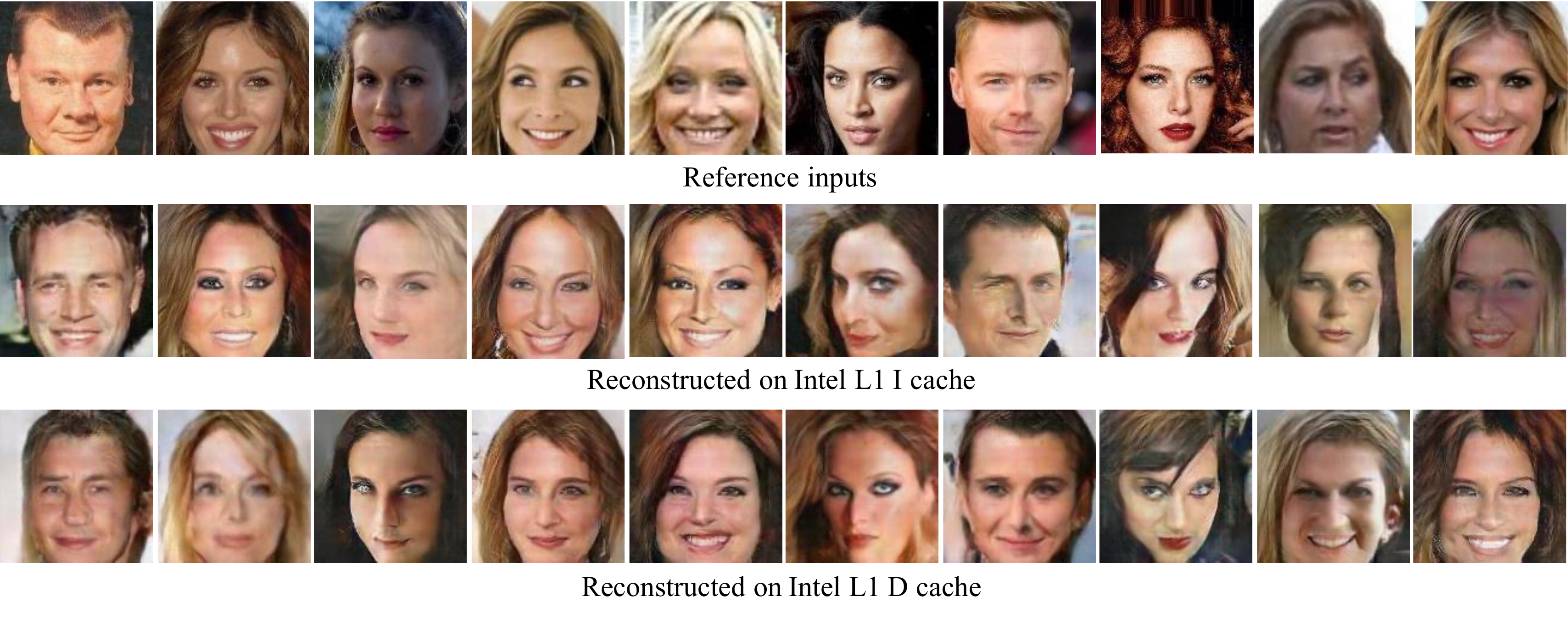}
    \caption{Reconstructed images on L1 cache of Intel Xeon CPU. We can observe highly
      correlated visual appearances, including gender, face orientation, skin
      color, nose shape, and hair styles.}
\label{fig:pp-celeba-intel}
\end{figure*}

\begin{figure*}[t]
  \centering 
      \includegraphics[width=1.0\linewidth]{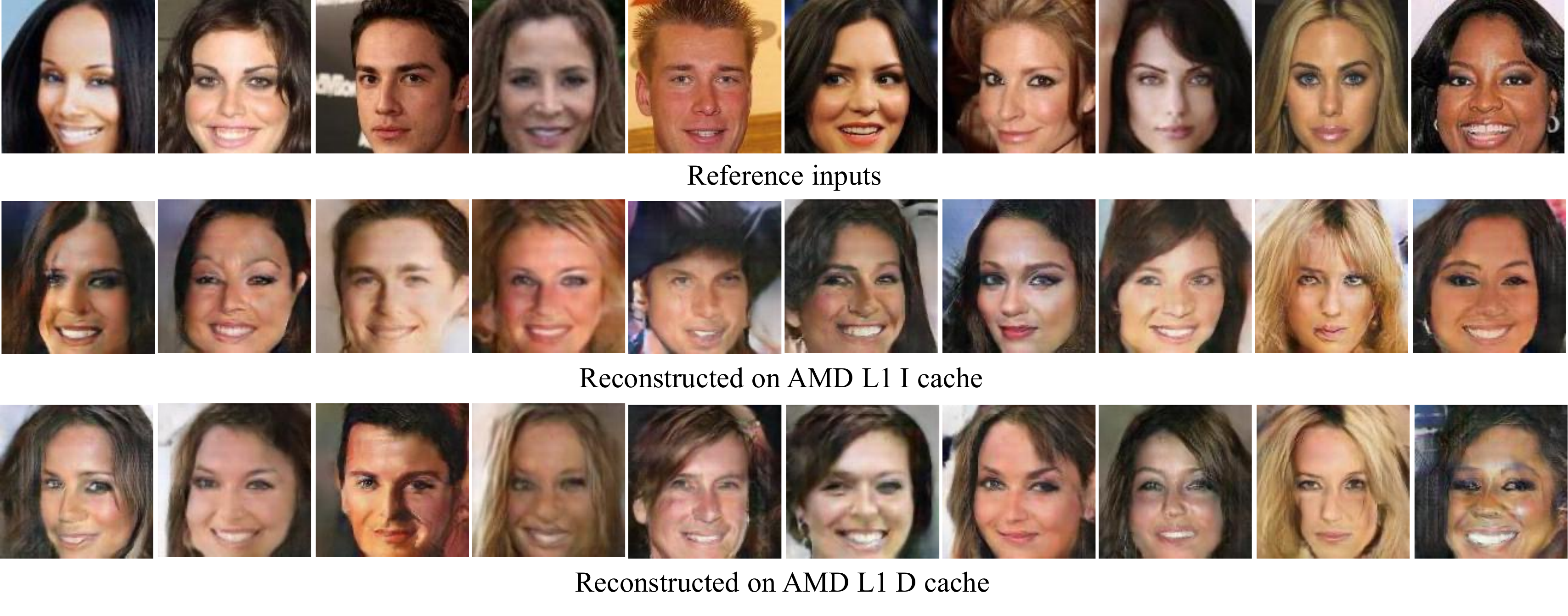}
      \caption{Reconstructed images on L1 cache of AMD Ryzen CPU. Again, visual
        appearances in the reconstructed images are correlated with reference
        inputs, including gender, face orientation, skin color, nose shape, and
        hair styles.}
  \label{fig:pp-celeba-amd}
  \end{figure*}

\noindent \textbf{Qualitative Evaluation Results.}~We have presented quantitative
evaluation results using \pp-recovered cache side channels in \T~\ref{tab:pp}.
Most extra (qualitative) evaluation results for attacks based on \pin-logged
traces are reported in \appx~\ref{sec:case}. For the seek of readability, we
discuss the qualitative results, in terms of reconstructed media data using
\pp\ attack, in this section. Particularly, \F~\ref{fig:pp-celeba-intel} and
\F~\ref{fig:pp-celeba-amd} report reconstructing CelebA face photos on different
CPUs. We interpret the reconstruction results as generally promising: a
considerable number of visual perceptions are faithfully retained in the
reconstructed images, which include gender, face orientation, skin color, nose
shape, and hair styles. There is a bit quality degradation on some images, for
instance, facial details (e.g., the appearance of teeth) are not always
precisely aligned with reference inputs. However, we emphasize that images
reconstructed under four settings all manifest comparable visual quality,
indicating high feasibility of applying our framework on the basis of
commonly-used side channels. The qualitative results are also consistent with
results reported in \T~\ref{tab:pp} --- though the attack is launched on
different CPUs and different caches, privacy leakage is steadily notable.

\begin{table}[t]
  \caption{Musical instrument type matching evaluation results.}
  \label{tab:audio1}
  \centering
\resizebox{0.8\linewidth}{!}{
  \begin{tabular}{l|c|c|c}
    \hline
            & \textbf{Cache bank} & \textbf{Cache line} & \textbf{Page table} \\
    \hline
     Content accuracy & 33.4\% & 32.8\% & 32.7\% \\
    \hline
  \end{tabular}
  }
\end{table}

\section{PathOHeap Setup}
\label{subsec:oram-details}

\S~\ref{subsec:oram} explored using PathOHeap to mitigate our proposed SCA.
PathOHeap, as a popular ORAM protocol, enforces probabilistic memory trace
obliviousness, such that it converts input-dependent memory access traces into
indistinguishable traces that provably hide input-dependent memory access
behavior and information leakage.

We study how ORAM can mitigate our SCA exploitation on \libjpeg. Our tentative
study shows that it takes over one hour to process the entire memory trace using
PathOHeap but cannot finish. Hence, we separately extract two key functions from
localized vulnerable modules, \texttt{IDCT} and \texttt{MCU}, that primarily
contribute to attacking \libjpeg\ (see \T~\ref{tab:libjpeg}). We thus collect
much shorter memory access traces corresponding to each function. We then pad
all traces to the same length (which is required by ORAM protocols) and use
PathOHeap to convert each logged memory access trace into an indistinguishable
memory trace, which takes about one minute to process. We then convert the
original memory traces and their corresponding ORAM outputs into cache line
access traces, and re-launch our SCA exploitation to recover \libjpeg\ input
images.

%% file: representation.tex
\section{Side Channel Representation \& Encoder Design}
\label{sec:input}

\begin{figure*}[t]
  \centering 
      \includegraphics[width=1\linewidth]{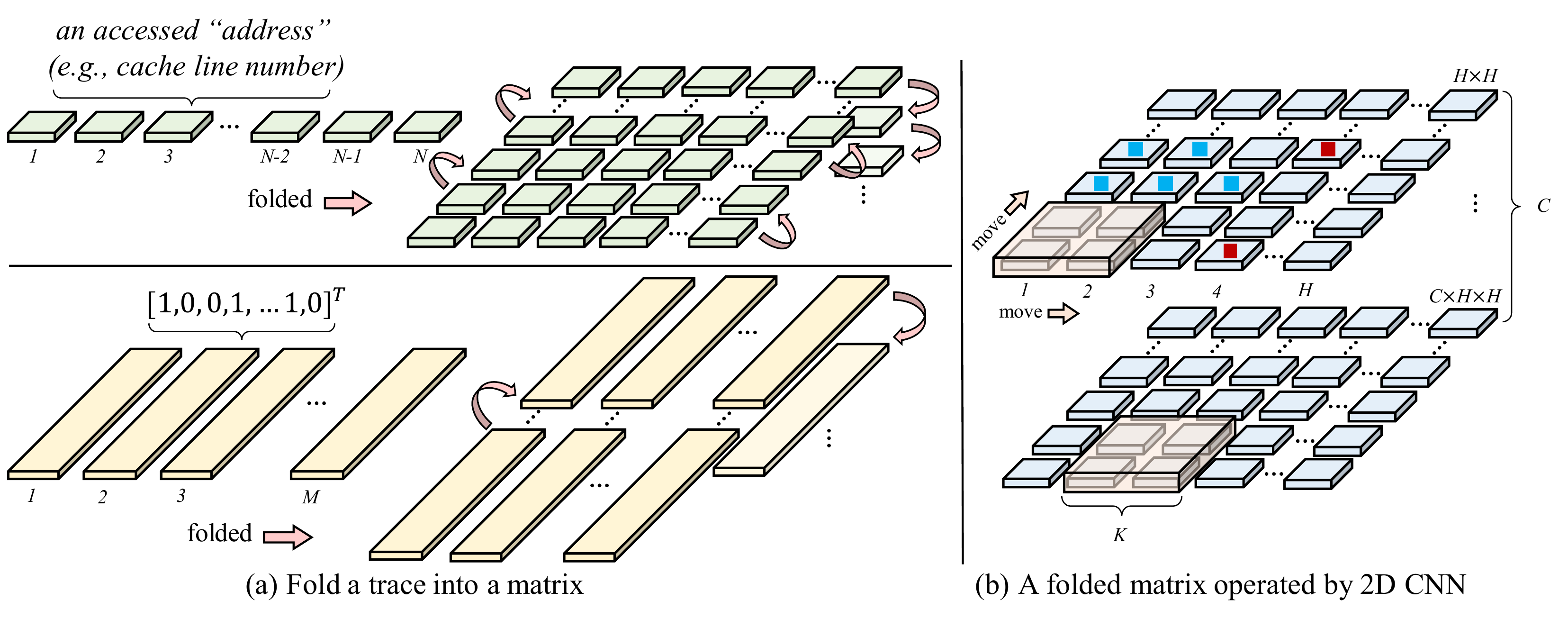}
      \caption{Representation and processing of side channel traces. Traces
        collected by \colorbox{pptgreen2}{\pin} and \colorbox{pptyellow1}{\pp}
        are marked in \colorbox{pptgreen2}{green} and
        \colorbox{pptyellow1}{yellow}, respectively. Two types of traces are
        processed by trace encoder of the same model structure.
        In general, our encoder enforces an important property named
        \textit{translation-invariance}. Suppose the secret-dependent records in the
        training traces only appear at the \colorbox{pptblue}{blue} locations,
        the $\phi_{\theta}$ can still capture secret-dependent records at the
        \colorbox{pptred}{red} locations in a testing trace. Therefore, even
        noise is introduced, the trace should still be reasonably informative unless
        all privacy-related records are completely vanished.}
  \label{fig:input}
  \end{figure*}

This section clarifies the representations of input side channels. We also
provide empirical results to explain why the representations work well in our
research context.

As aforementioned in the evaluation, we launch attacks using side channel traces
logged by \pin; this can mimic previous attacks where privileged system
software, e.g., OS kernels, are controlled by
adversaries~\cite{xu2015controlled,hahnel2017high}. We also launch standard
\pp\ attack to benchmark userspace-only scenarios~\cite{Tromer10,Zhang12}. As
shown in \F~\ref{fig:input}, each side channel trace collected by \pin\ contains
a sequence of records, and each record denotes one accessed cache unit or page
table entry index.
\F~\ref{fig:input} further shows how side channel logs are collected and organized using
\pp\ attack. Here, each trace composes a sequence of binary vectors $V$, where
the cardinality $|V|$ of each vector equals the number of cache sets (the value
is 64 in our CPU). $V[i] = 1$ indicates the $i$-th cache set is accessed by
victim.

\subsubsection*{Encoder Design \& Clarification on Its Noise Resiliency}

Our autoencoder framework is shown as effective to comprehend even lengthy side
channel records logged by either \pin\ or \pp: side channel traces are lengthy
albeit highly \textit{sparse}, where only a few elements are
\textit{secret-dependent}. Also, the secret-dependent records are not always
aligned, given that inputs of different values may lead to executing different
paths. Hence, the secret-dependent records may appear at different (relative)
locations on a trace in accordance with different inputs. In addition, our noise
resilience evaluation in \S~\ref{subsec:noise} illustrates highly encouraging
\textit{robustness} and \textit{resilience} of our autoencoder framework toward
perturbations on side channel traces.

In \S~\ref{subsec:design-attack}, we have clarified that manifold learning
itself manifests high capability of denoising. In this section, we further
analyze the noise resilience from the encoder model structure perspective.
Overall, before feeding a trace (logged by \pin\ or \pp) to our framework, each
trace is first folded into a squared matrix with zero-paddings.
\F~\ref{fig:input} depicts how a 2D CNN block of trace encoder $\phi_{\theta}$
operates on the folded matrix. A Conv2D block has $C$ kernels with each of shape
$K \times K$ ($K > 1$) and the input is a matrix of shape $C \times H \times H$.
Each kernel has its own parameters and operates on one $H \times H$ section of
the input. As the standard operation (see \F~\ref{fig:input}), the kernel moves
on the $H \times H$ section step by step, and therefore, all $K \times K$
regions in a section share the same kernel. In general, this classic design
ensures an important property named
\textit{translation-invariance}~\cite{lecun2004learning}. As shown in
\F~\ref{fig:input}, suppose the secret-dependent records in the training traces
only appear at the \colorbox{pptblue}{blue} locations, the $\phi_{\theta}$ can
still capture secret-dependent records at the \colorbox{pptred}{red} locations
in a testing trace. Therefore, even noise is introduced, the trace is still
sufficiently informative unless all privacy-related records are vanished. In
short, the well-known translation-invariance property enforced by our encoder can
properly improve the generalization and simultaneously guarantee the robustness
towards noise.

We are \textit{not} claiming credits from the model design; noise resilience
derived from translation-invariance is well studied in the AI
community~\cite{lecun2004learning,yuan2021private}. Nevertheless, we spend
efforts to explore other possible model structures at this step, which are seen
to manifest lower robustness toward noise. For instance, each record can be
attached to a unique weight (e.g., using a fully connected layer or a RNN to
process the trace). Consequently, any small perturbation, potentially due to
false cache hit or wrong orders, is seen to induce notable change in outputs.

%% file: case-study.tex
\begin{figure*}[!ht]
    \centering
    \includegraphics[width=0.8\linewidth]{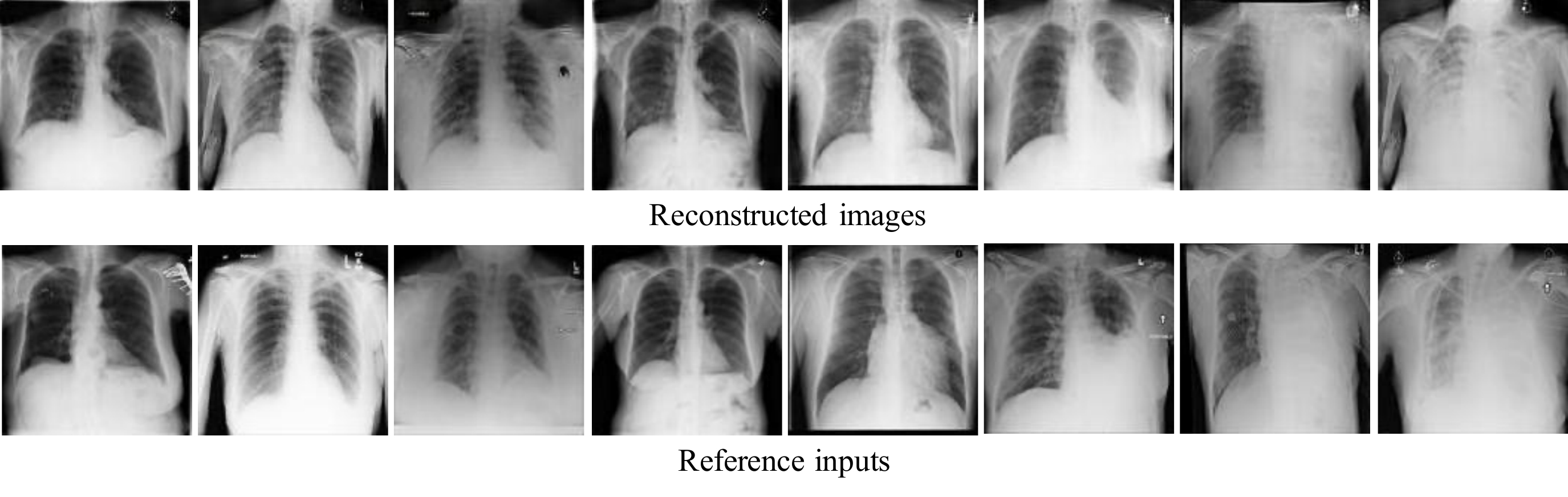}
    \caption{Qualitative evaluation results of Chest X-ray.}
    \label{fig:xray1}
\end{figure*}

\begin{figure*}[!ht]
    \centering
    \includegraphics[width=0.8\linewidth]{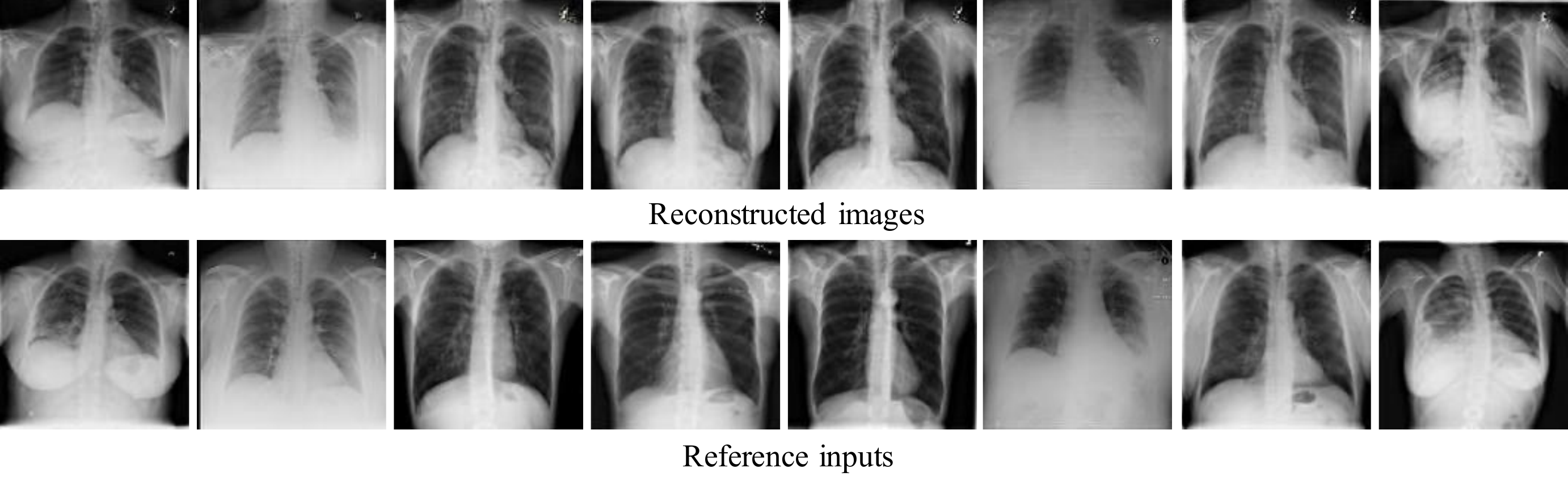}
    \caption{Qualitative evaluation results of Chest X-ray.}
    \label{fig:xray2}
\end{figure*}

\begin{figure*}[!ht]
  \centering
  \includegraphics[width=0.8\linewidth]{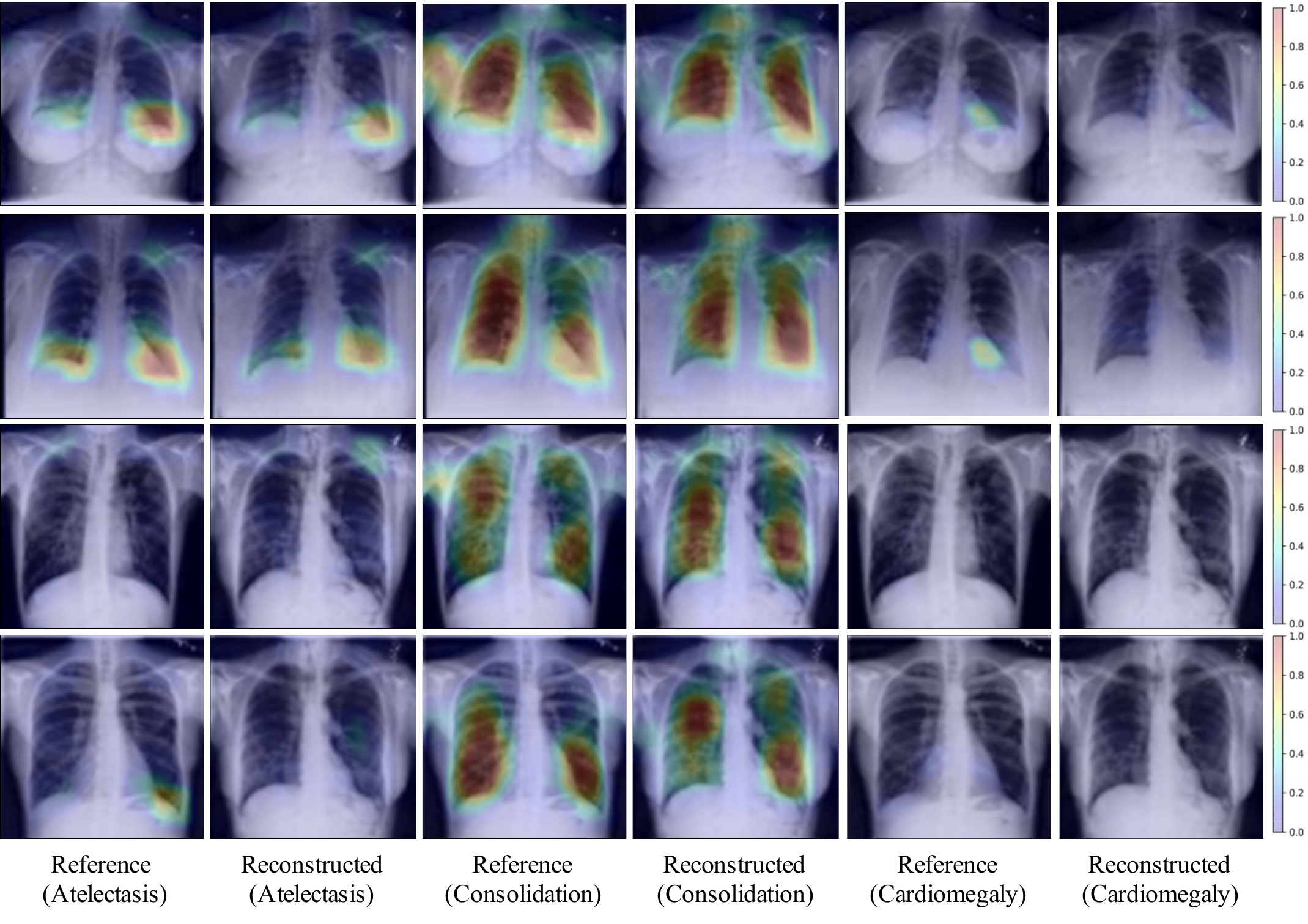}
  \caption{Localized lesions of Chest X-ray.}
  \label{fig:disease}
\end{figure*}

\begin{figure*}[!ht]
    \centering
    \includegraphics[width=0.8\linewidth]{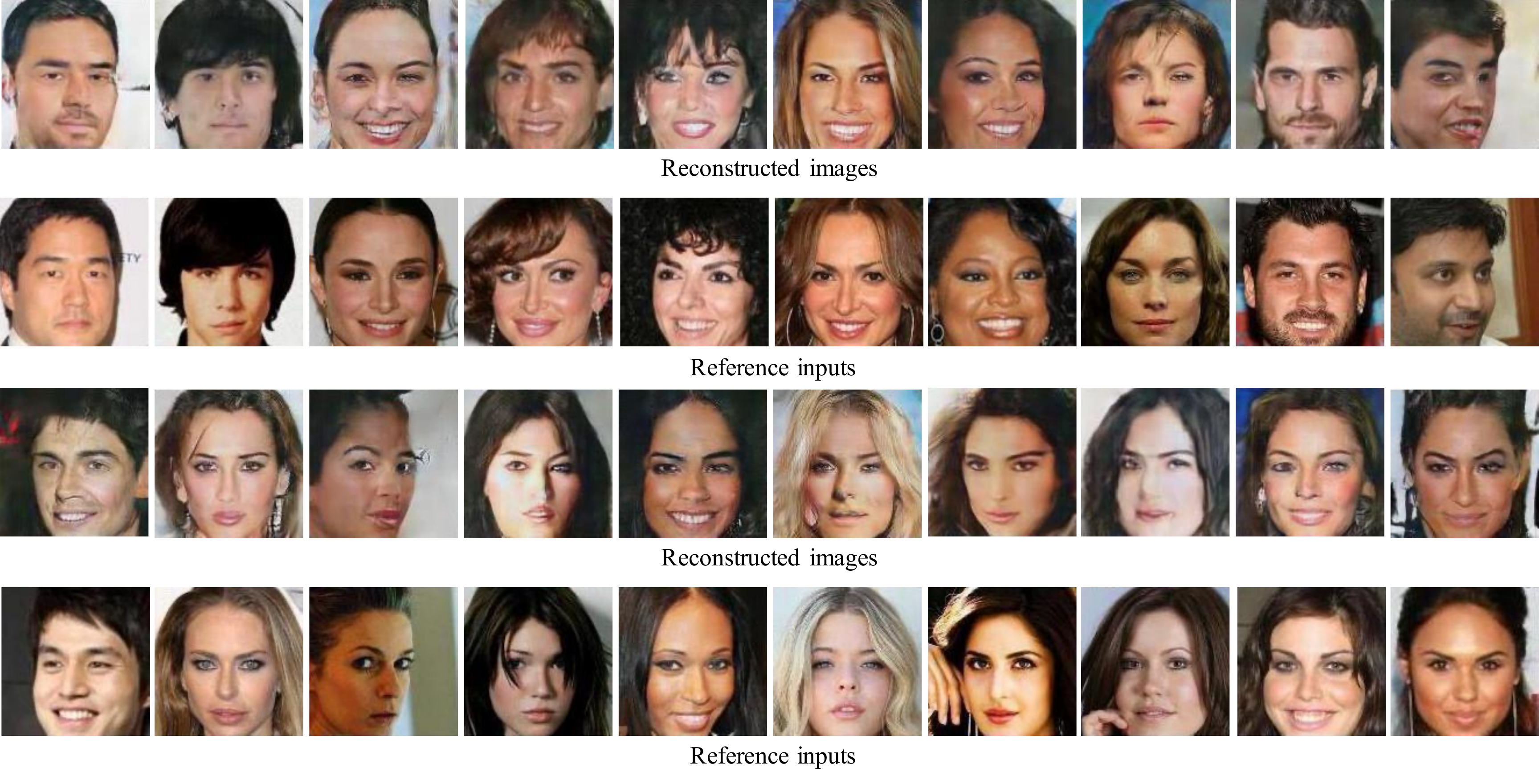}
    \caption{Qualitative evaluation of CelebA.}
    \label{fig:celeba1}
\end{figure*}

\begin{figure*}[!ht]
    \centering
    \includegraphics[width=1.0\linewidth]{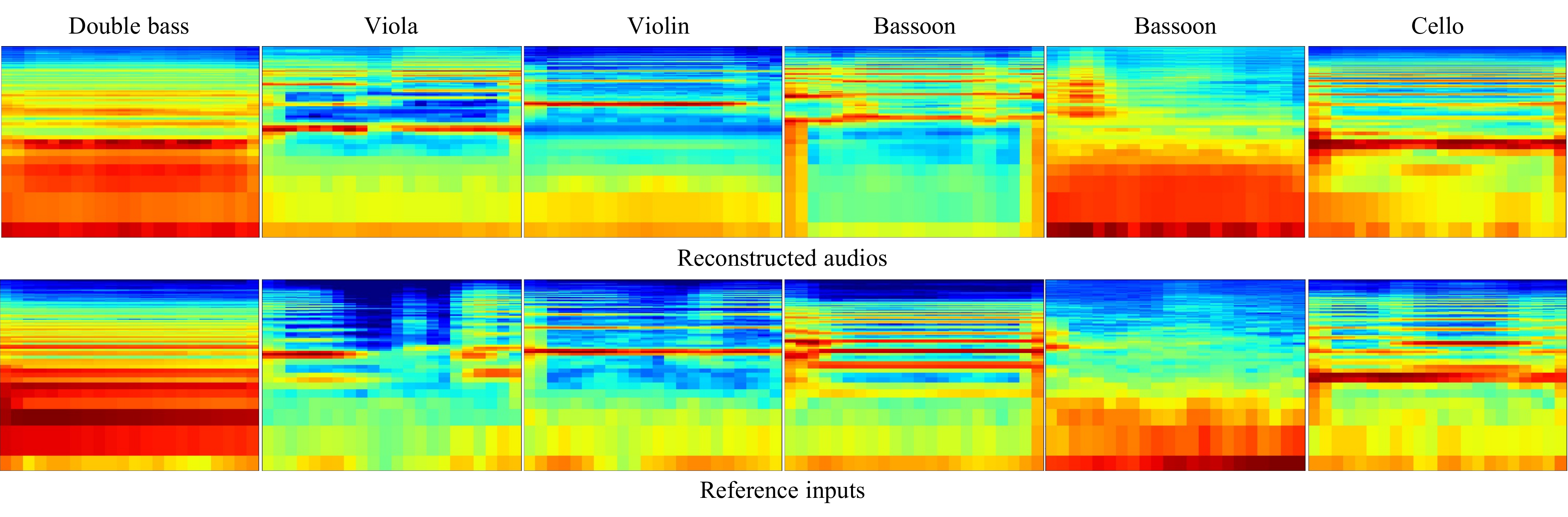}
    \caption{Qualitative evaluation of Sub-URMP. The horizontal axis of each LMS
    figure represents time from 0s to 0.5s, while the vertical axis represents
    frequency from 0Hz to 8192Hz.}
    \label{fig:audio1}
\end{figure*}

\begin{figure*}[!ht]
    \centering
    \includegraphics[width=0.8\linewidth]{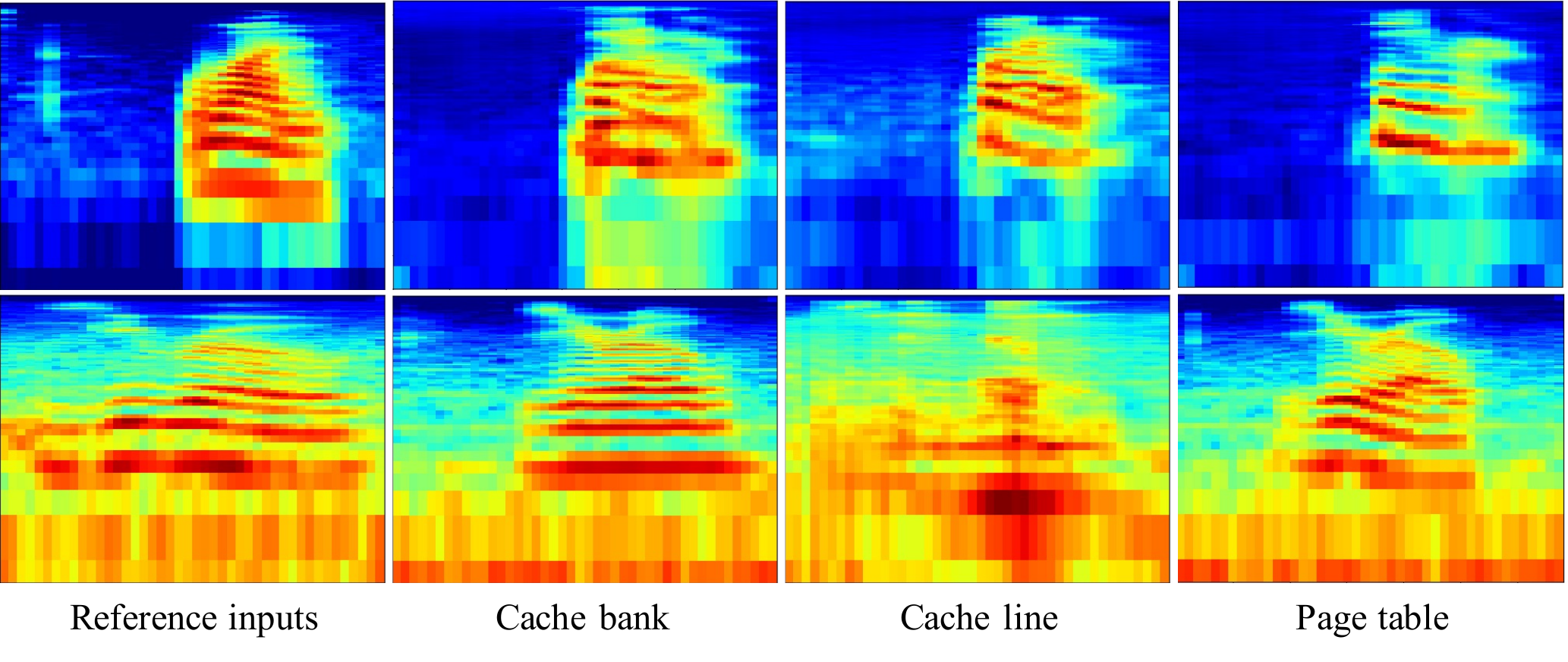}
    \caption{Qualitative evaluation of SC09. The horizontal axis of each LMS
      figure represents time from 0s to 1s, while the vertical axis represents
      frequency from 0Hz to 8192Hz. The first row is LMS figure of speaking
      ``one'', whereas the second row is speaking ``zero.''}
    \label{fig:audio}
\end{figure*}

\begin{figure*}[!ht]
  \centering
  \includegraphics[width=1.0\linewidth]{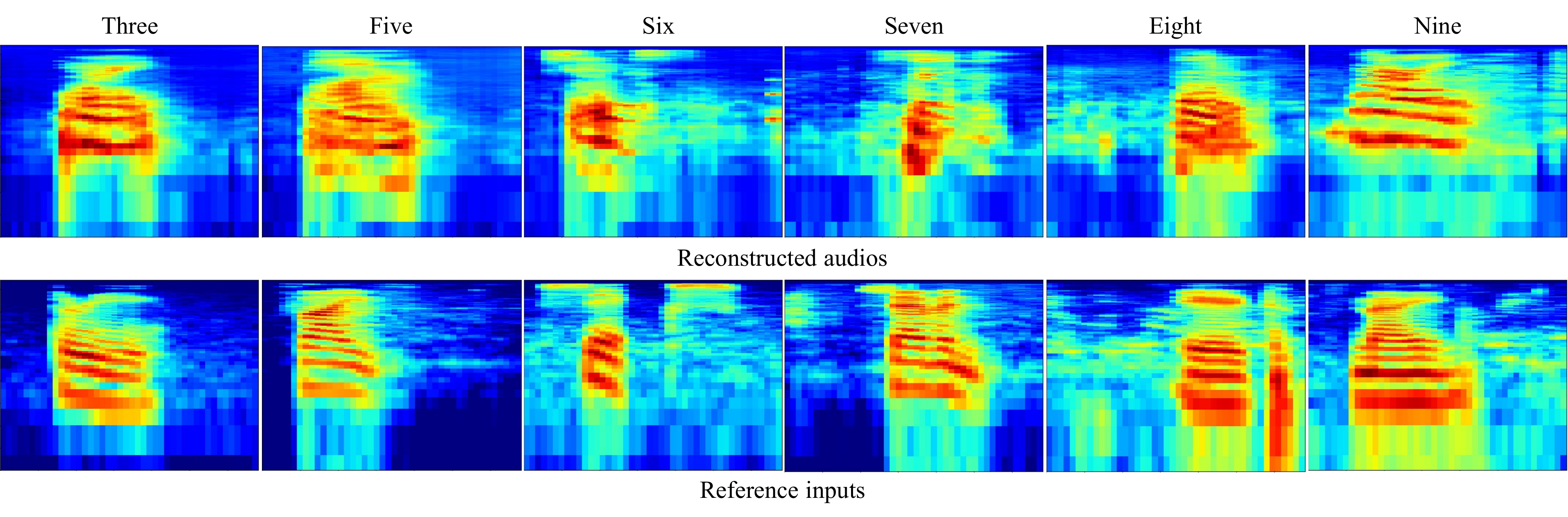}
  \caption{Qualitative evaluation of SC09. The horizontal axis of each LMS
  figure represents time from 0s to 1s, while the vertical axis represents
  frequency from 0Hz to 8192Hz.}
  \label{fig:audio2}
\end{figure*}

\begin{figure*}[!ht]
  \centering
  \includegraphics[width=1.0\linewidth]{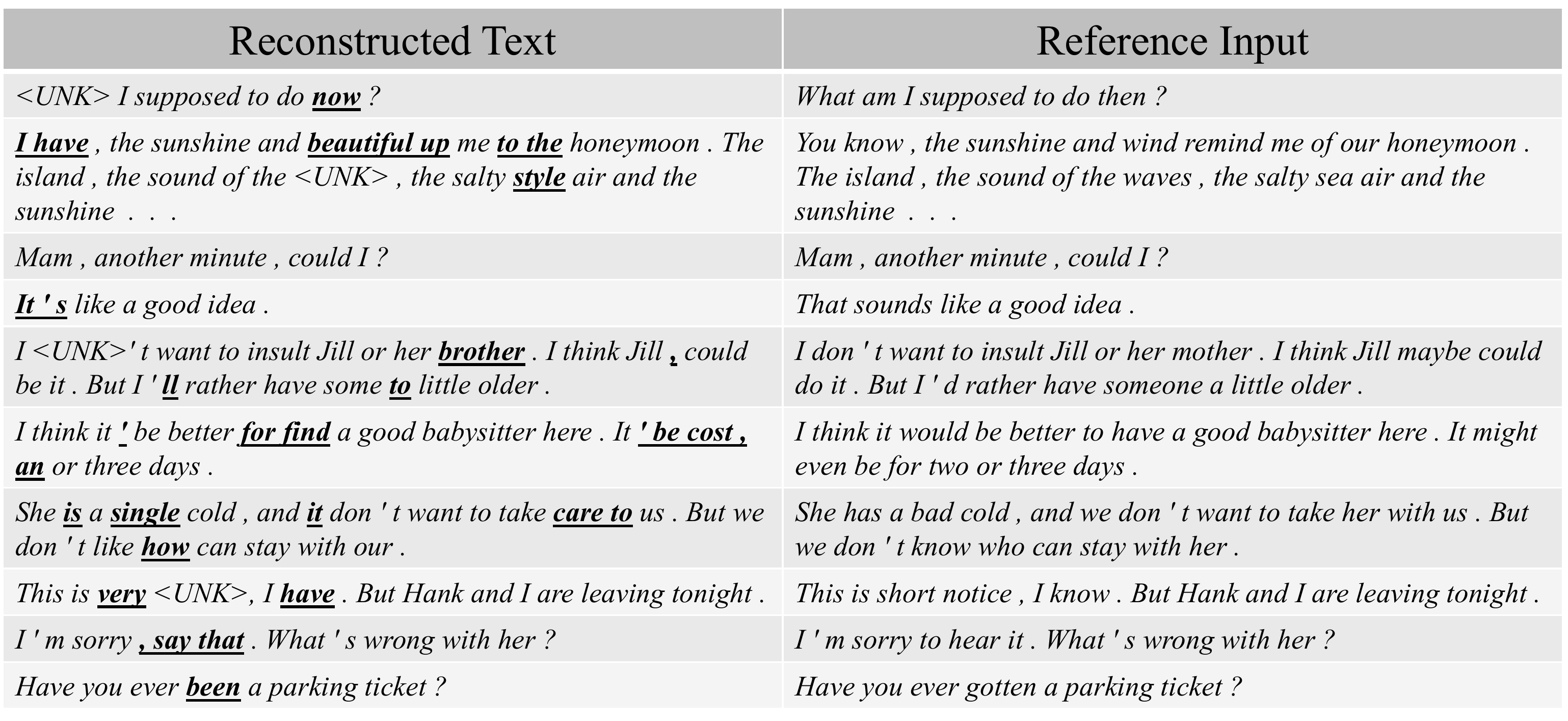}
  \caption{Qualitative evaluation of DailyDialog. We mark
  \underline{\textit{\textbf{inconsistent reconstructions}}}.}
  \label{fig:text2}
\end{figure*}

\begin{figure*}[!ht]
    \centering
    \includegraphics[width=1.0\linewidth]{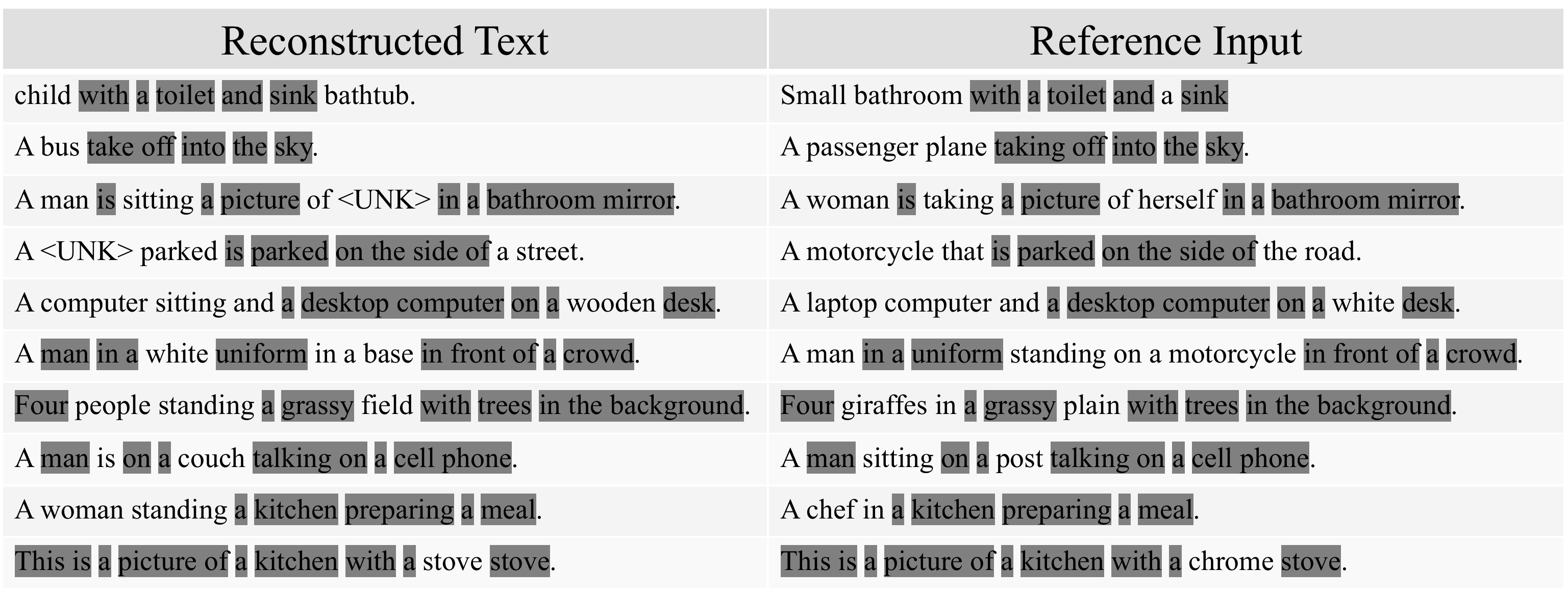}
    \caption{Qualitative evaluation results of COCO. We mark
      \colorbox{black!60}{consistent reconstructions}.}
    \label{fig:text1}
\end{figure*}


\begin{table}[t]
  \caption{Disease diagnosis matching rates of reconstructed chest X-ray images
    w.r.t.~(cache bank/cache line/page table) side channels.}
  \label{tab:xray}
  \centering
\resizebox{1.0\linewidth}{!}{
  \begin{tabular}{l|c|c|c}
    \hline
     Disease & \textbf{Cardiomegaly} & \textbf{Consolidation} & \textbf{Atelectasis} \\
    \hline
     Precision & 0.74/0.74/0.74    & 0.75/0.76/0.75 & 0.83/0.83/0.83           \\
    \hline
     Recall &    0.55/0.55/0.54       & 0.91/0.91/0.91 & 0.67/0.67/0.67           \\
    \hline
     F1 Score & 0.63/0.63/0.62     & 0.82/0.82/0.82 & 0.74/0.75/0.74           \\
    \hline
  \end{tabular}
  }
\end{table}

\begin{figure}[!ht]
    \centering
    \includegraphics[width=0.85\linewidth]{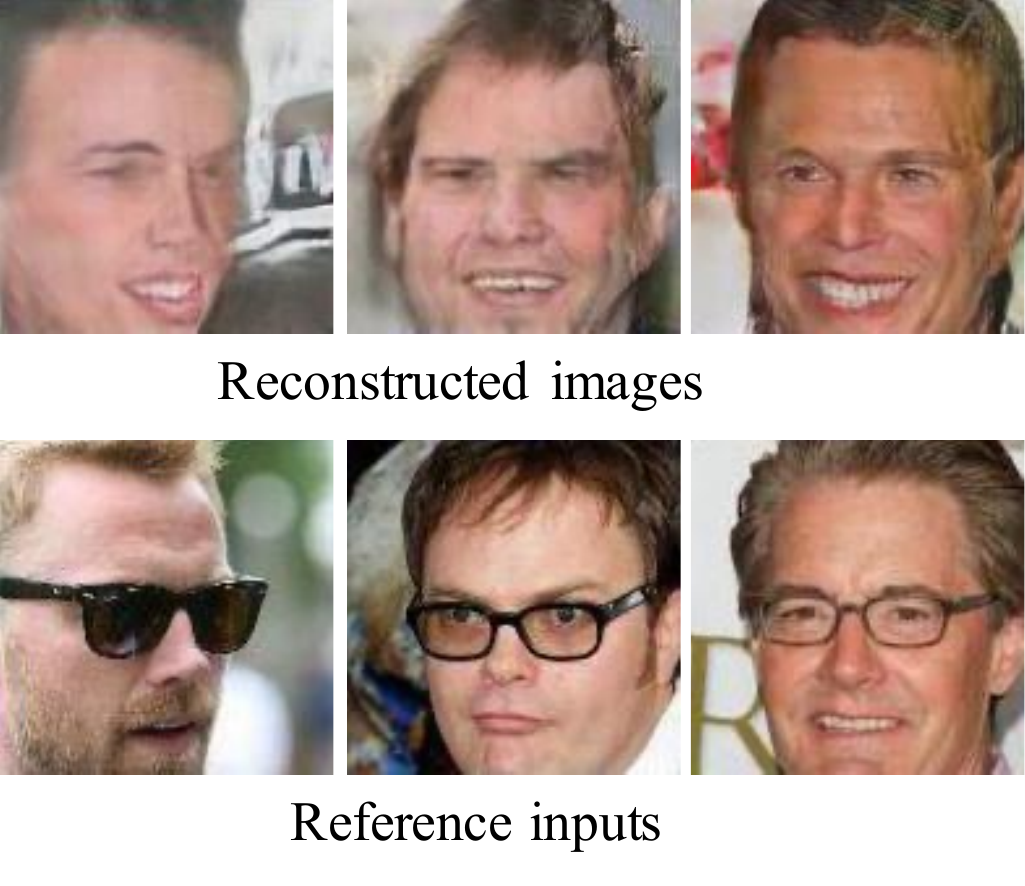}
    \caption{Reconstructed images with no glass attached.}
    \label{fig:glass}
\end{figure}


\section{SCA Results}
\label{sec:case}
This Appendix presents more reconstructed media data and compares them with the
reference inputs in \F~\ref{fig:xray1}, \F~\ref{fig:xray2},
\F~\ref{fig:celeba1}, \F~\ref{fig:audio1}, \F~\ref{fig:text2}, and \F~\ref{fig:text1}. We also
report the quantitative evaluation results of Sub-URMP dataset in
\T~\ref{tab:audio1}.

Similar to the promising results in reconstructing face
images reported in the evaluation section (\S~\ref{subsubsec:eval-qualitative}),
\F~\ref{fig:xray1} and \F~\ref{fig:xray2} show highly encouraging findings of
recovering chest X-ray images from side channel traces. Overall, we note that
the recovered X-ray images all manifest high visual quality and plausible
similarity with the reference inputs. Chest X-ray images contain generally less
perceptual features compared with CelebA, indicating an easier task for manifold
learning-based dimension reduction. On the other hand, we note that X-ray images
are of high resolution. To get the best results in disease diagnosis, it is
generally required that the recovered X-ray images preserve original details in
the reference inputs. \T~\ref{tab:xray} reports the quantitative evaluation of chest
X-ray images. We check whether the same kinds of diseases can be diagnosed from
the reference and reconstructed inputs, which would indicate serious privacy
leakage. Considering three diseases listed in \T~\ref{tab:xray}, we first train
a disease classifier $C$ (see \appx~\ref{sec:attack-appendix}) that achieves an F1 score above 98.0\% over the
reference inputs. Then, let true positive $TP$ be the diagnosis of the same
disease $d$ by $C$ from both reference and reconstructed X-ray images. False
positive $FP$ of a disease $d$ implies that $d$ is diagnosed from the
reconstructed chest X-ray image but not diagnosed from the input. Similarly,
false negative $FN$ indicates that a disease is diagnosed from the reference
input but it does not from the reconstructed image. We can thus compute the
precision ($\frac{TP}{TP + FP}$), recall ($\frac{TP}{TP + FN}$), and F1 score
(harmonic mean of the precision and recall), respectively. Overall,
\T~\ref{tab:xray} indicates that accurate disease information can be inferred
from the reconstructed X-ray images.

The high accuracy in conducting disease diagnosis over reconstructed images in
\T~\ref{tab:xray} indicates a serious leakage of patient's confidential information.
We also localize lesions (see
\appx~\ref{sec:attack-appendix} clarifying attack setup at this step) of
reconstructed X-ray images and corresponding reference inputs in
\F~\ref{fig:disease}: lesions in reconstructed images and reference inputs are
highly consistent.

\F~\ref{fig:audio} reports the evaluation results of reconstructing audio data
in LMS graphs, whose x-axis represents time and y-axis denotes log-amplitude of
frequencies. The references and reconstructed recordings have mostly consistent
LMS graphs in all cases. As mentioned in \T~\ref{tab:data-statistics} and
Appendix~\ref{sec:attack-appendix}, SC09 includes real-world ``speech
commands''; hence, recovering quality voice recording indicates a strong
likelihood of violating user privacy, such as voice commands. Given that the LMS
graphs of audio clips are usually too dense to read on paper, we have also
released reconstructed audio recordings~\cite{snapshot}; interested readers can
easily verify that the reconstructed audio clips exhibit high quality with
negligible noise.

\smallskip
\noindent \textbf{Neglecting Non-Privacy Factors.}~Our manual inspection on the
reconstructed images also reveal several interesting cases in As presented in
\F~\ref{fig:glass}, while the reconstructed images of celebrity faces are of
good quality and highly similar to the reference inputs, the reconstructed
images do not contain glasses. On one hand, we emphasize that the reconstructed
images can be matched to the reference inputs with above 99.9\% confidence score
using the commercial face recognition APIs provided by Face++. This indicates
that the ``privacy'' related features that can uniquely recognize human
identities are successfully preserved in the reconstructed photos. More
importantly, we interpret this evaluation has shown the effectiveness of
our customized objective functions which aim to retain specific privacy
indicators. Recall as defined in \T~\ref{tab:indicator}, we enforce our
autoencoder framework to particularly retain human privacy related contents like
gender. As a result, our autoencoder framework narrows the focus to critical
features and facilitates commercial APIs extracting key perceptual contents.
Non-privacy factors (e.g., glasses) are generally discouraged to distract the
attention of our attack framework.

\begin{figure}[!ht]
    \centering
    \includegraphics[width=1.0\linewidth]{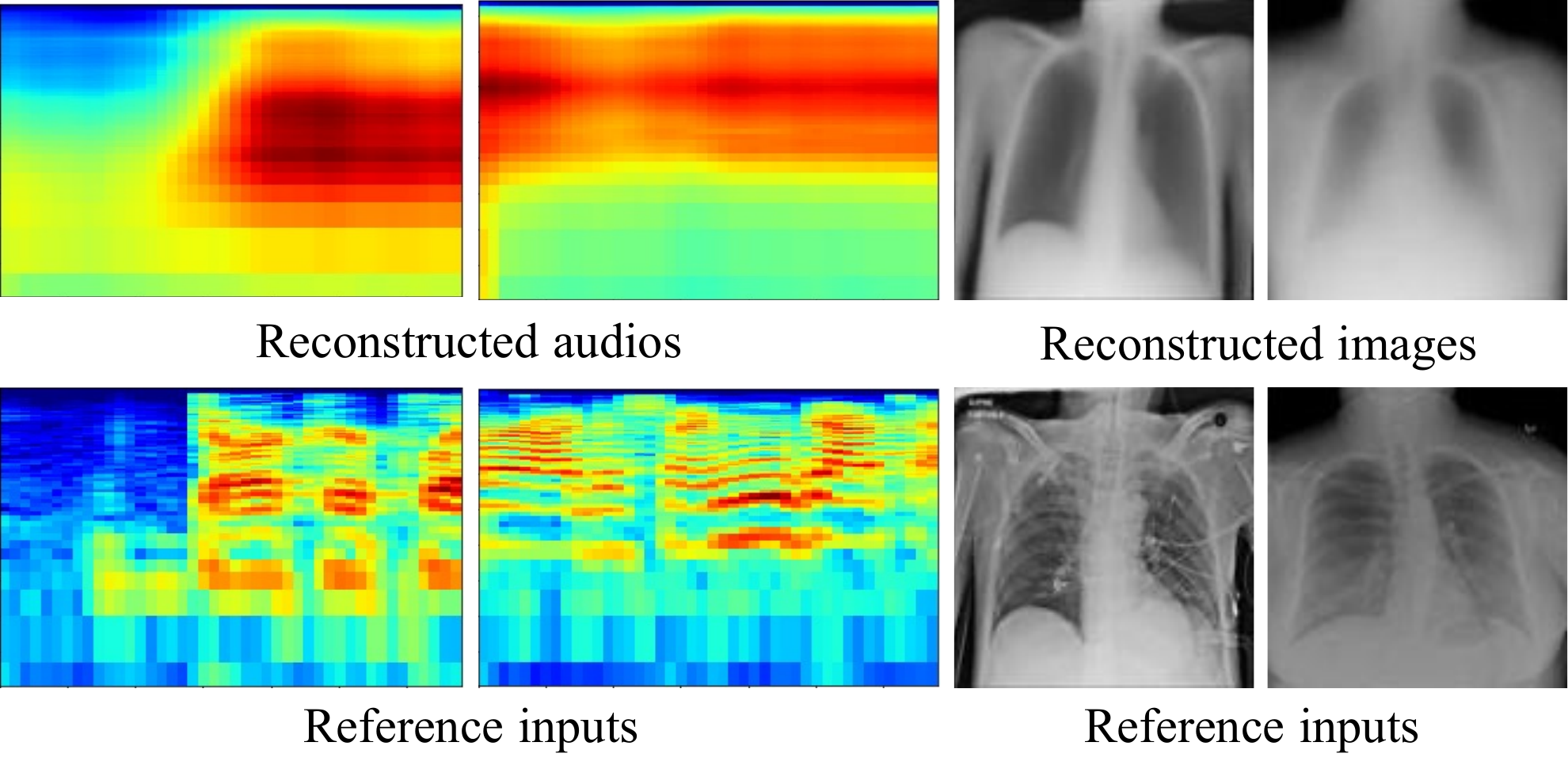}
    \caption{Reconstructed images suffering from over-smoothing.} 
    \label{fig:oversmooth}
\end{figure}

\smallskip
\noindent \textbf{Ablation Evaluations.}~Recall in designing objective
functions, we explain that using only pairwise distance metrics can likely
generate ``over-smoothing'' images, thus clarifying the necessity of composing
the objective function with discriminator to implicitly capture the ``semantics
similarity'' among reference and reconstructed images. \F~\ref{fig:oversmooth}
presents several reconstructed images by only using the pairwise distance. It is
easy to see that the reconstructed images exhibit low quality compared with
sample cases shown in \F~\ref{fig:xray1} and \F~\ref{fig:audio}. In particular,
it suffers from over-smoothing, where the \textit{details} in the LMS images and
chest X-ray photos become much blur, comparing with the reference inputs. We
view the results empirically demonstrate the necessity and strength of adopting
hybrid objective functions as the learning goal of our autoencoder framework.
Also, mode collapse may be potentially introduced by the implicit objective
function. Nevertheless, we clarify that high discriminability of the
reconstructed media data empirically demonstrates that mode collapse is
\textit{not} a major concern of our framework.

\smallskip
\noindent \textbf{Vulnerable Code.}~In addition to vulnerable code localized in
\ffmpeg\ and reported in \S~\ref{subsec:eval-localization}, we further report
vulnerable code in \libjpeg\ and \hunspell\ in \F~\ref{fig:libjpeg-vul} and
\F~\ref{fig:hunspell-vul}, respectively. Both code fragments contain obvious
input-dependent memory accesses (we mark program variables derived from inputs
in \textcolor{red}{red}). These input-dependent memory accesses further lead to
the accesses of different cache units or page table entries, thus enabling side
channel exploitations. To the best of our knowledge, \texttt{MCU} and
\texttt{putdic}, as vulnerable code fragments of \libjpeg\ and \hunspell, were
not pointed out by existing works~\cite{xu2015controlled,hahnel2017high}. In
addition, we identify these stealthy code fragments fully automatically using
neural attention mechanisms. This highlights the strength of our neural
approach. And to confirm our findings, we manually analyzed how inputs are
propagated into certain program variables, and how those program variables are
used to access memory (and further lead to side channels).

Also, we report all the localized assembly instructions that primarily
contribute to the reconstruction of private inputs. The corresponding line
number of localized assembly instructions are reported in~\cite{snapshot}. We
use the default configuration to compile each \ms, whose compiled executable
file can also be found in our released repository~\cite{snapshot}. Hence,
developers can use our reported information and released executable files to
further localize and patch relevant code fragments.

\begin{figure}[t]
\centering
\begin{lstlisting}
int HUFF_EXTEND(int x, int @s@) {
  // ``ex_test'' and ``ex_offset'' are
  // pre-calculated arrays
  <@\textbf{if (x < ex_test[\textcolor{red}{s}])}@>
    <@\textbf{return x + ex_offset[\textcolor{red}{s}];}@>
  else
    return x;
  }
  
boolean decode_mcu_fast(j_decompress_ptr @cinfo@,
  JBLOCKROW *MCU_data) {
  <@huff_entropy_ptr \textcolor{red}{entropy} = @>
  <@  (huff_entropy_ptr)\textcolor{red}{cinfo}->entropy;@>
  /* preprocessing */
  <@for (int i = 0; i < \textcolor{red}{cinfo}->blocks_in_MCU; i++) {@>
    <@d_derived_tbl *\textcolor{red}{dctbl} = \textcolor{red}{entropy}->dc_cur_tbls[i];@>
    int s, k, r, l;
    /* get index ``idx'' based on ``s'' */
    /* update ``r'' */
    <@\textbf{\textcolor{red}{s} = \textcolor{red}{dctbl}->lookup[idx];}@>
    // ``lookup'' is pre-calculated array
    <@if (\textcolor{red}{s}) {@>
      <@\textbf{s = HUFF_EXTEND(r, \textcolor{red}{s});}@>
    }
    /* do something */
  }
  /* do something and return */
} 
\end{lstlisting}
\caption{Vulnerable code components in \libjpeg. We mark variables depending on
  \libjpeg's input in \textcolor{red}{red}, and \textbf{bold} input-dependent
  memory accesses (e.g., line 4).}
\label{fig:libjpeg-vul}
\end{figure}

\begin{figure}[t]
  \centering
  \begin{lstlisting}
  boolean check(Hunspell** pMS, int* @d@, string& token) {
    // checking and transforming encoding of ``token''
    <@\textbf{if pMs[*\textcolor{red}{d}]->spell(token)}@>
      return true;
    return false;
    // ``pMs'' is a hash table storing the dictionary
    // spell() performs spell checking
  }
  
  int putdic(const std::string& @word@, Hunspell* pMS) {
    // checking and transforming encoding of ``word''
    size_t w = word.find(`/`, 1);
    if (w == std::string::npos) {
      if (word[0] == `*`)
        <@\textbf{ret = pMS->remove(\textcolor{red}{word}.substr(1));}@>
      else
        <@\textbf{ret = pMS->add(\textcolor{red}{word});}@>
    } else {
      std::string affix = word.substr(w + 1);
      word.resize(w);
      if (!affix.empty() && affix[0] == `/`)
          affix.erase(0, 1);
      <@\textbf{ret = pMS->add_with_affix(\textcolor{red}{word}, affix);}@>
    }
    // ``pMs'' is a hash table storing the dictionary
    return ret;
  }
  \end{lstlisting}
  \caption{Vulnerable code components in \hunspell. We mark variables depending on \hunspell's input in \textcolor{red}{red}, and \textbf{bold} input-dependent memory accesses (e.g., line 3).}
  \label{fig:hunspell-vul}
  \end{figure}

\begin{figure}[!ht]
  \centering
  \includegraphics[width=1.00\linewidth]{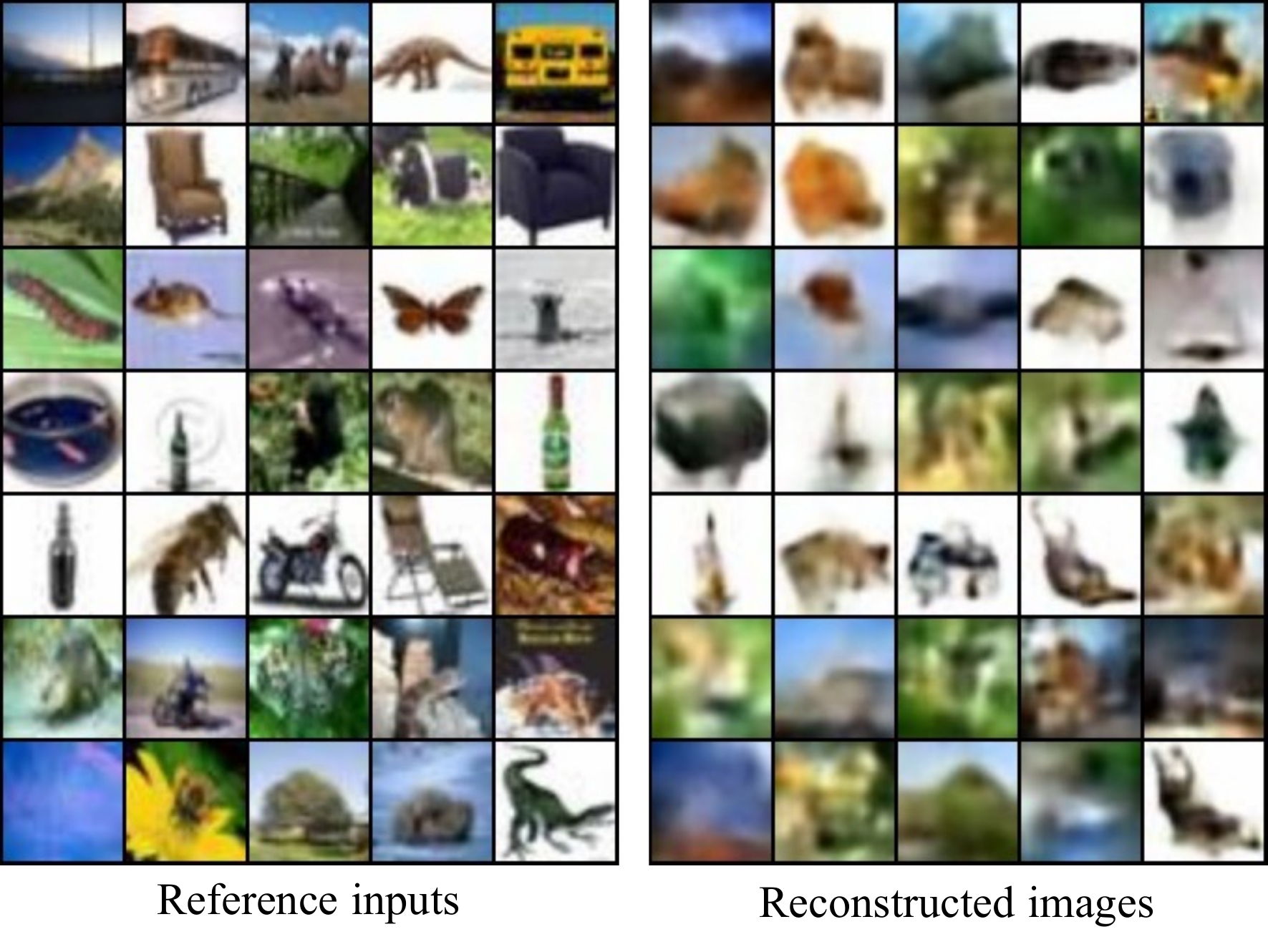}
  \caption{Reconstructed images on CIFAR-100.}
  \label{fig:cifar100}
\end{figure}

\begin{table}[t]
\caption{Generalization evaluation.}
\label{tab:general}
\centering
\resizebox{0.6\linewidth}{!}{
\begin{tabular}{c|c|c|c}
\hline
       ($k,N$)            & ($1,100$)  & ($5,100$) & ($20,100$) \\ \hline
 \texttt{libjpeg}  & 5.7\%      & 20.3\%    & 44.1\%     \\
 Baseline          & 1\%        & 5\%       & 20\%      \\ \hline
  \end{tabular}
}
\end{table}

\smallskip
\noindent \textbf{Generalization.}~As explained in \S~\ref{sec:manifold},
different types of media data (e.g., human face vs. vehicle photos) are
generally projected toward distinct manifold
spaces~\cite{belkin2001laplacian,roweis2000nonlinear,tenenbaum2000global}.
Hence, training a unified model to recover media data of different classes are
beyond the scope of our SCA.

To thoroughly explore the potential limitations, this section provides an
empirical assessment of the generalization of our SCA framework. To this end, we
collect side channel traces logged when using \libjpeg\ to process a general
dataset, CIFAR-100~\cite{krizhevsky2009learning}. These side channel traces are
used to train our autoencoder framework and to reconstruct unknown images in
CIFAR-100 from their induced side channel traces. The CIFAR-100 dataset
comprises 60K images of 100 classes. Each class of 600 images is divided into
500 training images and 100 testing images.

\F~\ref{fig:cifar100} reports the qualitative evaluation results, by comparing
the inference inputs with their reconstructed images. In general, we observed
that the reconstructed images manifest much worse visual quality than those
synthesized from specific datasets (e.g., CelebA and Chest X-ray). Nevertheless,
we still observed that some expressive features, particularly sketch and color,
are retained in the reconstructed images. 

For quantitative evaluation, we use SSIM~\cite{wang2004image} to assess
similarity between reconstructed images and reference inputs.
\T~\ref{tab:general} reports the evaluation results of determining whether the
reference input appears in the top-$k$ of $N$ (e.g., top-$5$ out of randomly
selected 100) images matched with the reconstructed images. The overall matching
rates are high and greatly outperform the baseline --- random guess. These
findings indicate the promising potential of our approach to exploit arbitrary
datasets and recover confidential user inputs. Though the reconstructed media
data was not visually vivid, it still notably facilities privacy stealth.

Note that a conventional approach in AI community is to provide the class label
of each image~\cite{brock2018large}. In short, the model can switch to proper
manifold according to image labels. Our experiment does not provide class labels
and mix all images together to faithfully explore model's generalization
capability. We leave it as one future work to enhance generalization with
labeled data and more advanced models.

%% file: mitigation.tex
\begin{figure*}[t]
  \centering \includegraphics[width=0.9\linewidth]{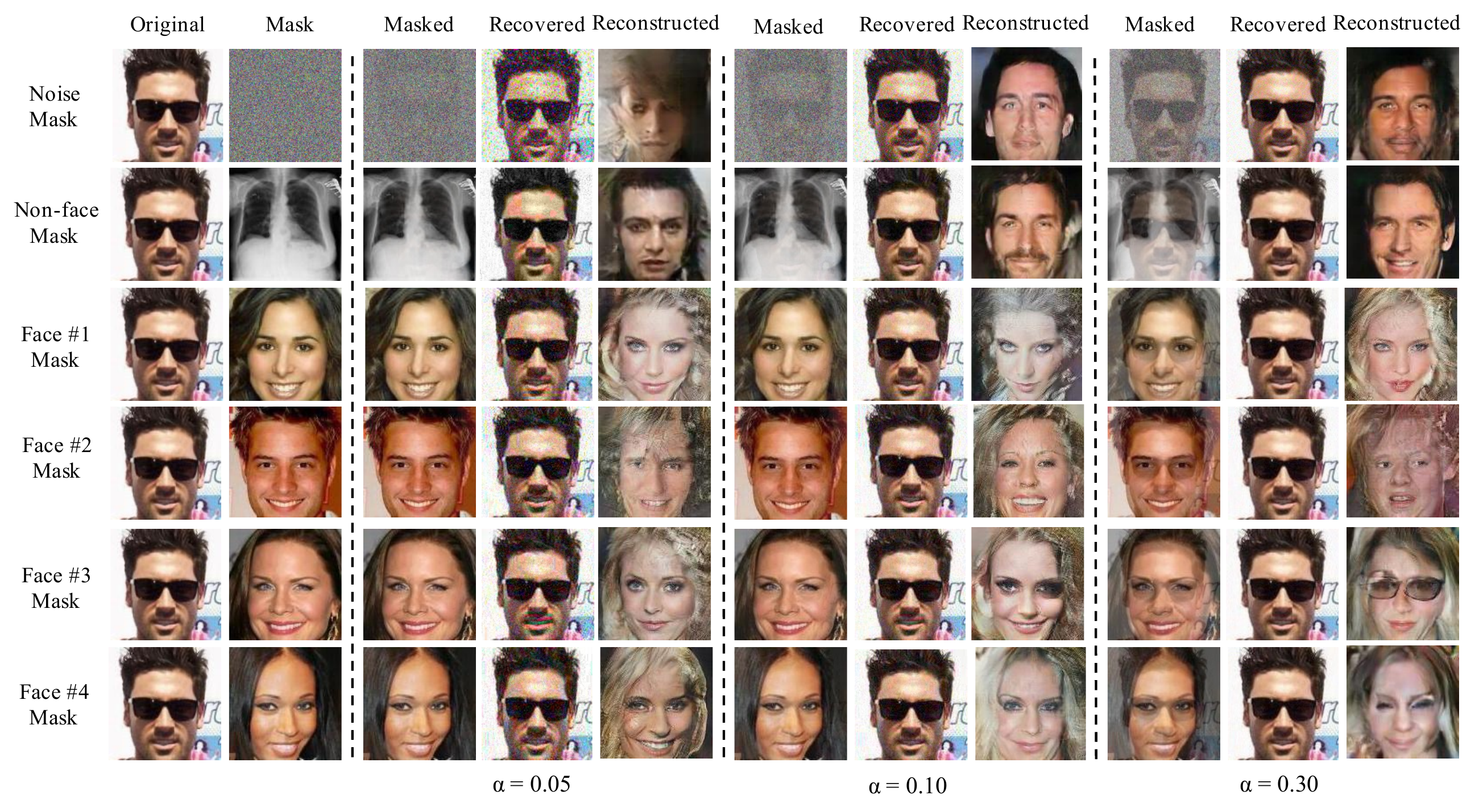}
  \caption{Qualitative evaluation results of perception blinding.}
  \label{fig:blinding1}
\end{figure*}

\begin{figure*}[t]
  \centering \includegraphics[width=0.9\linewidth]{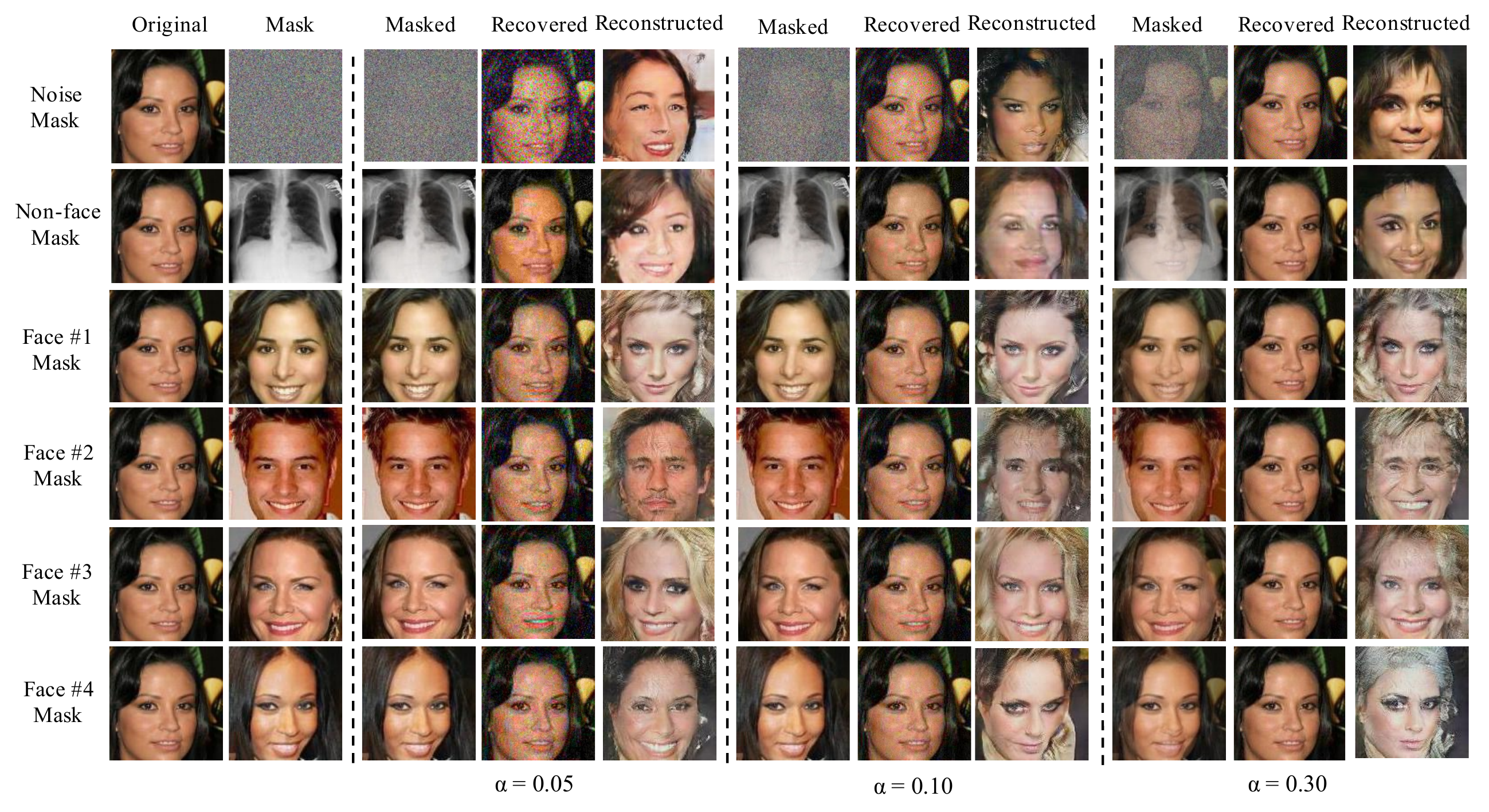}
  \caption{Qualitative evaluation results of perception blinding.}
  \label{fig:blinding2}
\end{figure*}

\section{Mitigation}
\label{sec:mitigation}
In accordance with \S~\ref{subsec:eval-mitigation}, \F~\ref{fig:blinding1}
further reports qualitative evaluation results of perception blinding-based
mitigation. In short, \F~\ref{fig:blinding1} manifests results and findings
mostly comparable with evaluations launched in \S~\ref{subsec:eval-mitigation}.
That is, ``noise mask'' (the first row) and ``non-face mask'' (the second row)
are generally not effective in blinding $i_{private}$; perceptual features can
be seen in the reconstructed images (e.g., face poses, hair style). Using real
face images from the CelebA dataset as the masks manifest highly encouraging
results to blind key perceptual-level features. In addition, to recover final
outputs (i.e., the ``Recovered'' columns), a larger $\alpha$, at least greater
than 0.10, should be desirable.

Regarding quantitative evaluation, \T~\ref{tab:xray-blinding} reports the
adversary disease diagnosis results after applying three blinding masks toward
the Chest X-ray dataset. Consistent with our findings from the CelebA dataset,
using ``X-ray\#1'' image to blind chest X-ray images can achieve a much better
result to reduce the disease diagnosis accuracy. We also report the quantitative
evaluation results of mitigating DailyDialog datasets in
\T~\ref{tab:mitigation-text2}, which manifest mostly consistent results with
\T~\ref{tab:mitigation-text1}.

\begin{table*}[t]
  \centering
  \caption{Mitigating adversary disease diagnosis attack on Chest X-ray dataset.
    We report text data inference accuracy in terms of cache bank/cache
    line/page table. We use three blinding masks as ``Noise'' ``Non-X-Ray'' and
    a real X-ray photo ``X-Ray \#1''.}
  \label{tab:xray-blinding}
  \resizebox{0.60\linewidth}{!}{
  \begin{tabular}{l|c|c|c|c}
    \hline
   \textbf{Mask} & Disease & $\alpha = 0.05$ & $\alpha = 0.1$ & $\alpha = 0.3$ \\ 
    \hline
      Noise     & Cardiomegaly  & 0.61/0.55/0.54 & 0.61/0.54/0.53 & 0.63/0.58/0.53 \\ 
                & Consolidation & 0.82/0.82/0.81 & 0.82/0.81/0.81 & 0.82/0.81/0.81 \\
                & Atelectasis   & 0.60/0.60/0.57 & 0.60/0.56/0.57 & 0.61/0.57/0.55 \\
    \hline
      Non-X-Ray & Cardiomegaly  & 0.61/0.54/0.58 & 0.63/0.55/0.56 & 0.63/0.52/0.51 \\ 
                & Consolidation & 0.81/0.81/0.82 & 0.82/0.82/0.82 & 0.81/0.81/0.81 \\
                & Atelectasis   & 0.61/0.55/0.58 & 0.65/0.55/0.58 & 0.61/0.57/0.54 \\
    \hline
      X-Ray \#1 & Cardiomegaly  & 0.15/0.19/0.14 & 0.20/0.21/0.19 & 0.20/0.27/0.20  \\ 
                & Consolidation & 0.73/0.83/0.79 & 0.74/0.82/0.80 & 0.74/0.85/0.83 \\
                & Atelectasis   & 0.08/0.07/0.08 & 0.12/0.11/0.15 & 0.19/0.12/0.14 \\
    \hline
  \end{tabular}
  }
\end{table*}

\begin{table}[t]
  \caption{Mitigating human voice matching attack on SC09 dataset with blinding.
    We report text data inference accuracy in terms of cache bank/cache
    line/page table. We use three blinding masks as ``Noise'' ``Non-Voice'' and
    a real data sample ``Voice''.}
  \label{tab:voice-blinding}
  \centering
\resizebox{1.0\linewidth}{!}{
  \begin{tabular}{l|c|c|c}
    \hline
     Mask & $\alpha = 0.05$ & $\alpha = 0.1$ & $\alpha = 0.3$ \\
    \hline
     Noise & 13.1/12.8/12.9\% & 15.2/15.3/15.1\% & 20.3/19.9/20.1\% \\
    \hline
     Non-voice & 14.2/13.9/14.1\% & 17.1/17.0/17.3\% & 20.2/20.0/19.8\%  \\
    \hline
     Voice & 7.2/7.0/7.0\% & 7.1/7.8/7.6\% & 8.7/8.5/9.1\%  \\
    \hline
  \end{tabular}
  }
\end{table}

\begin{table}[t]
  \caption{Mitigating DailyDialog text inference attack. We report text data
    inference accuracy in terms of cache bank/cache line/page table. $\alpha =
    0.05$ denotes word appended with total 19 masks. $\alpha = 0.1$ denotes word
    appended with total 9 masks, while $\alpha = 0.3$ denotes word appended with
    total 2 masks.}
  \label{tab:mitigation-text2}
  \centering
\resizebox{1.0\linewidth}{!}{
  \begin{tabular}{l|c|c|c}
    \hline
     \textbf{Mask} & \textbf{$\alpha = 0.05$} & \textbf{$\alpha = 0.1$} & \textbf{$\alpha = 0.3$} \\
    \hline
     ``I'' & 0.19/0.32/0.18\%   & 0.18/0.34/0.19\% & 0.50/0.88/0.38\% \\ 
    \hline
     ``you'' & 0.22/0.33/0.22\% & 0.24/0.34/0.26\% & 0.45/0.72/0.53\%  \\ 
    \hline
  \end{tabular}
  }
\end{table}

%% file: noise.tex
\section{Noise Resiliency Evaluation Setup}
\label{sec:noise}

\begin{figure}[t]
  \centering 
   \includegraphics[width=1.0\linewidth]{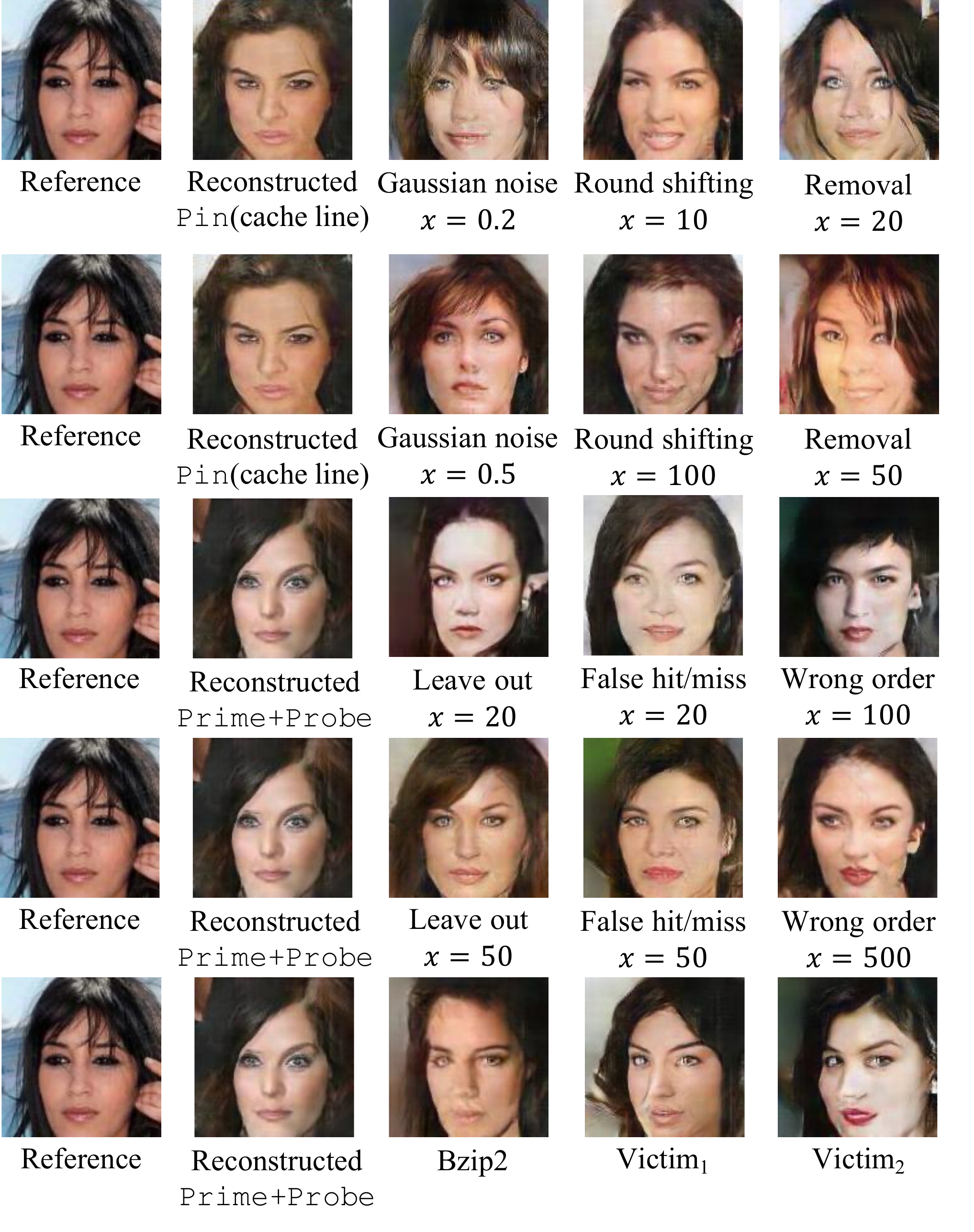}
   \caption{Evaluation results of \texttt{CelebA} dataset in terms of 9
     different noise settings.}
  \label{fig:noise}
\end{figure}

This section elaborates on the setup of noise resiliency evaluation launched in
\S~\ref{subsec:noise}. In sum, we evaluate adding noise to cache line access
traces logged by \pin\ and cache set access traces logged via \pp\ on Intel L1D cache. First, given a
\pin\ logged side channel trace, we leverage the following schemes to insert noise
into the trace.

\noindent \underline{Gaussian noise.}~We perturb every record $d$ on a side
channel trace using $d = x \times n + (1 - x) \times d$, where $x \in \{0.2
  ,0.5\}$. $n$ is randomly generated noise following Gaussian distribution.

\noindent \underline{Removal.}~We randomly remove $x\%$ of the data points on
the side channel, where $x \in \{20, 50\}$.

\noindent \underline{Round shifting.}~We round shift a side channel trace for
$x$ steps, where $x \in \{10, 100\}$.

Note that each noise insertion scheme has two configurations. For the ease of
presentation, we use \textbf{Low} and \textbf{High} to denote two
configurations, respectively. For instance, \textbf{Gaussian/High} denotes an
intensive setting such that we apply Gaussian noise to perturb side channel
traces when $x = 0.5$.
As for the cache set access traces logged via \pp, we also launch the following
three perturbation schemes:

\noindent \underline{Leave cache hit/miss out.}~We randomly drop $x\%$ of the
cache set hit/miss records on the logged trace, where $x \in \{20,
50\}$.

\noindent \underline{False cache hit/miss.}~We randomly flip $x\%$ of
records in the logged cache set access trace, where $x \in \{20, 50\}$.
This way, we create false cache hits/misses.

\noindent \underline{Wrong order of cache hit/miss.}~We randomly select $x$
non-repetitive cache set hit/miss records and compose $\sfrac{x}{2}$ pairs of
records, where $x \in \{100, 500\}$. We then exchange records in each pair.

Again, each noise insertion scheme has two configurations, and we use
\textbf{Low} and \textbf{High} to denote the intensity of two configurations. In
addition, we also mimic real-world noise, by introducing extra workload over the
CPU core when launching \pp. Particularly, in our exploitation launched
\S~\ref{subsec:real-world}, only victim, \spy, and coordinator processes occupy
the CPU core. At this step, however, we launch extra processes on the same CPU
core, which can likely introduce a considerable amount of noise on the trace.
Particularly, we consider the following three scenarios to systematically
explore noise introduced by real-world workload.

\noindent \underline{\textbf{Bzip2}.}~We pick the \textbf{bzip2} software from
SPEC CPU 2006 testsuite to compress a large file (100MB) simultaneously when the
\spy\ is launching \pp\ attack toward victim. SPEC CPU 2006 is a standard
CPU-intensive benchmark suite that is frequently used in security research.

\noindent \underline{\victima.}~We launch another victim software (e.g., another
\libjpeg) on the same core. This victim software will process the same input
simultaneously when the \spy\ is launching \pp\ attack toward victim.

\noindent \underline{\victimb.}~We launch another victim software (e.g., another
\libjpeg) on the same core. This victim software will process different inputs
simultaneously when the \spy\ is launching \pp\ attack toward victim.

To certain extent, noise introduced by these three workloads could have been
subsumed by our inserted noise (e.g., Gaussian or false cache hits/misses).
Nevertheless, we still launch this evaluation to thoroughly benchmark the noise
resiliency of our SCA.

\begin{table}[t]
  \caption{Quantitative evaluation results of human voice
  reconstructed from noisy side channels.}
  \label{tab:audio-noise}
  \centering
\resizebox{0.8\linewidth}{!}{
  \begin{tabular}{l|c|c|c}
    \hline
        Noise  & NA          & Low & High \\
    \hline
      \textbf{Gaussian} & 28.8\% & 20.4\%  & 17.0\% \\ 
      \textbf{Shift} & 28.8\% & 28.6\%  & 26.3\% \\
      \textbf{Removal} & 28.8\% & 25.0\%  & 24.5\% \\ 
    \hline
      \textbf{Leave hit/miss out} & 81.8\% & 80.7\%  & 80.0\% \\ 
      \textbf{False hit/miss} & 81.8\% & 78.5\%  & 3.4\% \\ 
      \textbf{Wrong order} & 81.8\% & 82.1\%  & 79.9\% \\ 
    \hline
      \textbf{Bzip2} & 81.8\% & \multicolumn{2}{c}{66.0\%} \\ 
      \victima\ & 81.8\% & \multicolumn{2}{c}{67.5\%} \\ 
      \victimb\ & 81.8\% & \multicolumn{2}{c}{55.8\%} \\ 
    \hline
  \end{tabular}
  }
\end{table}

\begin{table}[t]
  \caption{Quantitative evaluation results of dialog text
  reconstructed from noisy side channels.}
  \label{tab:text-noise}
  \centering
\resizebox{0.8\linewidth}{!}{
  \begin{tabular}{l|c|c|c}
    \hline
        Noise  & NA          & Low & High \\
    \hline
      \textbf{Gaussian} & 37.4\% & 35.5\%  & 30.8\% \\ 
      \textbf{Shift} & 37.4\% & 25.6\%  & 27.0\% \\
      \textbf{Removal} & 37.4\% & 32.8\%  & 32.6\% \\ 
    \hline
      \textbf{Leave hit/miss out} & 32.2\% & 32.1\%  & 32.1\% \\ 
      \textbf{False hit/miss} & 32.2\% & 32.2\%  & 32.2\% \\ 
      \textbf{Wrong order} & 32.2\% & 32.0\%  & 32.2\% \\ 
    \hline
      \textbf{Bzip2} & 32.2\% & \multicolumn{2}{c}{26.4\%} \\ 
      \victima\ & 32.2\% & \multicolumn{2}{c}{26.2\%} \\ 
      \victimb\ & 32.2\% & \multicolumn{2}{c}{25.9\%} \\ 
    \hline
  \end{tabular}
  }
\end{table}

We have reported key results in \S~\ref{subsec:noise} on exploiting
\libjpeg\ and reconstructing CelebA face photos. \F~\ref{fig:noise} reports the
corresponding qualitative evaluation results. The first column is the reference
input and the second column has images reconstructed from \pin-logged side
channel traces and \pp-logged cache side channels with no noise inserted. Each
row represents several configurations with the same intensity.
Despite the challenging noise insertion schemes, visually consistent contents
(e.g., gender, face orientation, eyes, mouth) between the reconstructed images
and reference inputs can still be observed. When perturbing \pin-logged traces,
reconstructed images under the \textbf{Round shifting} scheme retain better
visual appearances, indicating better noise resilience capability. The
\textbf{Removal} scheme, which extensively removes records in a trace, triggers
obvious quality degradation --- some perceptual contents become unaligned.
Similarly, \textbf{Gaussian noise}, particularly when $x = 0.5$, changes the
visual appearances. Nevertheless, as we clarified in the paper, visual
appearances change do not necessarily indicate attack accuracy degradation: as
reported in \T~\ref{tab:face-noise}, we still achieve a reasonably high attack
success rates even in front of Guarantee noise insertion.
After introducing noise into cache side channels collected by \pp\ (with three
schemes presented in the third and fourth rows of \F~\ref{fig:noise}), features
(e.g., hair style) in reconstructed face images are still primarily aligned with
images reconstructed without manual noise. Moreover, increasing noise intensity
only leads to negligible changes of facial features. We interpret that these
evaluations demonstrate the high resilience of our autoencoder framework toward
noisy settings. See our discussion below in \textbf{Noise Resilience Analysis}.
In addition, it is shown that stressing \pp\ with extra workload can induce
noticeable effect on the reconstructed images. Nevertheless, many perceptual
features are still retained in the reconstructed images. Again, we view the
results are generally consistent with our quantitative evaluation results in
\T~\ref{tab:face-noise}. See \textbf{Noise Resilience Analysis} below for further
discussion.

\T~\ref{tab:audio-noise} and \T~\ref{tab:text-noise} further present evaluation
results on exploiting \ffmpeg\ and \hunspell. Round shifting is more effective
in mitigating SCA toward text data. We note that round shifting (especially with
the \textbf{High} scheme) can presumably change the ``preceding words'' of a
to-be-predicted word and is thus more effective in disturbing our decoder of
discrete data (see \F~\ref{fig:workflow}). In contrast, the reconstructed image
and audio data are more resilient toward round shifting. Similar to the
\libjpeg\ evaluation reported in \T~\ref{tab:face-noise}, attack on \ffmpeg\ is
notably undermined in front of the \textbf{Removal} and \textbf{Gaussian noise}
schemes. As clarified in \S~\ref{subsec:noise}, these two schemes extensively
leave out or perturb data points on the logged trace (e.g.,
\textbf{Removal/High} removes 50\% of the records on a trace), show greater
influence on data reconstruction. Nevertheless, reasonably high attack accuracy
can still be achieved in the presence of perturbed side channel traces. Manifold
learning shows encouraging resilience toward noisy inputs. In addition, as
clarified in \appx~\ref{sec:input}, the logged side channel traces are lengthy
and highly \textit{sparse}, where only a few elements are informative and
secret-dependent. Therefore, the inserted noise does not necessarily break
informative data points. We also report promising results that noise on trace
collected by \pp\ (i.e., the middle three rows) imposes small influence in
undermining our exploitation, except the scheme \textbf{False hit/miss \& High}
on \ffmpeg. The similar encouraging observations can be found from noise
introduced by three real-world workload schemes as well.

\smallskip
\noindent \textbf{Noise Resilience Analysis}~In line with our discussion on
noise resilience offered by manifold learning concept
(\S~\ref{subsec:design-attack}) and neural trace encoder
(\appx~\ref{sec:input}), \S~\ref{subsec:noise} and this appendix section
empirically demonstrate the noise resilience of our attack. We now discuss the
noise resilience from the empirical perspective.
First, trace collected by \pp\ is very noisy with high stddev; when training
with such noisy traces, the autoencoder framework is ``enforced'' to obtain
higher generalization, but may sacrifice some accuracy. That is, the robustness
and noise resilience is indeed improved when training with side channel records
logged via \pp.
Second, as discussed in \appx~\ref{sec:input}, the logged side channel trace is
generally sparse. Suppose only ``1'', denoting a cache hit $\rightarrow$ miss
flip, in a logged trace (collected by \pp) contributes to reconstructing media
data, ``Leave out'' only drops a small portion of 1. For ``Wrong order'', since
CNN is translation-invariant (as we have introduced in \appx~\ref{sec:input}), suppose
all parameters in a kernel are $1$, and two exchanged records are in a $K \times
K$ region, then the output will not change, because ``convolution'' is
element-wise multiplication followed by a sum function.
Third, the results of \ffmpeg\ increase a lot on \pp\ but \libjpeg\ and
\hunspell\ decrease; this indicates that traces collected using \pp\ toward
\ffmpeg\ is more informative. Therefore, flipping $50\%$ of the records (the
\textbf{High} scheme) will more likely violate patterns learned by our
autoencoder framework despite the translation-invariance. This explains the accuracy
drop for noise insertion scheme \textbf{False hit/miss \& High}. Again, suppose
all parameters in a kernel are $1$, since in most of the cases, the number of 0 is
way larger than 1, flipping records will induce a noticeable impact on the
output of convolution operations.